\documentclass[acmsmall,screen,nonacm]{acmart}

\usepackage{multirow}

\newcommand{\Sec}[1]{\S\ \ref{#1}}

\begin{teaserfigure}
    \includegraphics[width=\textwidth]{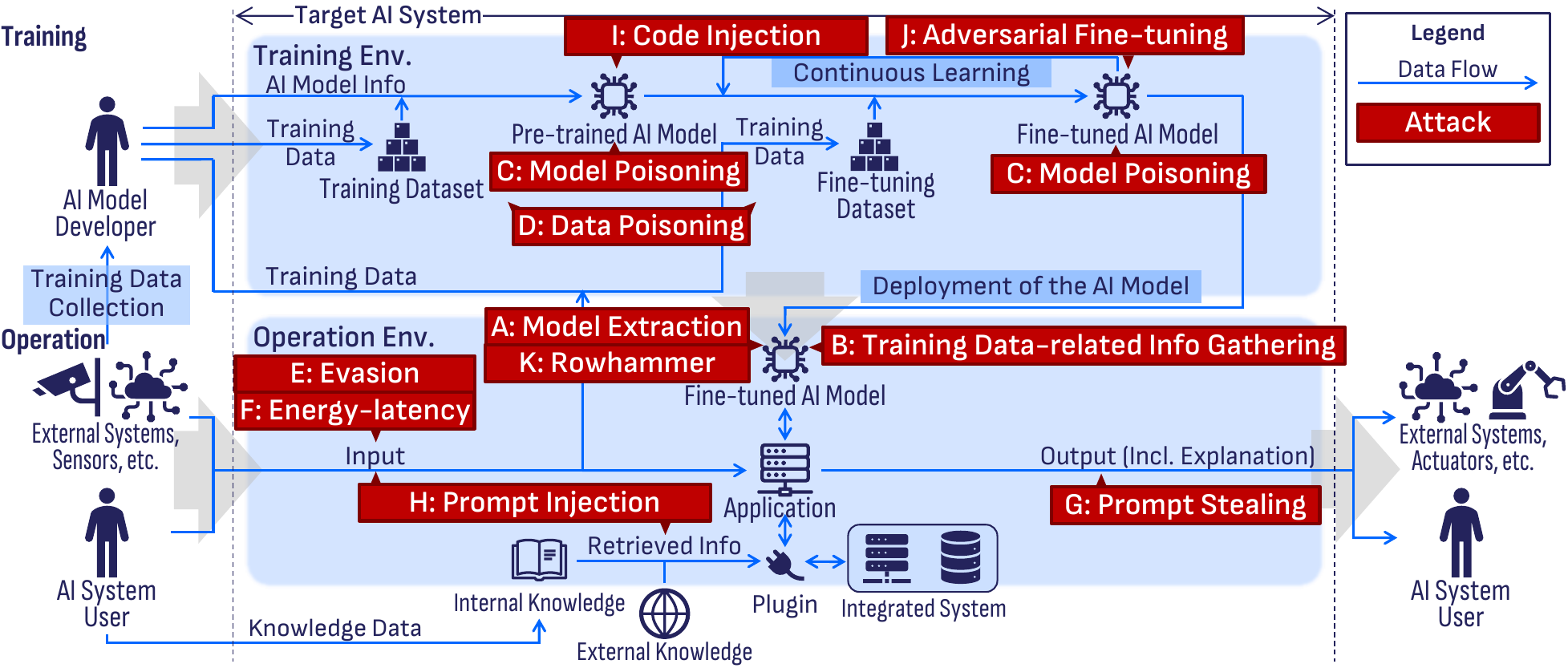}
    \caption{Overview of known eleven major attacks (A--K) on AI systems.}
    \label{fig:attacks}
    \Description{
    The figure shows a schematic of an AI system divided into training and operation environments. It identifies eleven attack types (labeled A to K), targeting components such as the training data, pre-trained and fine-tuned AI models, application inputs and outputs, internal and external knowledge sources, and integrated external systems. Data flow arrows illustrate interactions between these components, indicating precisely where each attack occurs within the AI system structure.
    }
\end{teaserfigure}
\begin{document}

\title{Securing AI Systems: A Guide to Known Attacks and Impacts}

\author{Naoto Kiribuchi}
\orcid{0009-0001-8843-5548}
\author{Kengo Zenitani}
\orcid{0000-0001-9983-8513}
\author{Takayuki Semitsu}
\orcid{0009-0001-4220-1092}
\affiliation{%
  \institution{Japan AI Safety Institute (J-AISI)}
  \city{Tokyo}
  \country{Japan}
}
\affiliation{%
  \institution{Information-technology Promotion Agency (IPA)}
  \city{Tokyo}
  \country{Japan}
}

\titlenote{This paper is an extended version of the document~\cite{JAISI25a} published by Japan AI Safety Institute.}

\begin{abstract}
Embedded into information systems, artificial intelligence (AI) faces security threats that exploit AI-specific vulnerabilities.
This paper provides an accessible overview of adversarial attacks unique to predictive and generative AI systems. 
We identify eleven major attack types and explicitly link attack techniques to their impacts---including information leakage, system compromise, and resource exhaustion---mapped to the confidentiality, integrity, and availability (CIA) security triad.
We aim to equip researchers, developers, security practitioners, and policymakers, even those without specialized AI security expertise, with foundational knowledge to recognize AI-specific risks and implement effective defenses, thereby enhancing the overall security posture of AI systems.
\end{abstract}




\maketitle

\section{Introduction}
The field of artificial intelligence (AI) has undergone rapid advancement in recent years, resulting in revolutionary changes across diverse domains, including healthcare, finance, and content generation. 
Innovations have not only expanded the potential applications of AI but also significantly increased its societal influence. 
As AI systems become integrated into critical infrastructures and decision-making processes, there has been a notable increase in global awareness and demand for ensuring their safe, reliable, and ethical operation --- collectively referred to as {\bf AI safety}.

In response to these mounting global concerns, Japan has assumed a pioneering role by spearheading the Hiroshima AI Process~\cite{G723}, a comprehensive policy framework that was adopted in 2023.
Concurrently, the United Kingdom and the United States established national AI Safety Institutes\footnote{The UK AI Safety Institute was renamed the AI {\it Security} Institute on February 14, 2025~\cite{DIST25}.}\footnote{The plan to transform the US AI Safety Institute into the US Center for AI Standards and Innovation (CAISI) was announced on June 3, 2025~\cite{USD25}.} to address AI-related risks. 
In alignment with these global developments, the Japan AI Safety Institute (J-AISI) was established in February 2024. 
It functions as Japan's central hub for AI safety, with the mission of aggregating technical expertise and promoting both domestic and international collaboration toward the responsible deployment of AI technologies. 
The secretariat of J-AISI is situated within the Information-technology Promotion Agency (IPA), thereby ensuring effective coordination and support for its activities~\cite{JAISI24a}.

Within the broader domain of AI safety, {\bf AI security} has emerged as a crucial subset focused explicitly on protecting AI systems from malicious actions designed to compromise their confidentiality, integrity, or availability. 
As AI systems handle increasingly sensitive data and perform critical societal functions, failures in AI security could result in severe consequences, including privacy violations, operational disruptions, economic damages, and threats to public safety. 
Consequently, a clear understanding of AI security risks and practical strategies to mitigate them is vital, not only for AI specialists but also for developers, security practitioners, and stakeholders responsible for deploying, operating, and managing AI systems across various sectors.

This paper contributes to AI security by providing broad stakeholders, especially security practitioners, including those without specialized AI security, with a comprehensive yet highly accessible overview of attacks against AI systems shown by Figure~\ref{fig:attacks}.
A comprehensive survey of the latest and most advanced attacks is beyond the scope of this paper.
Due to the abundance of research on AI security, it can be challenging for practitioners, especially those who are new to the field, to develop a coherent and comprehensive understanding of its landscape and underlying context. 
Accordingly, a practical guide is needed to support a comprehensive understanding of AI security.

The distinguishing features of this work are its emphasis on {\bf the impacts of attacks on AI systems} and intuitive visual summaries illustrating their relationships. 
By explicitly linking each established attack type with specific impacts on AI systems, this paper furnishes readers with actionable insights useful for security management practices. 
Developers and practitioners can leverage this structured knowledge to perform risk assessments or red-teaming exercises and identify critical attack vectors specifically relevant to their own AI systems.

Furthermore, this paper summarizes the referenced attack literature in Table~\ref{tab:big} organized by modality (e.g., text, image, or tabular), model type (e.g., neural network, logistic regression, or decision tree), and attacker's capabilities.
This table offers readers practical guidance for evaluating realistic risk scenarios within broader cybersecurity contexts.

In summary, the primary contributions of this paper are as follows:
\begin{itemize}
    \item Identifying eleven attack types on AI systems by incorporating globally established references from various AI relevant stakeholders.
    \item Explicitly linking each attack type to its specific impacts on AI systems, supporting practical risk assessments, including red-teaming exercises.
    \item Identifying attacker's capabilities associated with each attack, supporting realistic evaluations of AI security scenarios and risks.
    \item Providing intuitive summaries that facilitate an easy understanding of attack mechanisms and their implications, suitable for developers and security practitioners without specialized AI security expertise.
    \item Organizing referenced attack literature according to data modalities and AI model types, serving as a clear roadmap for readers pursuing further in-depth study.
\end{itemize}

The remainder of this paper is organized as follows. 
Section~\ref{sec:related} reviews related works in AI security, contextualizing our contributions within existing frameworks and literature. 
Section~\ref{sec:scope} details the AI system architecture and defines the scope of the attacks considered in this paper. 
Section~\ref{sec:overview} introduces our categorization of eleven major attack types and discusses their potential impacts on AI systems, mapped to the CIA security triad. 
Section~\ref{sec:attacks} provides in-depth descriptions of each identified attack, including its mechanisms, attacker capabilities, and affected AI modalities and model types, supported by examples from the literature. 
Finally, we conclude the paper by summarizing our key findings and suggesting directions for future research in Section~\ref{sec:conclusions}.

\section{Related Works}
\label{sec:related}
We review related works on adversarial threats to AI systems, highlighting their contributions and limitations to contextualize this paper within AI security.
We deeply respect the significant contributions of the existing frameworks and literature discussed here, which have led the global conversation on AI security. 
The purpose of this paper is not to negate or replace these works but rather to provide additional insights that can contribute to their practical application.
By highlighting the differences and supplementary perspectives, we aim to promote understanding among these works, point out key attack characteristics especially from the viewpoint of security practitioners, and contribute to the future development of international discussions.

\subsection{Academic Surveys}
Several academic surveys have systematically analyzed adversarial threats and defense methods relevant to predictive AI systems. 

Kong et al. provide a comprehensive classification that focuses on modality-based attacks, specifically targeting image, text, and malware domains, detailing typical attack algorithms and their corresponding defenses~\cite{KXW+21}. 
Oseni et al. analyze adversarial attacks, providing an overview of machine learning and mathematical explanations of attacks~\cite{OMJ+21}.
Additionally, Hu et al. categorized adversarial attacks based on the AI system lifecycle, mapping threats to various developmental stages (e.g., data collection, preprocessing, model training, and inference) and highlighting lifecycle-specific vulnerabilities and defenses~\cite{HKQ+21}. 
Chen and Babar also adopt a lifecycle perspective but extend their analysis to software engineering practices and secure development methodologies, thereby offering guidance on the construction of robust AI systems~\cite{CB24}.

Despite these comprehensive surveys, they predominantly address predictive AI systems, lacking insights into generative AI systems. Recent surveys have begun to address this gap by focusing on large language models (LLMs), a prominent category of generative AI. 

Yao et al. categorized LLM vulnerabilities into AI-Inherent and Non-AI-Inherent, identifying risks such as model extraction and remote code execution~\cite{YDX+24}. 
Shayegani et al. systematically surveyed adversarial attacks on LLMs, including multi-modal attacks~\cite{SMF+23}, while Dong et al. specifically examined conversation safety, including attack methodologies and corresponding defenses~\cite{DZY+24}. 
Yan et al. provided a detailed analysis of privacy risks and countermeasures for LLMs~\cite{YLX+25}, and Das et al. further highlighted various security threats, such as jailbreaking and data poisoning~\cite{DAW25}. 
In contrast, Sun et al. recently surveyed generative adversarial networks (GANs) and variational auto-encoders (VAEs)~\cite{SZZ+23}, focusing primarily on model-centric vulnerabilities.
Their analysis addresses model-centric vulnerabilities, barely considering broader system integration issues.

As shown above, existing surveys either focus on predictive models or generative models at the model level, leaving open the need for comprehensive studies encompassing systemic vulnerabilities within generative AI systems.
Also, these academic surveys are intended for AI researchers rather than practitioners and do not explicitly contrast their findings with practical frameworks, such as NIST AI 100-2e2025 and MITRE ATLAS\texttrademark.

In the following subsections, we review the prevalent frameworks from a practitioner's perspective, highlighting their differences from and complementary roles to the academic surveys discussed above.

\subsection{NIST AI 100-2e2025}
\newcommand{\nistaml}[1]{\texttt{NISTAML.#1}}

A notable recent work is the updated report titled “Adversarial Machine Learning: A Taxonomy and Terminology of Attacks and Mitigations” (NIST AI 100-2e2025)~\cite{VOF+25} by the National Institute of Standards and Technology (NIST).
Compared to its predecessor~\cite{VOF+24}, this latest edition expands its literature review to over 400 references.
This literature analysis thoroughly explicates their historical evolution from initial concepts to the latest research. 
The report also introduces a glossary that provides definitions for AI security terminology.
Our paper refers to the NIST report, and the numerous sources cited therein have played a substantial role in our analysis.

Although the newly introduced index has improved navigation, the report's taxonomy remains complex.
NIST classifies attacks first by the targeted AI system type (i.e., predictive and generative AI systems) and then by the attacker's objectives (i.e., availability, integrity, privacy, misuse, and supply chain attacks). 
However, many attack methods overlap across these classifications, resulting in repetitive descriptions.
For instance, Indirect Prompt Injection (\nistaml{015}) and Misaligned Outputs (\nistaml{027}) describe identical attacks from different objectives.
The specific attack, Model Poisoning, has different IDs (\nistaml{011}, \texttt{026}, and \texttt{051}) across various AI system types and objectives.  
These overlaps can obscure the practical utility of taxonomy, especially for practitioners.

\begin{table}[bt]
    \centering
    \caption{Correspondence Between This Paper and NIST AI 100-2e2025}
    \label{tab:NIST25}
    \begin{tabular}{ll}
    \toprule
    {\bf Attacks in This Paper} & {\bf Attacks in NIST AI 100-2e2025~\cite{VOF+25}}\\
    \midrule
    \Sec{sec:a} A: Model Extraction & \nistaml{031} Model Extraction\\ \cmidrule(lr){1-2}
    \Sec{sec:b} B: Training Data-related & \\
    \ \ \ \ \ \ \ \ \ \ \ \ Information Gathering & \\    
    \ \ \Sec{sec:b1} B1: Membership Inference & \nistaml{033} Membership Inference\\
    \ \ \Sec{sec:b2} B2: Attribute Inference & \nistaml{032} Reconstruction\\
    \ \ \Sec{sec:b3} B3: Property Inference & \nistaml{034} Property Inference\\
    \ \ \Sec{sec:b4} B4: Model Inversion & \nistaml{032} Reconstruction\\
    \ \ \Sec{sec:b5} B5: Data Reconstruction & \nistaml{032} Reconstruction\\
    \ \ \Sec{sec:b6} B6: Data Extraction & \nistaml{038} Data Extraction\\ \cmidrule(lr){1-2}
    \Sec{sec:c} C: Model Poisoning & \nistaml{011,026,051} Model Poisoning\\ \cmidrule(lr){1-2}
    \Sec{sec:d} D: Data Poisoning & \nistaml{012} Clean-label Poisoning\\
     & \nistaml{013} Data Poisoning\\
     & \nistaml{021} Clean-label Backdoor\\
     & \nistaml{023} Backdoor Poisoning\\
     & \nistaml{024} Targeted Poisoning\\ \cmidrule(lr){1-2}
    \Sec{sec:e} E: Evasion & \nistaml{022} Evasion\\
     & \nistaml{025} Black-box Evasion\\ \cmidrule(lr){1-2}
    \Sec{sec:f} F: Energy-latency & \nistaml{014} Energy-latency\\ \cmidrule(lr){1-2}
    \Sec{sec:g} G: Prompt Stealing & \nistaml{038} Data Extraction\\ \cmidrule(lr){1-2}
    \Sec{sec:h} H: Prompt Injection & \nistaml{015} Indirect Prompt Injection\\
     & \nistaml{018} Prompt Injection\\
     & \nistaml{027} Misaligned Outputs\\
     & \nistaml{035} Prompt Extraction\\
     & \nistaml{036} Leaking Info from User Interactions\\
     & \nistaml{039} Compromising Connected Resources\\ \cmidrule(lr){1-2}
    \Sec{sec:i} I: Code Injection & \multicolumn{1}{c}{-} \\ \cmidrule(lr){1-2}
    \Sec{sec:j} J: Adversarial Fine-tuning & \multicolumn{1}{c}{-} \\ \cmidrule(lr){1-2}
    \Sec{sec:k} K: Rowhammer & \nistaml{031} Model Extraction\\
    \bottomrule
    \end{tabular}
\end{table}

In contrast, we deliberately adopts a simplified taxonomy.
Rather than categorizing attacks according to the type of AI system targeted or the attacker’s objectives, we classify attacks according to their underlying technical methods.
By explicitly separating each attack mechanism from the type of AI targeted or its intended objective, this paper remains intuitive and designed to avoid overlaps, thereby focusing on supporting easy comprehension and practical application.

Table~\ref{tab:NIST25} shows the mapping of the attacks in this paper to those in the NIST report.
This paper includes several attacks not covered in the report that we believe warrant attention, such as {\bf I: Code Injection} and {\bf J: Adversarial Fine-tuning}.
While we refer to the NIST report's literature review and detailed technical analysis, this paper introduces simplified categorizations and intuitive visualizations to complement the report's thorough yet duplicative approach.

\newcommand{\aml}[1]{\href{https://atlas.mitre.org/techniques/AML.#1}{\texttt{AML.#1}}}
\begin{table}[bt]
    \centering
    \caption{Correspondence Between This Paper and MITRE ATLAS\texttrademark}
    \label{tab:ATLAS}
    \begin{tabular}{ll}
    \toprule
    {\bf Attacks in This Paper} & {\bf Techniques in MITRE ATLAS\texttrademark}~\cite{Mit21} (Data v4.8.0)\\
    \midrule
    \Sec{sec:a} A: Model Extraction & \aml{T0024.002} Extract ML Model\\ \cmidrule(lr){1-2}
    \Sec{sec:b} B: Training Data-related & \aml{T0024} Exfiltration via ML Inference API\\
    \ \ \ \ \ \ \ \ \ \ \ \ Information Gathering & \\
    \ \ \Sec{sec:b1} B1: Membership Inference & \aml{T0024.000} Infer Training Data Membership\\
    \ \ \Sec{sec:b2} B2: Attribute Inference & \multicolumn{1}{c}{-} \\
    \ \ \Sec{sec:b3} B3: Property Inference & \multicolumn{1}{c}{-} \\
    \ \ \Sec{sec:b4} B4: Model Inversion & \aml{T0024.001} Invert ML Model\\
    \ \ \Sec{sec:b5} B5: Data Reconstruction & \aml{T0024.001} Invert ML Model\\
    \ \ \Sec{sec:b6} B6: Data Extraction & \aml{T0057} LLM Data Leakage \\ \cmidrule(lr){1-2}
    \Sec{sec:c} C: Model Poisoning & \aml{T0010.003} ML Supply Chain Compromise: Model\\
     & \aml{T0018.000} Backdoor ML Model: Poison ML Model\\
     & \aml{T0019} Publish Poisoned Datasets\\
     & \aml{T0058} Publish Poisoned Models\\ \cmidrule(lr){1-2}
    \Sec{sec:d} D: Data Poisoning & \aml{T0010.002} ML Supply Chain Compromise: Data\\
     & \aml{T0018.000} Backdoor ML Model: Poison ML Model\\
     & \aml{T0020} Poison Training Data\\
     & \aml{T0031} Erode ML Model Integrity\\
     & \aml{T0059} Erode Dataset Integrity\\ \cmidrule(lr){1-2}
    \Sec{sec:e} E: Evasion & \aml{T0015} Evade ML Model\\
     & \aml{T0031} Erode ML Model Integrity\\
     & \aml{T0043} Craft Adversarial Data\\ \cmidrule(lr){1-2}
    \Sec{sec:f} F: Energy-latency & \aml{T0034} Cost Harvesting\\
     & \aml{T0043} Craft Adversarial Data\\ \cmidrule(lr){1-2}
    \Sec{sec:g} G: Prompt Stealing & \multicolumn{1}{c}{-} \\ \cmidrule(lr){1-2}
    \Sec{sec:h} H: Prompt Injection & \aml{T0051} LLM Prompt Injection\\
     & \aml{T0054} LLM Jailbreak\\
     & \aml{T0056} Extract LLM System Prompt\\
     & \aml{T0070} RAG Poisoning\\ \cmidrule(lr){1-2}
    \Sec{sec:i} I: Code Injection & \aml{T0011.000} User Execution: Unsafe ML Artifacts\\
     & \aml{T0018.001} Backdoor ML Model: Inject Payload\\ \cmidrule(lr){1-2}
    \Sec{sec:j} J: Adversarial Fine-tuning & \multicolumn{1}{c}{-} \\ \cmidrule(lr){1-2}
    \Sec{sec:k} K: Rowhammer & \multicolumn{1}{c}{-} \\
    \bottomrule
    \end{tabular}
\end{table}

\subsection{MITRE ATLAS\texttrademark}
Another relevant resource for categorizing adversarial threats against AI systems is MITRE ATLAS\texttrademark ~\cite{Mit21}, a structured knowledge base designed specifically for adversarial attacks on AI systems. 
Based on MITRE ATT\&CK\textsuperscript{\textregistered}~\cite{Mit13}, the well-known framework in the cybersecurity community, ATLAS systematically categorizes real-world attack scenarios according to hierarchical concepts of tactics (attackers' goals) and techniques (methods used to achieve those goals).

A key benefit of ATLAS is its practical orientation, which explicitly maps attacks to specific threat scenarios observed in real-world deployments. 
It provides actionable insights, particularly for cybersecurity professionals, by linking detailed descriptions of adversarial tactics and techniques to documented examples. 
In addition, ATLAS effectively illustrates attacker capabilities and objectives, facilitating realistic risk assessments.

Table~\ref{tab:ATLAS} illustrates the correspondence between our attack categorization and ATLAS. 
This work elucidates attacks not shown in ATLAS, including {\bf B2: Attribute Inference}, {\bf B3: Property Inference}, {\bf G: Prompt Stealing}, {\bf J: Adversarial Fine-tuning}, and {\bf K: Rowhammer} attacks.
Our work complements ATLAS's practical nature by systematically integrating relatively recent findings, thereby providing broader coverage of emerging threats.

\clearpage
\subsection{OWASP Top 10 for LLM Applications}
A further influential resource addressing security risks in LLMs is the OWASP Top 10 for LLM Applications (Version 2025)~\cite{WD24}. 
This document is published by the Open Web Application Security Project (OWASP), an international organization recognized worldwide for its contributions to cybersecurity, most notably the famous OWASP Top 10 for web application security~\cite{SGS+21}.

In the context of AI security, the OWASP Top 10 for LLM identifies the most critical vulnerabilities, including Prompt Injection, Sensitive Information Disclosure, Supply Chain Vulnerabilities, System Prompt Leakage, and emerging concerns related to multimodal interactions. 
The primary benefit of the OWASP approach is in its practicality and clarity, effectively translating complex AI security concepts into straightforward, actionable guidance.

While the OWASP framework provides clear and practical insights into the immediate security risks of LLM applications, it does not aim to provide a detailed categorization of attacker tactics and techniques. 
Consequently, it provides limited insight into attacker capabilities and the technical foundations of various attack methods. 
Additionally, OWASP primarily focuses on LLM-specific vulnerabilities, which may limit its applicability to broader AI system architectures and modalities.

Our categorization complements OWASP by providing a more systematic structure spanning a broader range of AI modalities, thereby supporting practical and analytical use cases.

\subsection{AI Quality Management Guideline}
A further relevant resource is the AI Quality Management Guideline (AIQM)~\cite{AIQM4}, whose fourth edition significantly expanded its coverage of AI security, particularly in Section 10.3. 
The AIQM focuses on the systematization of machine learning-specific threats, vulnerabilities, and security controls, providing substantial insights based on an extensive review of the academic literature and elaborating technical details in the paper~\cite{KMK+23}.
Our paper also cites many of the same foundational sources and acknowledges the contributions of the AIQM.

Due to its lifecycle or process-oriented description style, AIQM structures its discussions through deeply nested subsections that spread explanations of individual attacks across multiple contexts, such as system impacts, attacker goals, and specific techniques. 
This structure could make it difficult for readers to gain a coherent understanding of specific attack scenarios. 

In contrast, our work provides a simpler representation of attacks and thus complements the detailed but dispersed approach by AIQM.

\subsection{Other Guidelines and Frameworks}
While various organizations have published guidelines and frameworks addressing risk management and secure development practices for AI systems~\cite{Nat23a,Nat23b,Nat24,NCI+23,APP+23,NPB+23}, specific AI attack methodologies are generally beyond the scope of these publications.

Finally, another relevant resource is the Guide to Red Teaming Methodology on AI Safety~\cite{JAISI24b}, published by J-AISI. 
This guide is a preliminary effort by our institute to provide practical and structured recommendations for planning, preparing, and conducting red teaming exercises for LLM systems. 
It briefly introduces representative attacks such as prompt injection  as illustrative examples to support practical system evaluation and risk mitigation.
The guide's primary objective is to outline the overall red teaming process. 
Accordingly, its coverage of adversarial threats remains limited, offering neither detailed explanations nor a comprehensive taxonomy of attacks.
This paper complements the guide by providing deeper technical explanations, a broader range of adversarial threats, and detailed insights into attack mechanisms and impacts. 
Our categorization serves as a handy resource for identifying and prioritizing adversarial threats, further enhancing the practical red teaming methodologies proposed by J-AISI.

\section{AI System and Scope}
\label{sec:scope}
To establish a foundation for the subsequent discussion, this section defines a generalized architecture for AI systems and outlines the scope of attacks that will be examined in this paper.
\subsection{AI System Architecture}
\begin{figure}[tb]
    \centering
    \includegraphics[width=\textwidth]{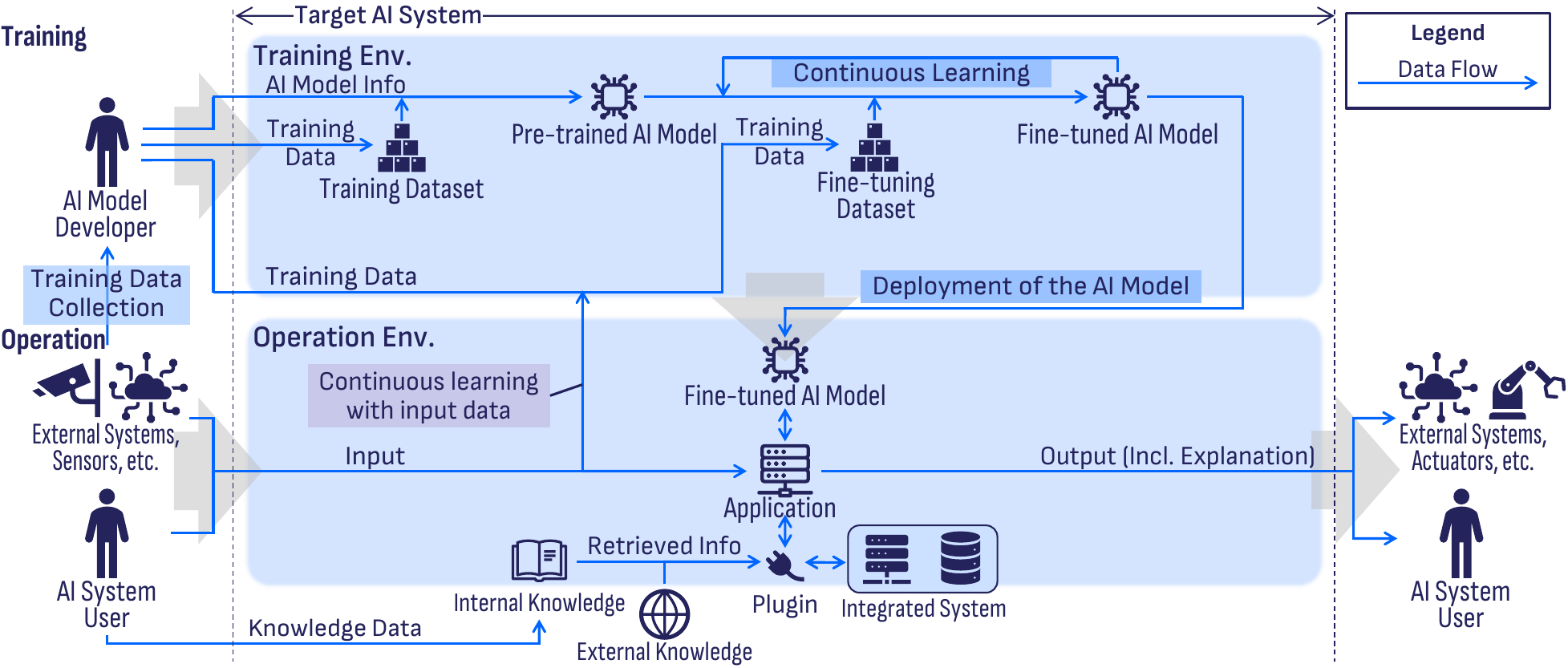}
    \caption{Assumed AI system architecture. 
    The diagram illustrates the seamless data flow from model training to deployment and ongoing operation. 
    In the training environment, we assume there is continuous or periodic fine-tuning that leverages data obtained from operational activities as well as training data provided by AI model developers. 
    In the operation environment, we assume the use of internal or external knowledge for information retrieval and integrated systems connected via plugins. While these elements are often used with LLMs, the AI model is not limited to them.}
    \Description{
    The figure illustrates an AI system structured into two environments: training and operation. The training environment includes a pre-trained AI model that is fine-tuned using collected datasets and data from continuous learning during operation. The operation environment comprises the deployment of the fine-tuned AI model, which interacts with user inputs, external sensors, actuators, and knowledge sources (both internal and external), integrated via plugins. Arrows represent data flows from training data through model fine-tuning, deployment, and operational input-output processes.
    }
    \label{fig:system}
\end{figure}
An AI system is an information system that incorporates AI models as components.
Figure~\ref{fig:system} illustrates the generalized AI system architecture assumed in this paper.
The system is conceptually divided into two primary environments:
\begin{itemize}
    \item {\bf Training Environment (upper part of Figure~\ref{fig:system}):} In this environment, AI models undergo training and fine-tuning using datasets that include data collected during real-world operations. 
    We also consider scenarios involving continuous or periodic fine-tuning, leveraging data obtained from operational activities.
    \item {\bf Operation Environment (lower part of Figure~\ref{fig:system}):} Here, the trained AI model is deployed and integrated into an application framework. 
    Data inputs from external systems, sensors, or human users are processed by the model, producing outputs that could be directed toward the same or different external entities.
\end{itemize}
Data generally flows from left to right. 
Detailed descriptions of each component illustrated in Figure~\ref{fig:system} are provided in Appendix~\ref{appendix:components}.
The system architecture considers the integration of internal and external knowledge through plugins or retrieval-augmented generation (RAG), primarily associated with LLMs. 
Note that our discussion applies beyond LLM-based systems.

This abstracted model simplifies complex real-world architectures while retaining enough detail to analyze vulnerabilities realistically and practically. 
Nonetheless, it is not a one-size-fits-all framework. 
Specific advanced machine learning architectures, such as federated learning, do not fit into this abstraction and necessitate dedicated analyses of unique vulnerabilities.

\subsection{Scope of the Attacks Considered in This Paper}
Securing AI systems inherently involves traditional cybersecurity measures to mitigate conventional cyber threats. 
In this paper, we presuppose that general cybersecurity practices such as protection against unauthorized network access are already in place. 
Thus, we explicitly focus on {\it AI-specific attacks involving interference with AI model training or inference, as well as the manipulation or misuse of AI model inputs and outputs on the targeted system}.

We do not address attacks that exploit standard cybersecurity vulnerabilities, such as unauthorized access resulting in the theft of AI models unless these attacks explicitly interfere with the AI-specific processes.

The volume of research on attacks against AI systems has grown considerably, making it impractical to survey all existing literature exhaustively. 
Thus, this paper does not assert completeness in that sense. 
Instead, we have incorporated both well-known foundational works, as delineated in Section \ref{sec:related}, and recent advancements from academic conferences. 

\section{Attack Types and Their Impacts}
\label{sec:overview}
This section describes our classification of attacks targeting AI systems and explains the relationship between these attacks and their impacts on AI systems.

\subsection{Methodology}
We derived the types of attacks and impacts summarized in this paper from a review and integration of the sources discussed in Section~\ref{sec:related}, following the procedure outlined below:
\begin{enumerate}
    \item {\bf Identification and Extraction:} We identified and extracted relevant descriptions corresponding to potential attacks and impacts from the referenced materials in Section~\ref{sec:related}. 
    We also extracted unique attack methods identified by the authors in recent papers~\cite{CTZ+24,LWX+24,NMF+24,YYC+24,ZWZ+24} from international conferences, such as the ACM CCS.
    \item {\bf Mapping to the AI System Architecture:} We mapped the attack methods extracted above to the generalized AI system architecture presented in Figure~\ref{fig:system}, taking a technical perspective, such as the attack surface and targeted components.
    \item {\bf Grouping:} We grouped similar or related attacks along with the mapping in (2) into attack types (see Figure~\ref{fig:attacks}). 
    The impacts extracted in (1) were also grouped based on the components affected in the AI system architecture and the perspective of the CIA triad (see Figure~\ref{fig:impacts}). 
    \item {\bf Linking Attacks to Impacts:} We linked each attack type to its impacts on the AI system based on referenced materials (see Figure~\ref{fig:relations} and Table ~\ref{tab:big}).
\end{enumerate}
The following subsections explain the final categorization of attacks and impacts obtained through this methodology.

\subsection{Categorization of Attacks Against AI Systems}
As shown in Figure~\ref{fig:attacks}, this paper classifies attacks against AI systems into eleven types, labeled A through K. 
Each type represents a unique attack vector linked to AI-specific vulnerabilities in AI systems.
Detailed explanations of each attack type are provided in Section~\ref{sec:attacks}.

The naming convention of attacks adopted in this work largely adheres to established academic and industry terminology, thereby enhancing readability and comprehension.
Among the attack categories delineated in this work, ``Training Data–related Information Gathering'' (Attack B) is newly introduced as a class that encompasses attacks intended to extract or infer information about a model's training dataset.

For multiple established terms describing the same attack from different perspectives, we primarily choose a naming based on the technical mechanism of the attack.
For instance, while both ``Prompt Injection'' and ``Jailbreak'' frequently denote the same attack, we adopt ``Prompt Injection'' because it reflects the underlying mechanism, whereas ``Jailbreak'' emphasizes the result bypassing safeguards.
This naming convention provides a precise understanding by recognizing that a single attack technique can fulfill multiple adversarial objectives. 
We reflect this insight in our taxonomy, which separates attacks and their impacts.

In certain instances, prevailing industry terminology does not adequately address attacker methodologies, and there is a lack of established alternatives.
We retain these terms rather than introducing new ones to ensure familiarity and maintain compatibility with existing literature.

\subsection{Relationships between Attacks and Impacts on AI Systems}

\begin{figure}[tb]
    \centering
    \includegraphics[width=\textwidth]{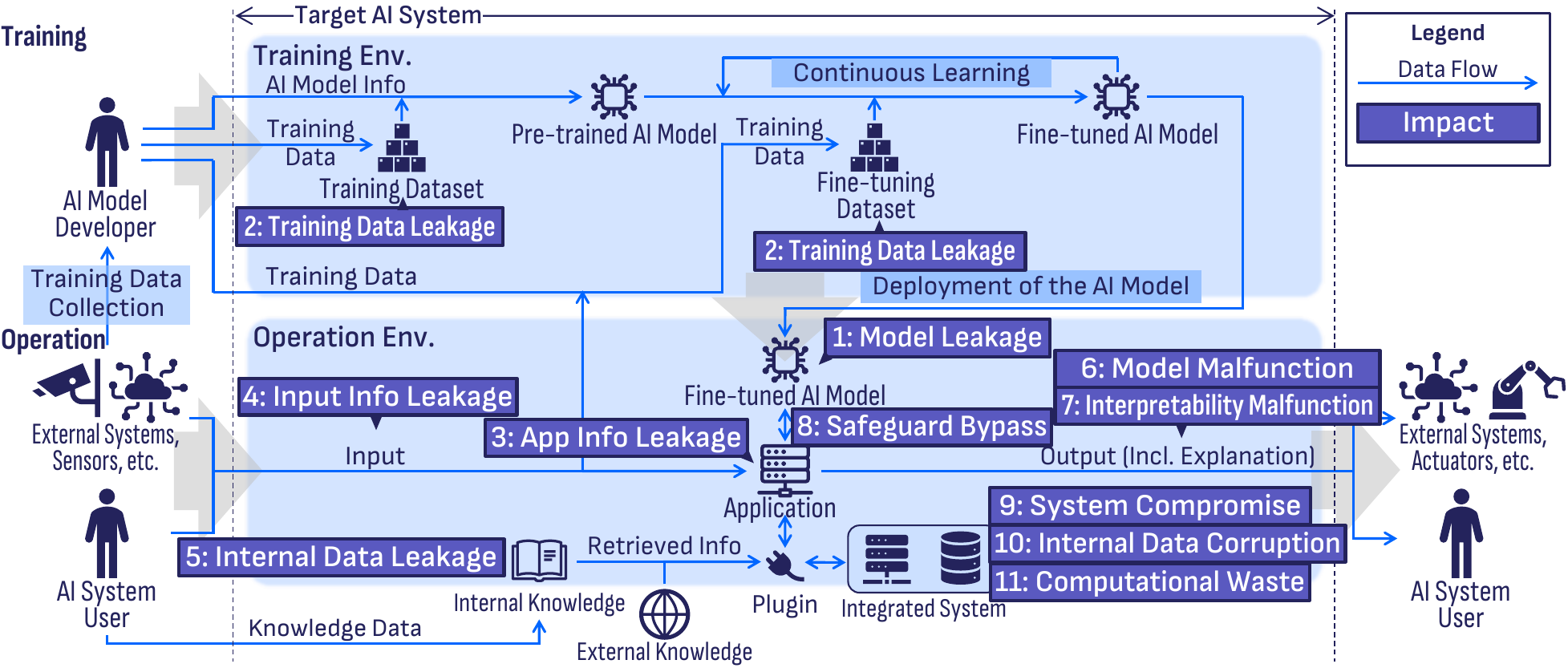}
    \caption{Overview of impacts caused by known attacks (described in Figure~\ref{fig:attacks}) on AI systems.}
    \Description{
    The figure depicts eleven distinct impacts caused by adversarial attacks within an AI system. Impacts include leakage of the AI model, training data, application information, input data, and internal data, as well as malfunctions in model functionality and interpretability. Additional impacts shown are bypassing of safeguards, system compromise, corruption of internal data, and computational waste. Each impact is located on the diagram according to the specific component it affects, clearly distinguishing between the training and operational environments. Arrows represent data flows, linking system components and illustrating the points at which these impacts occur.
    }
    \label{fig:impacts}
\end{figure}

\begin{figure}[tb]
    \centering
    \includegraphics[width=\textwidth]{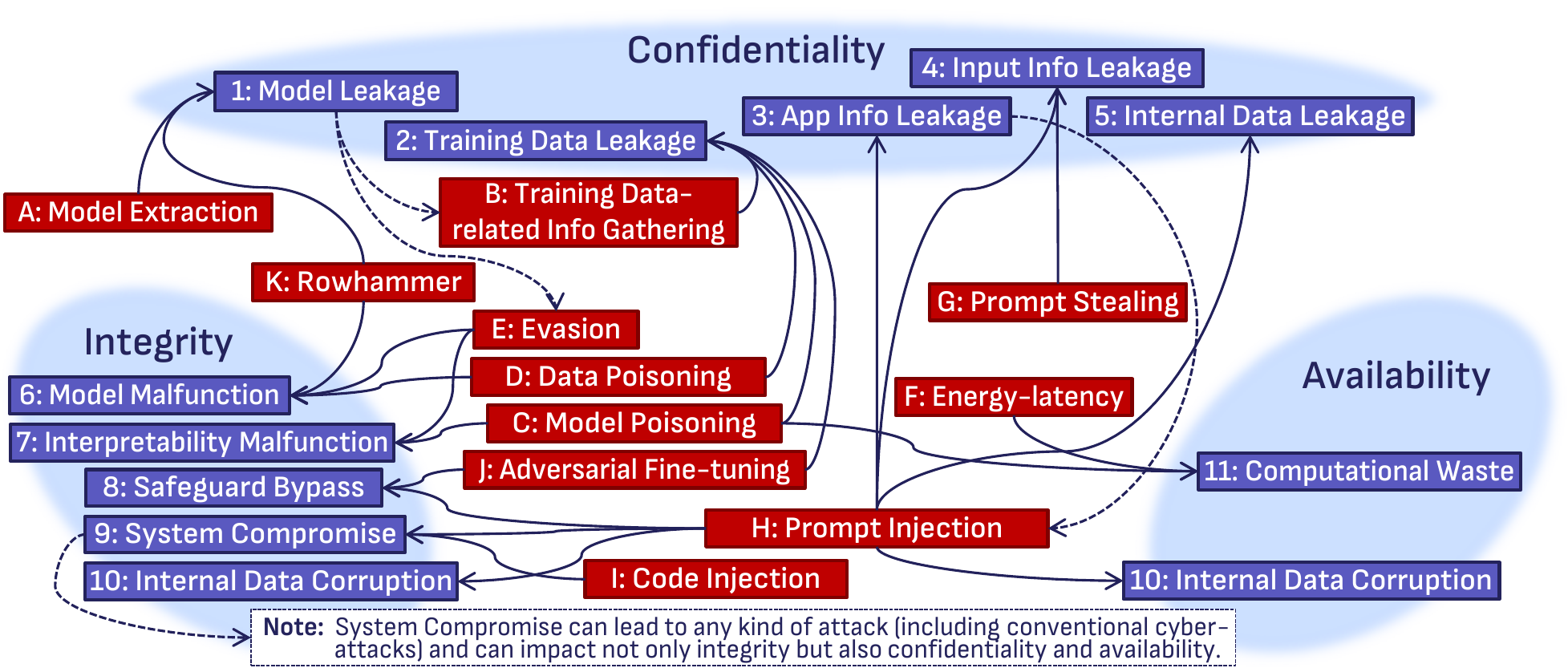}
    \caption{Relationship between the known attacks and their impacts on AI systems. The dashed line indicates the potential for use in an attack at the arrow's endpoint.}
    \label{fig:relations}
    \Description{
    The figure maps eleven identified attacks (labeled A to K) to their resulting impacts on AI systems, categorized under confidentiality, integrity, and availability. Attacks such as model extraction and data poisoning are linked to specific impacts including model leakage, training data leakage, model malfunction, and computational waste. This visual mapping clearly associates each attack type directly with its potential consequences.
    }
\end{figure}

In Figure~\ref{fig:attacks}, we delineated a wide array of attacks on AI systems.
Figure~\ref{fig:impacts} illustrates the various impacts that these attacks can have on AI systems. 
Eleven types of impacts have been identified, ranging from model leakage and data leakage to operational failures, safeguard bypasses, and system compromise. 
Appendix~\ref{appendix:impacts} gives additional explanations for these eleven impacts.

Figure~\ref{fig:relations} visually maps the relationships between attack types (shown in Figure~\ref{fig:attacks}) and their corresponding impacts (shown in Figure~\ref{fig:impacts}) on AI systems from the perspectives of confidentiality, integrity, and availability, known as the CIA triad of security. 
Dashed arrows indicate impacts that can facilitate subsequent attacks, thus capturing potential cascading risks. 
Notably, impact ``9: System Compromise'' is highlighted due to its potential to enable various subsequent attacks, including conventional cybersecurity threats, affecting all aspects of the CIA triad.

\begin{table}[tb]
    \centering
    \caption{Reverse Lookup Table to Identify Attacks from Their Impacts}
    \label{tab:I2A}
    \begin{tabular}{clll}
    \toprule
    {\bf CIA} \hspace{-0.8em} & {\bf Impact} & {\bf Attack} & {\bf Section} \\
    \midrule
    \multirow{7}{*}{\rotatebox[origin=c]{90}{{\bf C}onfidentiality}$\left\{ 
        \begin{tabular}{l} \\ \\ \\ \\ \\ \\ \\ \end{tabular}
    \right.$ \hspace{-2.8em}}
     & 1: Model Leakage & A: Model Extraction & \Sec{sec:a} \\
     &  & K: Rowhammer & \Sec{sec:k} \\
     & 2: Training Data Leakage & B: Training Data-related Info Gathering & \Sec{sec:b} \\
     & 3: App Information Leakage & H: Prompt Injection & \Sec{sec:h} \\
     & 4: Input Information Leakage & G: Prompt Stealing & \Sec{sec:g} \\
     &  & H: Prompt Injection & \Sec{sec:h} \\
     & 5: Internal Data Leakage & H: Prompt Injection & \Sec{sec:h} \\
    \multirow{10}{*}{\rotatebox[origin=c]{90}{{\bf I}ntegrity}$\left\{ 
        \begin{tabular}{l} \\ \\ \\ \\ \\ \\ \\ \\ \\ \\ \end{tabular}
    \right.$ \hspace{-2.8em}}
     & 6: Model Malfunction & D: Data Poisoning & \Sec{sec:d} \\
     &  & E: Evasion & \Sec{sec:e} \\
     &  & K: Rowhammer & \Sec{sec:k} \\
     & 7: Interpretability Malfunction & C: Model Poisoning & \Sec{sec:c} \\
     &  & E: Evasion & \Sec{sec:e} \\
     & 8: Safeguard Bypass & H: Prompt Injection & \Sec{sec:h} \\
     &  & J: Adversarial Fine-tuning & \Sec{sec:j} \\
    \multirow{5}{*}{\rotatebox[origin=r]{90}{{\bf A}vailability}} \ 
     & 9: System Compromise & H: Prompt Injection & \Sec{sec:h} \\
     &  & I: Code Injection & \Sec{sec:i} \\
    \multirow{3}{*}{\ $\left\{ 
        \begin{tabular}{l} \\ \\ \\ \end{tabular}
    \right.$ \hspace{-2.3em}}
     & 10: Internal Data Corruption & H: Prompt Injection & \Sec{sec:h} \\
     & 11: Computational Waste & C: Model Poisoning & \Sec{sec:c} \\
     & & F: Energy-latency & \Sec{sec:f} \\
    \bottomrule
    \end{tabular}
\end{table}

Table~\ref{tab:I2A} provides a reverse-lookup capability to identify attack types from their impacts on AI systems. 
Each listed attack is hyperlinked to detailed discussions in Section \ref{sec:attacks}. 
This table can be used for risk assessment and red-teaming activities, as it directly links observed impacts to relevant attacks that need to be addressed.

\section{Mechanisms of Each Attack Type}
\label{sec:attacks}
\newcommand{\AtkTabNotice}{{\rm This is summary from literature and does not limit modality or target. IDs and abbreviations are defined at the beginning of Section \ref{sec:attacks}}}
\begin{table}[tbp]
    \centering
    \caption{Known attacks and their impacts on AI systems. \AtkTabNotice}
    \label{tab:big}
    \scriptsize
    \begin{tabular}{llll}
    \toprule
    {\bf Attack} & {\bf Impact} & {\bf Attacker's Capability} & {\bf Modality and AI Model Type}\\
    \midrule
    \Sec{sec:a} A: Model & 1: Model Leakage & Massive queries to the model & Image: NN~\cite{OSF19a,OSF19b,JSM+19,CBB+18}, LR~\cite{TZJ+16} \\
    \ \ \ \ \ \ \ \ \ \ \ \ \ Extraction & & & Tabular: RR~\cite{WG18}, DT~\cite{TZJ+16}, \\
    & & & \ \ \ \ \ \ \ RL, NN, SVM~\cite{TZJ+16,WG18}\\
        \cmidrule(lr){1-4}
    \multicolumn{2}{l}{\Sec{sec:b} B: Training Data-related Info Gathering} & & \\ 
        \cmidrule(l{8mm}r){1-4}
    \Sec{sec:b1} B1: Membership & 2: Training Data Leakage & Input to the model & Image, Tabular: LR, DT~\cite{YGF+18}, NN~\cite{AYM+19,YGF+18}\\
    \ \ \ \ \ \ \ \ \ \ \ \ \ \ \ \ \ \ Inference & & & Text: NN~\cite{AYM+19}\\
    & & Massive queries to the model & Image, Tabular: NN~\cite{SSS+17,LBW+18}\\
    & & & Text: LM~\cite{LSS+23}\\
    & & Access to the model's internal info & Image, Tabular: NN~\cite{NSH19}\\
        \cmidrule(l{8mm}r){1-4}
    \Sec{sec:b2} B2: Attribute & 2: Training Data Leakage & Input to the model & Image, Tabular: LR, DT~\cite{YGF+18} \\
    \ \ \ \ \ \ \ \ \ \ \ \ \ \ \ \ \ \ Inference & & Access to the model's internal info & Image, Text: NN~\cite{SS20} \\
        \cmidrule(l{8mm}r){1-4}
    \Sec{sec:b3} B3: Property & 2: Training Data Leakage & Access to the model's & Image, Tabular: NN~\cite{GWY+18}, \\
    \ \ \ \ \ \ \ \ \ \ \ \ \ \ \ \ \ \ Inference & & \ \ internal information & Audio: NN, SVM, HMM, DT~\cite{AMS+15} \\
    & & & Network Traffic: \\
    & & & \ \ \ NN, SVM, HMM, DT~\cite{AMS+15} \\
        \cmidrule(l{8mm}r){1-4}
    \Sec{sec:b4} B4: Model & 2: Training Data Leakage & Massive queries to the model & Image, Tabular: DT, NN~\cite{FJR15}\\
    \ \ \ \ \ \ \ \ \ \ \ \ \ \ \ \ \ \ Inversion & & Access to the model's internal info & Text: Transformer~\cite{ZHK22}\\
        \cmidrule(l{8mm}r){1-4}
    \Sec{sec:b5} B5: Data & 2: Training Data Leakage & Access to the model's internal info & Image: NN~\cite{BCH22,HVY+22,BHY+23} \\
    \ \ \ \ \ \ \ \ \ \ \ \ Reconstruction & & Massive queries to the model & Text: LM~\cite{LSS+23}\\
        \cmidrule(l{8mm}r){1-4}
    \Sec{sec:b6} B6: Data & 2: Training Data Leakage & Massive queries to the model & Text: LSTM, RNN~\cite{CLE+19}, LM~\cite{CTW+21,LSS+23}\\
    \ \ \ \ \ \ \ \ \ \ \ \ Extraction & & & Text to Image: Diffusion~\cite{CHN+23} \\
        \cmidrule(lr){1-4}
    \Sec{sec:c} C: Model & 2: Training Data Leakage & Training program tampering & Image: NN~\cite{SRS17}\\
    \ \ \ \ \ \ \ \ \ \ \ \ \ Poisoning & & & Text: SVM, LR~\cite{SRS17} \\
    & 7: Interpretability & Model tampering & Tabular: NN, EM~\cite{SHJ+20} \\
    & \ \ \ \ Malfunction & & \\
    & 11: Computational Waste & Model tampering & Image: NN~\cite{CDB+23} \\
        \cmidrule(lr){1-4}
    \Sec{sec:d} D: Data & 2: Training Data Leakage & Training data injection \& & Tabular: LR, NN~\cite{MGC22}, \\ 
    \ \ \ \ \ \ \ \ \ \ \ \ \ Poisoning & & \ \ Massive queries to the model & Text: LR~\cite{MGC22} \\
    & 6: Model Malfunction & Training data injection & Image: NN~\cite{Car21}, Text: LM~\cite{Sch19} \\
    & & & Tabular: Linear Regression~\cite{JOB+18} \\
        \cmidrule(lr){1-4}
    \Sec{sec:e} E: Evasion & 6: Model Malfunction & Massive queries to the model & Image: NN, LR, SVM, DT, kNN~\cite{PMG+17}, \\
    & & & Text: LM~\cite{BSA+22} \\
    & & Access to the model's & Image: NN~\cite{SZS+14, GSS15}\\
    & & \ \ internal information & Text: LSTM~\cite{ERL+18} \\
    & & Input to the model & Text: Transformer~\cite{GSJ+21}, LSTM, DA~\cite{WFK+19} \\
    & 7: Interpretability & Massive queries to the model & Image: NN~\cite{DAA+19} \\
    & \ \ \ \ Malfunction & & \\
        \cmidrule(lr){1-4}
    \Sec{sec:f} F: Energy- & 11: Computational Waste & Input to the model & Image, Text: NN~\cite{SZB+21} \\
    \ \ \ \ \ \ \ \ \ \ \ \ \ latency & & Massive queries to the model & Text: LM~\cite{BSA+22} \\
        \cmidrule(lr){1-4}
    \Sec{sec:g} G: Prompt & 4: Input Info Leakage & Access to the model's output & Text to Image: Diffusion~\cite{SQB+24}\\
    \ \ \ \ \ \ \ \ \ \ \ \ \ \ Stealing & & & \\
        \cmidrule(lr){1-4}
    \Sec{sec:h} H: Prompt & 2: Input Info Leakage & User-referenced info poisoning & Text: A Specific Chat Service~\cite{Sam23} \\
    \ \ \ \ \ \ \ \ \ \ \ \ \ \ Injection & 3: App Info Leakage & Input to the model via system API & Text: LM~\cite{ZCI24} \\
    & 5/10: Internal Data & Input to the model via system API & Text to Tabular: LM~\cite{PEC+25} \\
    & \ \ Leakage/Corruption & & \\
    & 8: Safeguard Bypass & Input to the model & Text: LM~\cite{LDL+24,SCB+24,ZWC+23,WHS23,CRD+24,MZK+24,PHS+22,WFK+19}\\
    & & Access to the model's internal info & Text, Image: NN~\cite{CNC+23} \\
    & 9: System Compromise & AI system-referenced info poisoning & Text, Image to Text: LM~\cite{GAM+23} \\
        \cmidrule(lr){1-4}
    \Sec{sec:i} I: Code Injection & 9: System Compromise & Malicious model distribution & Any: Any~\cite{ZWZ+24} \\
        \cmidrule(lr){1-4}
    \Sec{sec:j} J: Adversarial & 2: Training Data Leakage & Access to fine-tuning functionality & Text: LM~\cite{CTZ+24} \\
    \ \ \ \ \ \ \ \ \ \ \ \ \ \ Fine-tuning & 8: Safeguard Bypass & Access to the model's internal info & Text: LM~\cite{YYC+24} \\
        \cmidrule(lr){1-4}
    \Sec{sec:k} K: Rowhammer & 1: Model Leakage & Access to physical memory & Image: NN~\cite{RCY+22} \\
    & 6: Model Malfunction & Access to physical memory & Image: NN~\cite{LWX+24} \\
    & & Access to model's internal info, & Image, Text: LM, Transformer~\cite{NMF+24} \\
    & & \ \  physical memory & \\
    \bottomrule
    \end{tabular}
\end{table}
In this section, we provide detailed descriptions of the eleven attack types (A--K) previously introduced in Figure~\ref{fig:attacks}.
For each attack type, we present a figure of the high-level flow from the initiation of the attack to its eventual impacts on the system. 

Table~\ref{tab:big} summarizes known attack types and their corresponding impacts on AI systems. 
For each attack, the assumed capabilities of the attacker are listed, along with the modalities and the types of AI models demonstrated in the literature, accompanied by references to the relevant papers.

It is important to note that the modalities and targeted AI model types listed in Table~\ref{tab:big} are summaries based on experimental demonstrations documented in the literature and do not necessarily indicate inherent limitations of the attack methods themselves.

\paragraph{Note:} In Table~\ref{tab:big}, 
the following abbreviations are used for AI model types: 
NN (Neural Network), LR (Logistic Regression), RR (Ridge Regression), SVM (Support Vector Machine), DT (Decision Tree), kNN (Nearest Neighbor), HMM (Hidden Markov Model), LM (Language Model), LSTM (Long-short Term Memory), RNN (Recurrent Neural Network), EM (Ensemble Method), and DA (Decomposable Attention).

\subsection{A: Model Extraction}\label{sec:a}
\begin{figure}[b]
    \centering
    \includegraphics[width=\textwidth]{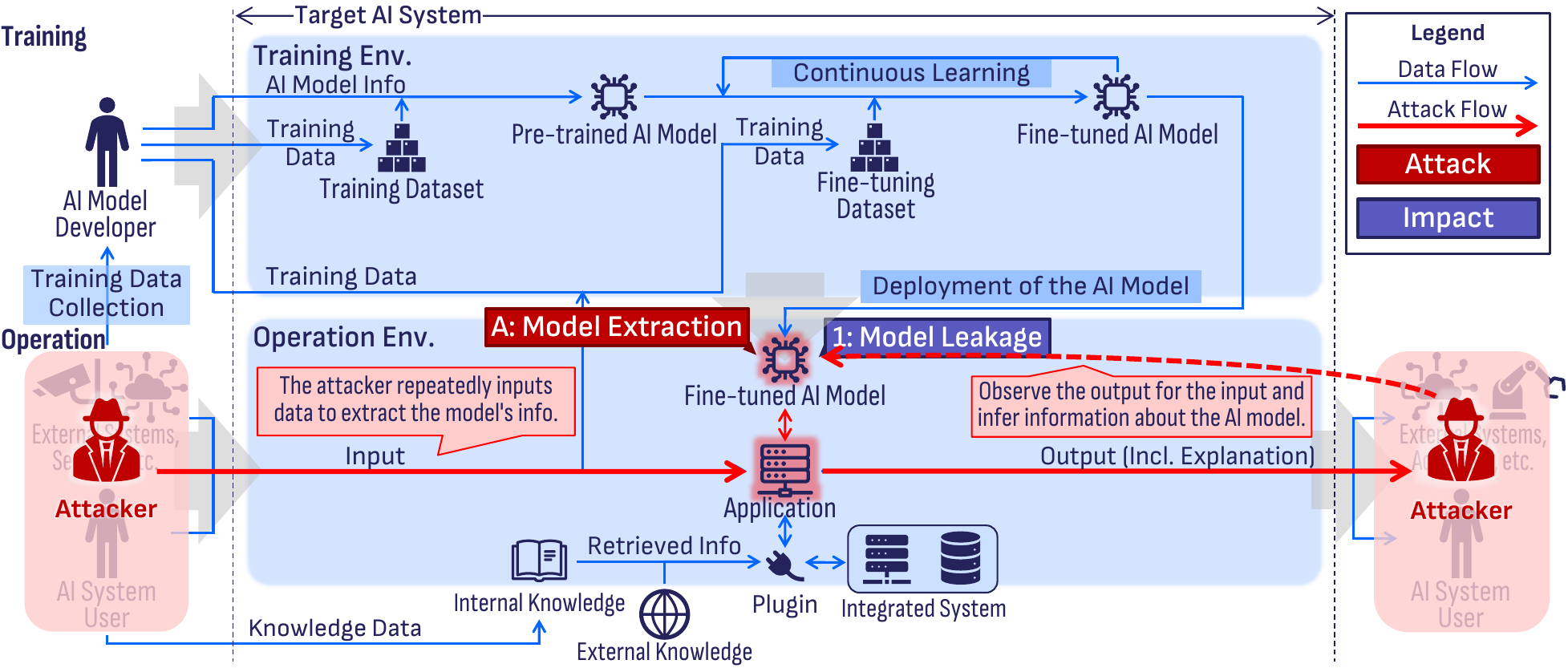}
    \caption{Attack A: Model Extraction (Overview)}
    \Description{
    The figure illustrates the process of a model extraction attack on an AI system. An attacker repeatedly inputs data into the AI model during its operation phase. By analyzing the outputs, the attacker infers confidential information about the model. Arrows indicate the data flows and attack path, clearly identifying the targeted model component and the resultant model leakage impact.
    }
    \label{fig:atk_a}
\end{figure}

Model extraction attacks involve adversaries who do not have direct access to the internal details of an AI model (e.g., users interacting via a publicly accessible API). 
By systematically querying the model with various inputs and analyzing the corresponding outputs, an attacker attempts to infer sensitive information about the underlying model structure and parameters (Figure~\ref{fig:atk_a}). 

The motivations for model extraction have grown alongside AI's increasing economic and societal importance.
State-of-the-art AI models typically embody extensive training investments, both in terms of computational resources and proprietary datasets. 
As a result, such an attempt could be highly valuable for attackers, for instance, who might sell the extracted model in black markets.
Moreover, model extraction often acts as a stepping stone for more sophisticated attacks, such as Training Data-related Information Gathering (Attack B in Section \ref{sec:b}) and Evasion (Attack E in Section \ref{sec:e}).

Previous research has demonstrated the feasibility of extracting various forms of model-specific information, such as:
\begin{itemize}
    \item Model Architecture (e.g., the structure and configuration of neural networks)~\cite{OSF19a}
    \item Hyperparameters (e.g., regularization parameters to prevent overfitting during training)~\cite{WG18}
    \item Parameters (e.g., neuron weights within neural networks)~\cite{TZJ+16}
    \item Decision Boundaries (boundaries where model predictions change class in input space)~\cite{JSM+19}
    \item Functionality (the model’s input--output relationships without internal structures)~\cite{CBB+18,OSF19b}
\end{itemize}
Table~\ref{tab:big} summarizes the modalities and AI model types examined by these representative studies.

Several approaches have been proposed to mitigate model extraction attacks. 
A basic defense principle is to minimize unnecessary information disclosure, such as providing only discrete classification results without associated confidence scores. 
Detection-oriented methods, including tools such as PRADA~\cite{JSM+19}, monitor query patterns to identify suspicious extraction attempts. 
In addition, input perturbation or obfuscation techniques have been proposed to disrupt attackers’ attempts to accurately infer model information without significantly impairing legitimate usage~\cite{Gra20}. 
Frameworks such as ML-Doctor~\cite{LWH+22} have also been introduced to assess an AI model’s vulnerability to extraction attacks. 
Ensemble learning can also provide robustness by obscuring or diversifying the model’s behavior, thus complicating the extraction process.

While not preventing the initial attack, watermarking~\cite{LWW+23} and fingerprinting~\cite{CJG21} methods have been explored as reactive measures, providing mechanisms to detect or prove unauthorized copies of models after extraction has occurred.

\subsubsection{Model Extraction vs. Knowledge Distillation.}
It is worth noting the conceptual similarity and critical distinction between model extraction and the well-known machine learning practice of knowledge distillation. 
{\bf Knowledge distillation} refers to a technique intentionally used by developers to create smaller, computationally efficient (``student'') models that mimic the behavior of a larger, more complex (``teacher'') model. 
The fundamental difference lies in intent and legitimacy: knowledge distillation is explicitly intended and conducted by authorized entities for efficiency and scalability purposes, not for unauthorized reproduction or redistribution of intellectual property. 
Thus, it is typically not classified as an adversarial attack. 
Nevertheless, due to concerns about unauthorized model duplication, many AI service providers explicitly prohibit knowledge distillation in their terms of service.

\subsection{B: Training Data-related Information Gathering}\label{sec:b}
\begin{figure}[b]
    \centering
    \includegraphics[width=\textwidth]{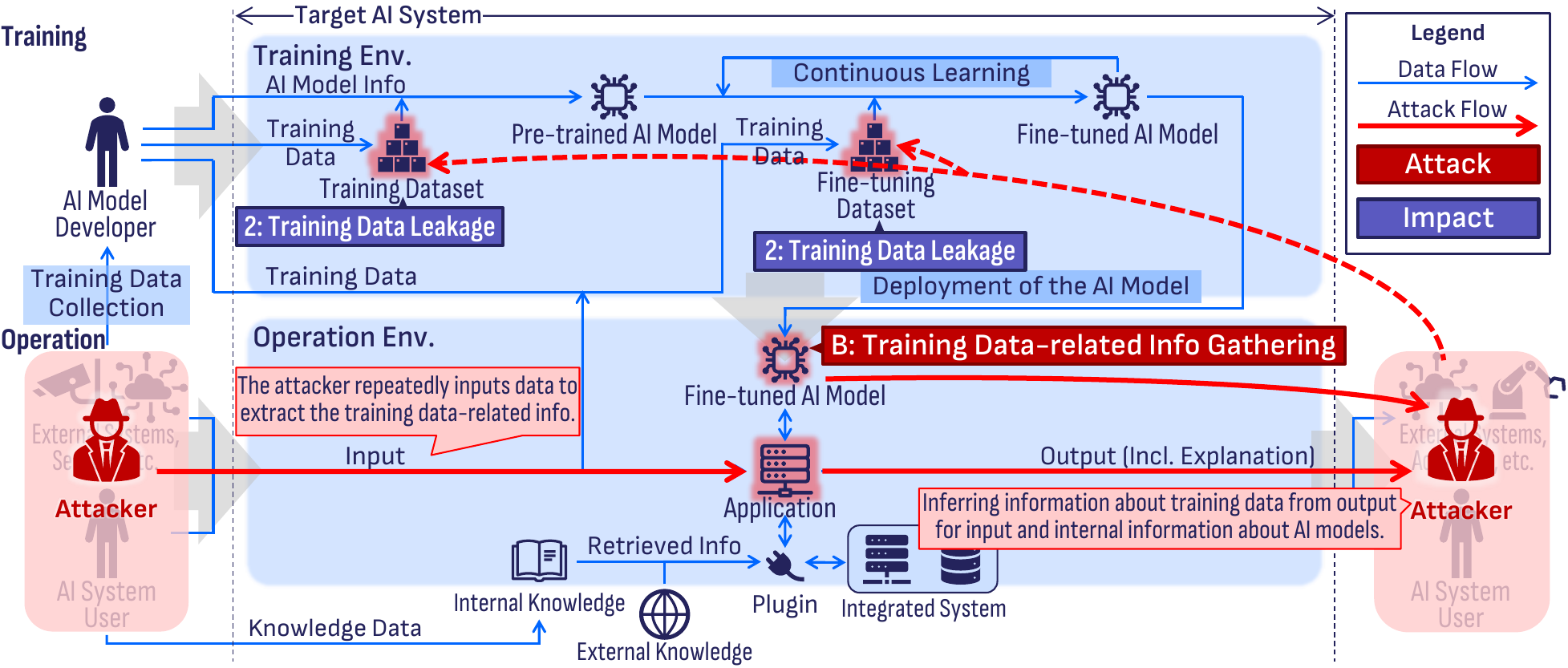}
    \caption{Attack B: Training Data-related Information Gathering (Overview)}
    \Description{
    The figure shows the process of an attack aimed at gathering sensitive information related to training data. An attacker inputs data repeatedly into the AI model during operation and uses the model's outputs and internal details to infer confidential training data information. The attack affects both pre-trained and fine-tuned AI models, resulting in training data leakage. Arrows depict the data flow and illustrate the paths utilized in this attack.
    }
    \label{fig:atk_b}
\end{figure}
In contrast to direct data theft, the category of attacks referred to as {\bf Training Data-related Information Gathering} involves extracting or inferring information related to the training data of an AI model by systematically observing its inputs and corresponding outputs, and in some cases by exploiting internal information about the model itself (see Figure~\ref{fig:atk_b}).

The term ``related information'' here encompasses a broad spectrum of scenarios in which information about the training data is revealed indirectly, extending beyond the direct extraction of original training examples.
Typical attacks include {\bf Membership Inference}, which infers whether certain individuals or records were part of the training dataset; 
{\bf Attribute Inference}, which infers unknown sensitive attributes of particular records in the dataset;  
{\bf Property Inference}, which aims to infer general statistical characteristics of the dataset; 
{\bf Model Inversion}, which reconstructs representative class-level samples by iteratively querying model outputs; 
{\bf Data Reconstruction}, where attackers leverage internal model parameters or intermediate representations to theoretically reconstruct precise, specific training data samples; 
and {\bf Data Extraction}, which retrieves exact memorized training examples directly from the model outputs, especially in generative contexts.

The following subsections provide detailed explanations of each attack type.

\subsubsection{B1: Membership Inference}
\label{sec:b1}
This type of attack involves inferring whether specific data samples were included in the training dataset by observing the model's input--output behavior. 
Such attacks pose serious privacy threats, particularly when models are trained on sensitive data. 
For example, an AI system trained on financial records of loan defaults, could reveal sensitive personal information simply by leaking whether a specific individual's history was part of its training dataset.

A typical implementation of this attack involves the use of shadow models that mimic the behavior of the target model. 
The attacker first trains shadow models using datasets whose membership status (included or excluded) of each sample is known. 
Then, another model (often referred to as the {\it attack model)} learns the differences in shadow model behavior for data samples that are included or excluded in their training datasets. 
Using this attack model, the attacker can infer the membership status of specific data points from the output of the target model.
Table~\ref{tab:big} summarizes the literature on membership inference attacks across various modalities and model types.

To mitigate membership inference attacks, several countermeasures have been proposed:
\begin{itemize}
    \item {\bf in Training Environment}
    \begin{itemize}
            \item {\bf Preventing Overfitting:} Employing regularization techniques during model training to minimize unintended memorization of training data.
            \item {\bf Assessment Tools:} Utilizing specialized analysis tools (e.g., ML Privacy Meter~\cite{MS20} or ML-Doctor~\cite{LWH+22}) to proactively assess and mitigate membership inference attacks.
    \end{itemize}
    \item {\bf in Operation Environment}
    \begin{itemize}
        \item {\bf Output Control:} Reducing the detail of outputs. Examples include reporting only top predictions without confidence scores or reducing the numerical precision of outputs.
        \item {\bf Differential Privacy:} Incorporating differential privacy mechanisms to mathematically limit information leakage about any individual data point.
    \end{itemize}
\end{itemize}

\subsubsection{B2: Attribute Inference}\label{sec:b2}
This type of attack refers to inferring unknown sensitive attributes (e.g., a specific genotype) of data samples used to train an AI model by exploiting publicly available or partially known information (e.g., age, gender obtained from social media) and the input--output behavior of the model~\cite{YGF+18}. 
In other words, with partial knowledge of certain attributes, an attacker can infer sensitive attributes that should remain confidential, potentially leading to the unintended disclosure of highly private personal information.
Table~\ref{tab:big} summarizes the literature on attribute inference attacks across various modalities and model types.

To mitigate attribute inference attacks, measures similar to those used for membership inference attacks can be employed, including regularization to prevent overfitting and differential privacy to limit leakage of sensitive attributes.

\subsubsection{B3: Property Inference}\label{sec:b3}
This type of attack infers global statistical properties of the training dataset by analyzing information derived from the model. 
In contrast to membership or attribute inference, which target specific data samples or attributes, property inference attacks target general dataset-level characteristics (e.g., distributional properties, proportion of a particular demographic group).

Figure~\href{https://dl.acm.org/doi/pdf/10.1145/3243734.3243834#page=5}{1} in the foundational paper~\cite{GWY+18} provides an overview of how a property inference attack can be performed. 
First, an attacker trains multiple shadow classifiers, some on datasets containing a certain statistical property $P$, and others on datasets lacking that property (denoted as $\overline{P}$). 
Then, features such as internal parameters or outputs are extracted from these shadow classifiers and used to construct a meta-training dataset. 
This dataset is subsequently used to train a meta-classifier capable of distinguishing between models trained with and without the property $P$. 
Finally, by extracting similar features from a target model, the attacker can infer whether the target model's training dataset possesses the property $P$.
Table~\ref{tab:big} summarizes the literature on property inference attacks across various modalities and model types.

Proposed mitigation strategies of property inference attack include introducing noisy data into training datasets~\cite{GWY+18}.

\subsubsection{B4: Model Inversion}\label{sec:b4}

This type of attack generates inputs that strongly activate a model's particular class prediction, thereby reconstructing representative data of a certain class.
The attacker typically iteratively adjusts the input data to maximize the classifier's confidence in a target class based on the model's feedback (e.g., confidence scores), thus converging on data that characterizes the class. 
The foundational paper~\cite{FJR15} offers an example of such an attack in its Figure \href{https://dl.acm.org/doi/pdf/10.1145/2810103.2813677#page=2}{1}, which illustrates both an image reconstructed by the attack and the corresponding original image from the model's training dataset.
While successful reconstruction can be highly conditional, this type of attack highlights potential privacy risks.

Table~\ref{tab:big} summarizes the relevant literature on model inversion attacks. 

A simple mitigation mentioned in the literature is masking or limiting the confidence scores returned by the model.

\subsubsection{B5: Data Reconstruction}\label{sec:b5}
This type of attack theoretically reconstructs specific data samples from internal representations or outputs of a trained AI model.
In contrast to model inversion, which produces generic representations for a given class, data reconstruction attacks reconstruct concrete, specific examples that closely resemble the original training samples.

Detailed examples of the literature demonstrating how attackers leverage model parameters or outputs to precisely reconstruct individual training samples are summarized in Table~\ref{tab:big}.

These attacks raise significant privacy concerns by showing how sensitive or personal information contained in the training dataset could be accurately exposed, potentially violating confidentiality or regulatory compliance.
Typical countermeasures include differential privacy, output randomization, and strict control of access to intermediate model representations.

\subsubsection{B6: Data Extraction}\label{sec:b6}
This type of attack is specifically designed to identify and retrieve exact memorized training data from the output of a trained model, especially in generative AI systems such as language models or diffusion models.
In contrast to model inversion, which targets generic class representations, and in contrast to  data reconstruction, which theoretically derives specific data points from internal representations, data extraction directly targets memorized data samples through repeated queries to the model, typically by exploiting unintended memorization phenomena.

A notable example is the attack on LLMs demonstrated by Carlini et al.~\cite{CTW+21}. 
Their method involves providing an initial prompt that could plausibly precede private or sensitive information, generating a large number of candidate text completions, and then using membership inference techniques to identify which of these completions actually appeared in the training data.

Table~\ref{tab:big} summarizes the literature on data extraction attacks. 

Mitigation strategies discussed in the literature include the use of differential privacy and regularization techniques to reduce unintended memorization.

\subsection{C: Model Poisoning}\label{sec:c}
Model poisoning attacks directly manipulate the AI model itself or its training program, rather than altering the training data, to compromise the model's behavior.
In contrast to data poisoning, where malicious inputs are introduced into the dataset, model poisoning modifies the underlying structure, parameters, or training process of the model.

\begin{figure}[tb]
    \centering
    \includegraphics[width=\textwidth]{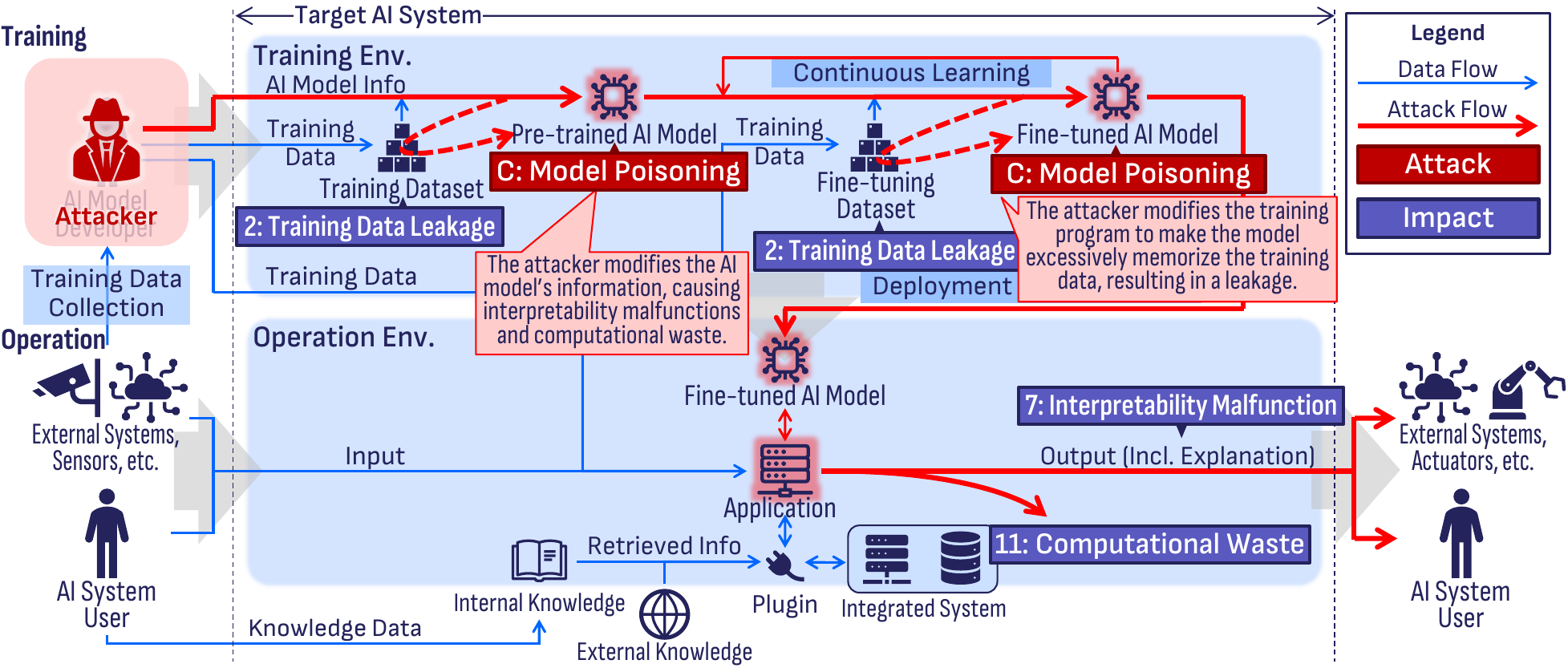}
    \caption{Attack C: Model Poisoning (Overview)}
    \Description{
    The figure depicts a model poisoning attack within an AI system. An attacker modifies the AI model or its training program, causing excessive memorization of training data. This manipulation results in interpretability malfunction, computational waste, and increased risk of training data leakage. Arrows illustrate the flow of data and highlight the specific points of attack within both the training and operational environments.
    }
    \label{fig:atk_c}
\end{figure}

A typical example of model poisoning, as documented in the prior literature~\cite{SRS17}, involves an attacker maliciously modifying the training algorithm or the configuration parameters to induce excessive memorization of training data.
This increased memorization significantly increases the risk of sensitive training data leakage, as the model could inadvertently reproduce exact data points from its training set during inference.
Such attacks could ultimately lead to severe interpretability malfunctions, excessive computational waste due to inefficient model performance, or data leakage when interacting with external queries during operation (illustrated in Figure~\ref{fig:atk_c}).

A summary of specific experimental demonstrations, attack capabilities, and their impacts on various modalities and model types is detailed in Table~\ref{tab:big}.

To mitigate the risks posed by model poisoning attacks, robust verification processes for both models and training procedures are critical.
Verification strategies include rigorous code audits, model integrity checks, and employing secure, tamper-resistant hardware or software frameworks.
Furthermore, employing cryptographic integrity verification methods and secure, traceable versioning of AI artifacts can effectively reduce the attack surface.

\subsection{D: Data Poisoning}\label{sec:d}
\begin{figure}[tb]
    \centering
    \includegraphics[width=\textwidth]{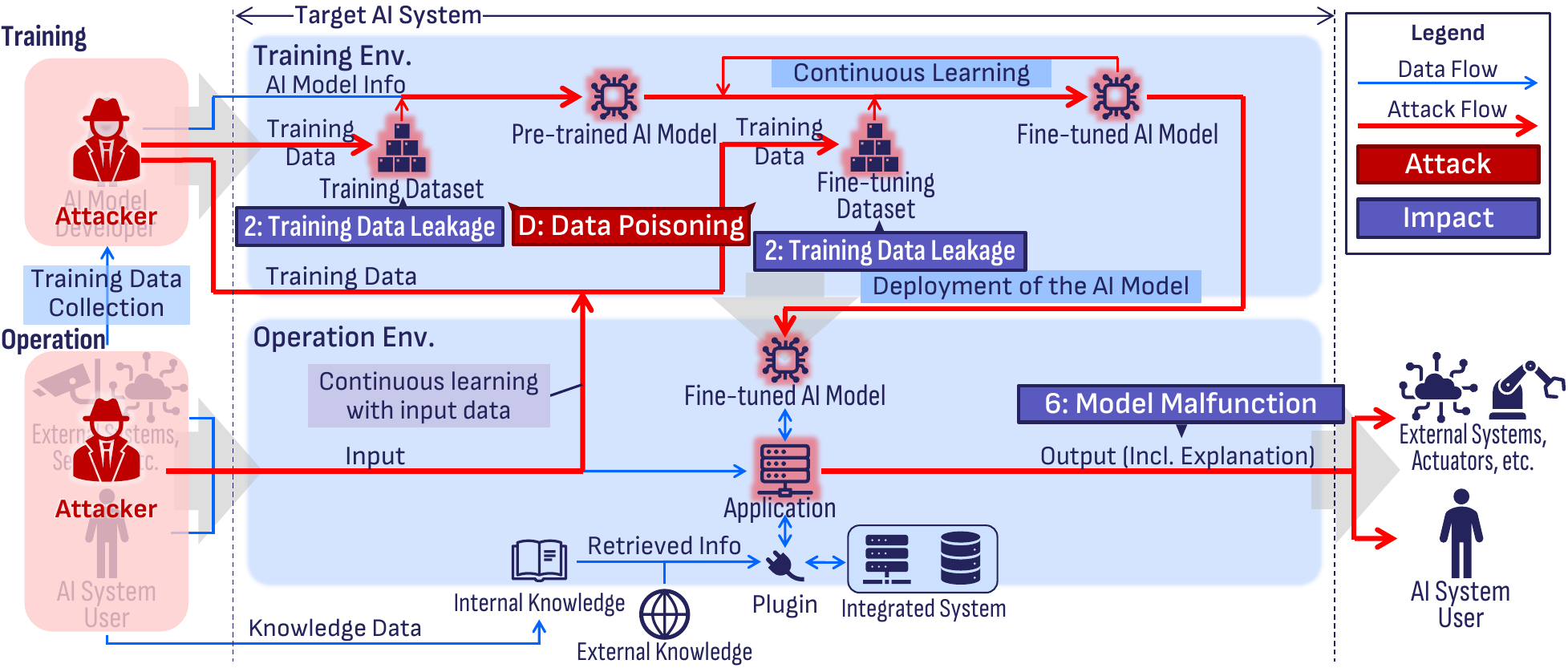}
    \caption{Attack D: Data Poisoning (Overview)}
    \Description{
    The figure illustrates a data poisoning attack within an AI system. The attacker injects malicious data into the training dataset during the training phase. This compromised data affects the AI model, causing model malfunction and potential leakage of training data. Arrows represent data flows and clearly indicate where malicious data insertion occurs within the training environment and how it impacts model operations.
    }
    \label{fig:atk_d}
\end{figure}
\begin{figure}[tb]
    \centering
    \includegraphics[width=\textwidth]{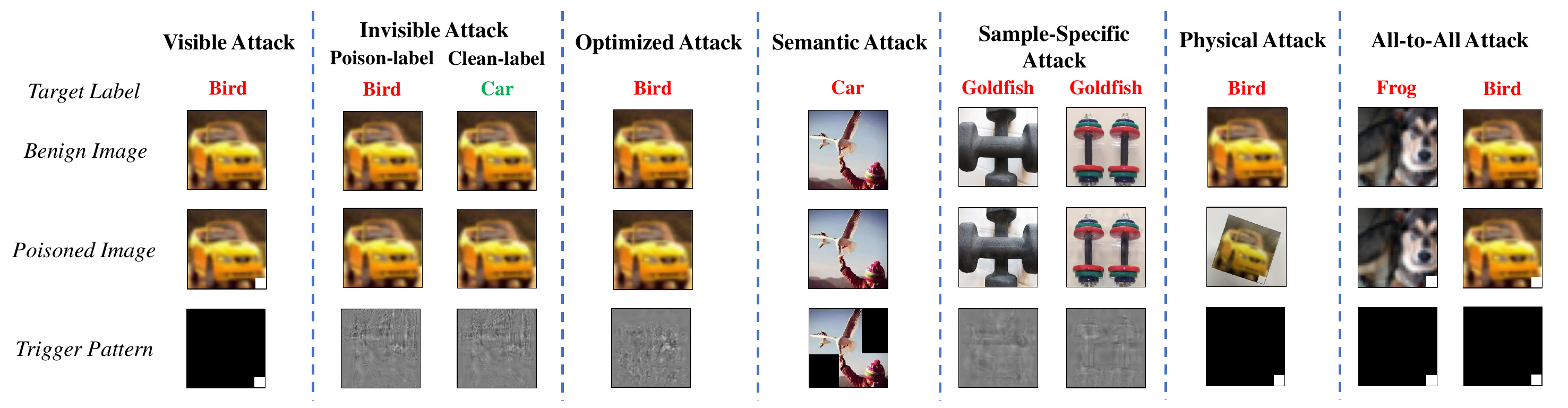}
    \caption{
     An example of poisoned samples generated by different types of backdoor poisoning.
     Quotation from Figure 4 in ``Backdoor Learning: A Survey''~\cite{LJL+24} by Yiming Li, Yong Jiang, Zhifeng Li, and Shu-Tao Xia (arXiv:\href{https://arxiv.org/abs/2007.08745}{2007.08745}[cs.CR]) [licensed under \href{https://creativecommons.org/licenses/by/4.0/}{CC BY 4.0}]}
     \Description{
     The figure provides visual examples of different types of backdoor poisoning attacks using images. It shows benign images alongside their poisoned versions. Poisoned examples include visible markers, invisible perturbations, optimized perturbations, semantic combinations, sample-specific triggers, physical-world alterations, and multiple targeted misclassifications ("all-to-all"). Each poisoned sample is paired with its original benign counterpart to highlight the subtle or obvious manipulations applied to cause misclassification by AI models.
     }
    \label{fig:LJL+24}
\end{figure}
Data poisoning attacks involve intentionally inserting malicious data into an AI model's training dataset, aiming either to compromise the model's accuracy or induce specific misbehavior during inference. 
Figure~\ref{fig:atk_d} illustrates the high-level operational flow of data poisoning attacks, showing how malicious data injected into the training environment affects model behavior during operation.
Broadly speaking, these attacks can be categorized into three main types:
\begin{itemize}
    \item {\bf Targeted Poisoning:} These attacks change the model’s behavior for specific inputs or data points, causing the model to make incorrect predictions for particular targeted samples.
    \item {\bf Untargeted Poisoning:} In contrast to targeted attacks, untargeted poisoning degrades the overall accuracy of the model indiscriminately. By introducing misleading or noisy data broadly across the training dataset, attackers cause a generalized reduction in model performance, resulting in unreliable predictions across various inputs.
    \item {\bf Backdoor Poisoning:} By embedding subtle triggers in selected training samples, backdoor poisoning causes the model to produce incorrect outputs only when these triggers are present during inference. In other words, the poisoned model performs normally on benign inputs, but deviates maliciously when triggered inputs are encountered.
\end{itemize}
Backdoor poisoning can be considered a specialized subset of targeted poisoning due to its conditional nature on specific inputs.
However, it is often treated separately due to its characteristic requirement for a trigger mechanism.

\subsubsection{Detailed Categorization of Backdoor Poisoning}
The existing survey~\cite{LJL+24} categorizes backdoor poisoning attacks as follows (see Figure~\ref{fig:LJL+24}):
\begin{enumerate}
    \item {\bf Visible Attack:} Uses explicitly visible triggers (e.g., a distinct marker such as a white square at the corner of an image), leading the model to associate the trigger with incorrect labels.
    \item {\bf Invisible Attack:} Incorporates subtle, imperceptible perturbations as triggers, making detection more difficult.
    \item {\bf Optimized Attack:} Utilizes sophisticated optimization methods, such as universal adversarial perturbations, to maximize poisoning effectiveness.
    \item {\bf Semantic Attack:} Relies on specific semantic combinations rather than visual perturbations, causing the model to misclassify semantically triggered inputs (e.g., images containing both ``bird'' and ``human'' may be misclassified as ``car'').
    \item {\bf Sample-specific Attack:} Involves unique trigger patterns individually tailored to specific poisoned samples.
    \item {\bf Physical Attack:} Implements poisoning patterns in the physical world, captured by sensors, demonstrating real-world applicability.
    \item {\bf All-to-All Attack:} Assigns different target labels to different poisoned samples for easier evasion of target-oriented defenses compared to traditional all-to-one backdoor attacks.
\end{enumerate}

Table~\ref{tab:big} summarizes the literature on data poisoning attacks, highlighting their impact on various modalities and AI model types.

Mitigation measures against data poisoning typically include rigorous validation of training data, anomaly detection methods to flag potentially malicious data points, and training procedures specifically designed to improve model resilience to poisoned inputs.

\subsection{E: Evasion}\label{sec:e}
\begin{figure}[b]
    \centering
    \includegraphics[width=\textwidth]{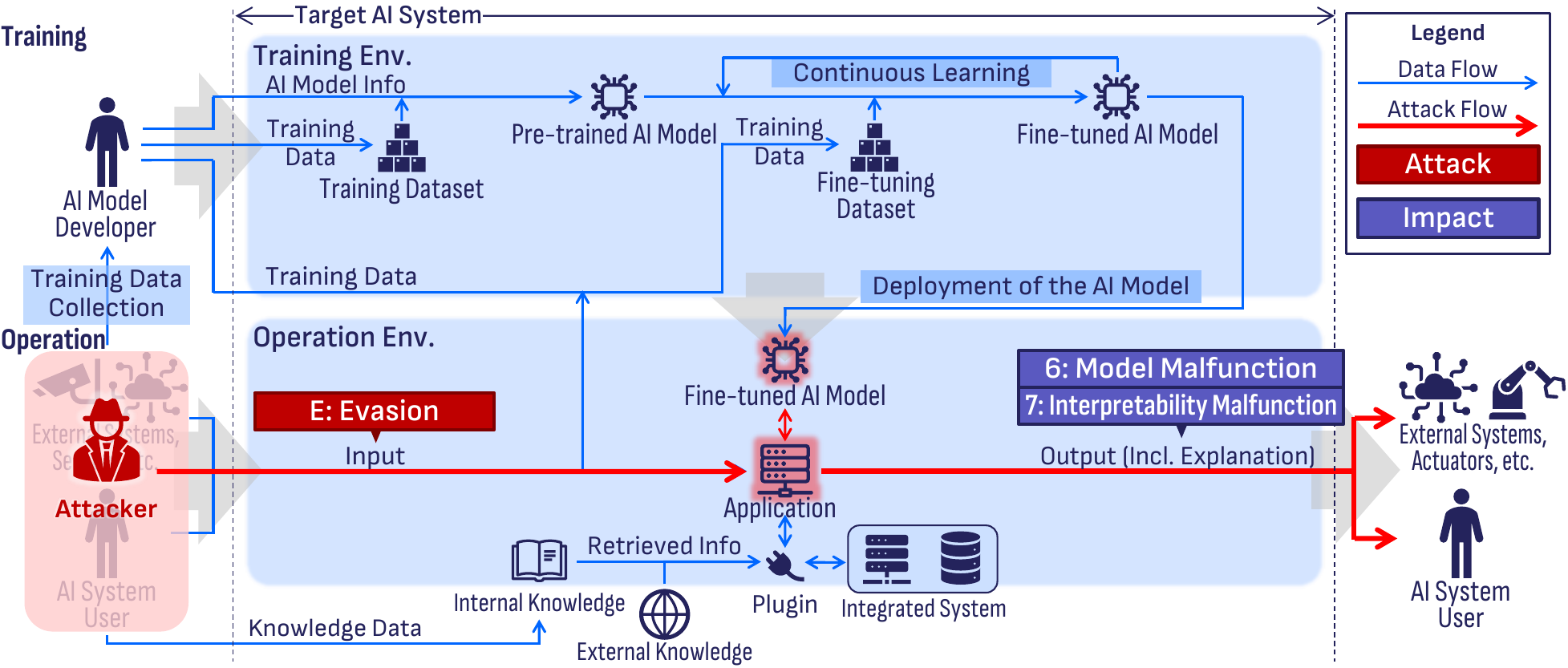}
    \caption{Attack E: Evasion (Overview)}
    \Description{
    The figure illustrates an evasion attack within an AI system. An attacker manipulates input data provided to the deployed AI model during its operational phase, causing the model to malfunction or produce incorrect outputs. Additionally, such attacks can compromise the interpretability of the model's decisions. Arrows indicate the data flow and show the points where manipulated inputs affect model operations.
    }
    \label{fig:atk_e}
\end{figure}

Evasion attacks represent a threat in which an adversary manipulates the behavior of AI models during their operational phase by providing carefully crafted inputs, known as {\bf Adversarial Examples}, without contaminating the training data or directly modifying the model itself (see Figure~\ref{fig:atk_e}).
Adversarial examples are generated by introducing subtle perturbations into legitimate data, designed to trick the model into making incorrect predictions.

Adversarial examples can be generated under two primary settings: 
\begin{itemize}
    \item {\bf White-box setting:} An attacker has complete knowledge of the model's internal architecture and parameters.
    \item {\bf Black-box setting:} An attacker observes only the model's inputs and outputs without access to internal details.
\end{itemize}
In both settings, small but purposeful modifications to the input are applied to exploit the model's decision boundaries, resulting in significant deviations in the model's output.

The seminal paper~\cite{GSS15} shows a demonstration of such an evasion attack in its Figure~\href{https://arxiv.org/pdf/1412.6572#page=3}{1}.
In this example, a small, visually imperceptible perturbation is applied to an image of a panda.
Despite the minimal visible difference, this adversarial example successfully fools the model into misclassifying the image as a gibbon, demonstrating that evasion attacks can induce significant {\it model malfunction}.

Another critical consequence of evasion attacks, {\it interpretability malfunction}, is illustrated in Figure~\href{https://proceedings.neurips.cc/paper_files/paper/2019/file/bb836c01cdc9120a9c984c525e4b1a4a-Paper.pdf#page=1}{1} of another paper~\cite{DAA+19}.
Here, the attack specifically targets the model's explanation mechanisms, altering regions identified by explainability methods (e.g., heatmaps) as influential in the decision-making process.
Such manipulated explanations can dangerously mislead stakeholders who rely on these visual interpretations for validation or decision-making.

Table~\ref{tab:big} provides a summary of the literature on evasion attacks that cause these impacts for different modalities, and AI model types.

Mitigation strategies against evasion attacks generally involve increasing model robustness and implementing evaluation and monitoring frameworks. 
Notable approaches include:
\begin{itemize}
    \item {\bf Adversarial Training:} Enhancing model robustness by proactively incorporating adversarial examples into the training dataset.
    \item {\bf Robustness Evaluation Tools:} Employing specialized frameworks such as ART~\cite{NSN+19}, RobustBench~\cite{CAS+21}, and CleverHans~\cite{PFC+18} to systematically benchmark model resilience against adversarial perturbations.
    \item {\bf Input Access Controls:} Limiting the number and frequency of inputs an AI system processes, reducing opportunities for attackers to probe  model behavior.
    \item {\bf Adversarial Data Detection Techniques:} Leveraging advanced detection methods to identify and mitigate adversarially manipulated inputs~\cite{XEQ18,MLT+19,AHF+22}.
\end{itemize}
Some mitigation methods focus on interpretability malfunctions. 
$\beta$-smoothing achieves smoother gradient propagation to increase robustness against adversarial manipulations~\cite{DAA+19}. 

\subsection{F: Energy-latency}\label{sec:f}
\begin{figure}[b]
    \centering
    \includegraphics[width=\textwidth]{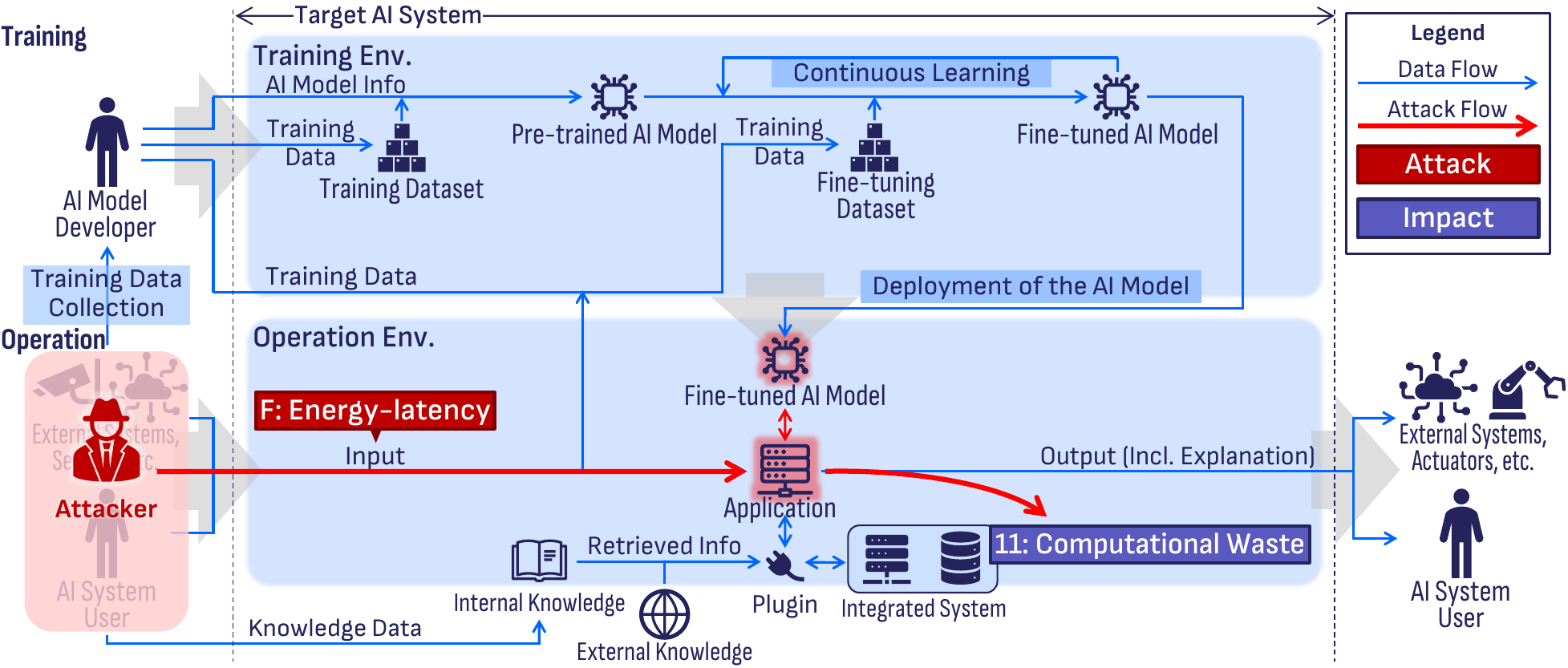}
    \caption{Attack F: Energy-latency (Overview)}
    \Description{
    The figure illustrates an energy-latency attack on an AI system. An attacker inputs specially crafted data (sponge examples) into the operational AI model, causing excessive computational load and resource consumption. Arrows indicate data flow and identify the model component targeted, highlighting the resulting computational waste.
    }
    \label{fig:atk_f}
\end{figure}
Energy-latency attacks, illustrated in Figure~\ref{fig:atk_f}, exploit vulnerabilities in AI models by providing specially crafted inputs, known as {\bf Sponge Examples}, to increase the  computational load during inference. 
In contrast to adversarial examples, which are designed to mislead predictions, sponge examples focus on resource exhaustion, significantly increasing energy consumption and response latency.

As summarized in Table~\ref{tab:big}, energy-latency attacks have been demonstrated across various modalities, including images and text. 
Prior studies highlight that these attacks do not necessarily require internal knowledge of the AI model; even external interaction with a public API may be sufficient to execute successful energy-latency exploits.

The impact of these attacks is severe, particularly in scenarios where AI systems operate within resource-constrained environments or latency-critical applications. 
Systems subjected to repeated or sustained sponge examples may experience degraded service availability, higher operational costs, or even complete denial-of-service (DoS).

Mitigation strategies include setting cutoff thresholds to terminate processing when energy consumption or latency exceeds predefined limits, resulting in errors that prevent system overload. 
In addition, introducing fallback mechanisms that activate under excessive computational load, such as switching to simpler processing, can help maintain system stability and availability.

\subsection{G: Prompt Stealing}\label{sec:g}
\begin{figure}[b]
    \centering
    \includegraphics[width=\textwidth]{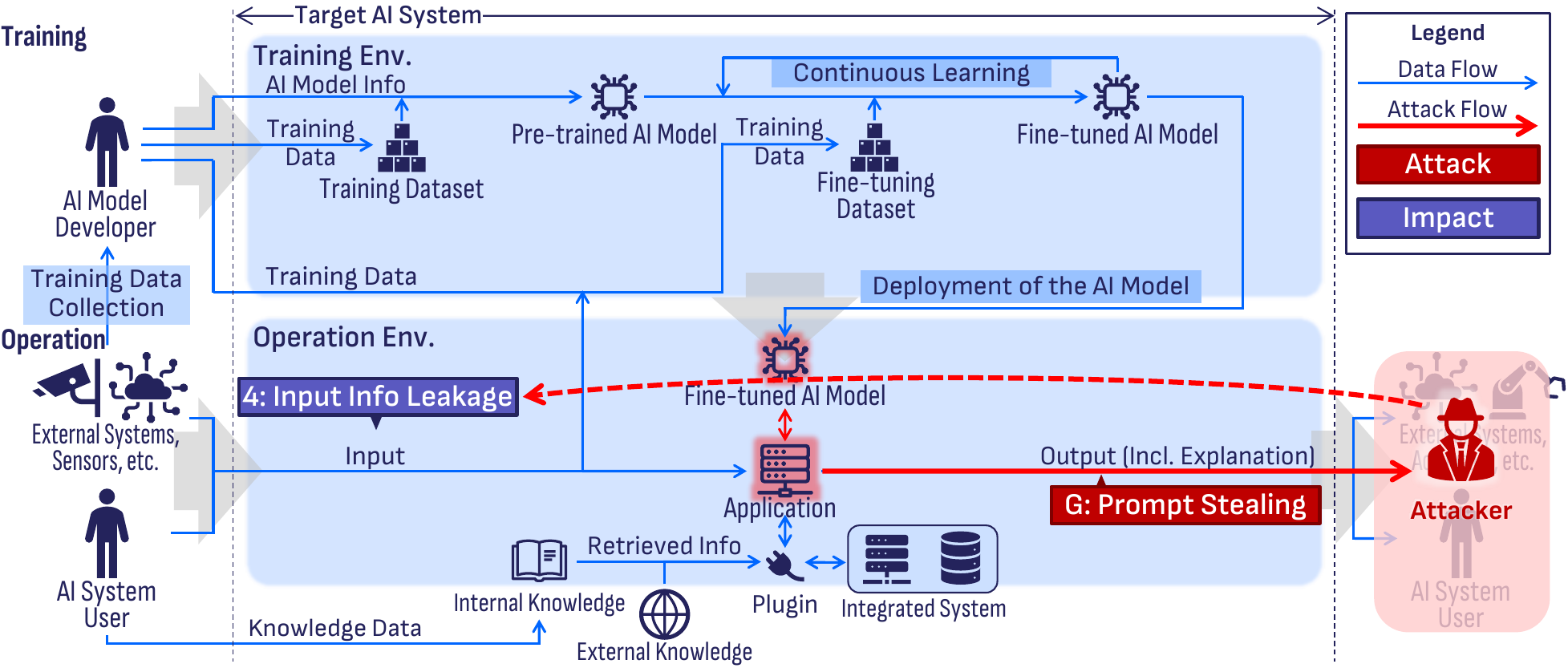}
    \caption{Attack G: Prompt Stealing (Overview)}
    \Description{
    The figure illustrates a prompt stealing attack within an AI system. The attacker analyzes outputs from an operational AI model to reconstruct the original input prompts, resulting in the leakage of confidential input information. Data flow arrows show the path of the attack and identify the components involved.
    }
    \label{fig:atk_g}
\end{figure}
Prompt stealing attacks reconstruct the original text prompts used for AI-generated content (see Figure\ref{fig:atk_g}). 
In contrast to prompt injection, which compromises the behavior of generative models by manipulating inputs, prompt stealing involves analyzing model outputs to reverse-engineer the original inputs. 
Such attacks particularly target diffusion-based text-to-image models, posing a notable threat to the intellectual property embedded in carefully designed prompts, as well as disrupting marketplaces where these prompts are commercially traded (see Table~\ref{tab:big}).

A recent study~\cite{SQB+24} presented a practical prompt stealing method, that uses two complementary predictive and generative AI components: a {\it subject generator} and a {\it modifier detector}. 
The {\it subject generator}, typically implemented as a generative image-captioning model, inferentially generates a textual description of the primary subject depicted in the generated image. 
The {\it modifier detector}, implemented as a predictive multi-label classifier, identifies keywords or stylistic modifiers that determine finer visual details and artistic styles. 
By combining these predicted subjects and modifiers, this method effectively reconstructs detailed prompts that closely resemble the originals.

The prompt stealing attacks do not steal the generated images, since the attackers already have these outputs. 
Instead, these attacks target the underlying text prompts due to their practical and economic value. 
Access to the original prompts enables attackers to reproduce images of a particular artistic style, freely adapt prompts for related creative endeavors, or circumvent fees and intellectual property restrictions associated with prompt marketplaces.

The study~\cite{SQB+24} also proposed a countermeasure to prompt stealing attacks.
The process involves the incorporation of adversarial perturbations into images. 
These imperceptible alterations to images impede the attackers' ability to accurately infer the original prompts. 
However, it was also noted that adaptive adversaries employing advanced methods might partially overcome these defenses, emphasizing the ongoing need for robust and adaptive mitigation strategies.

\subsection{H: Prompt Injection}\label{sec:h}
\begin{figure}[b]
    \centering
    \includegraphics[width=\textwidth]{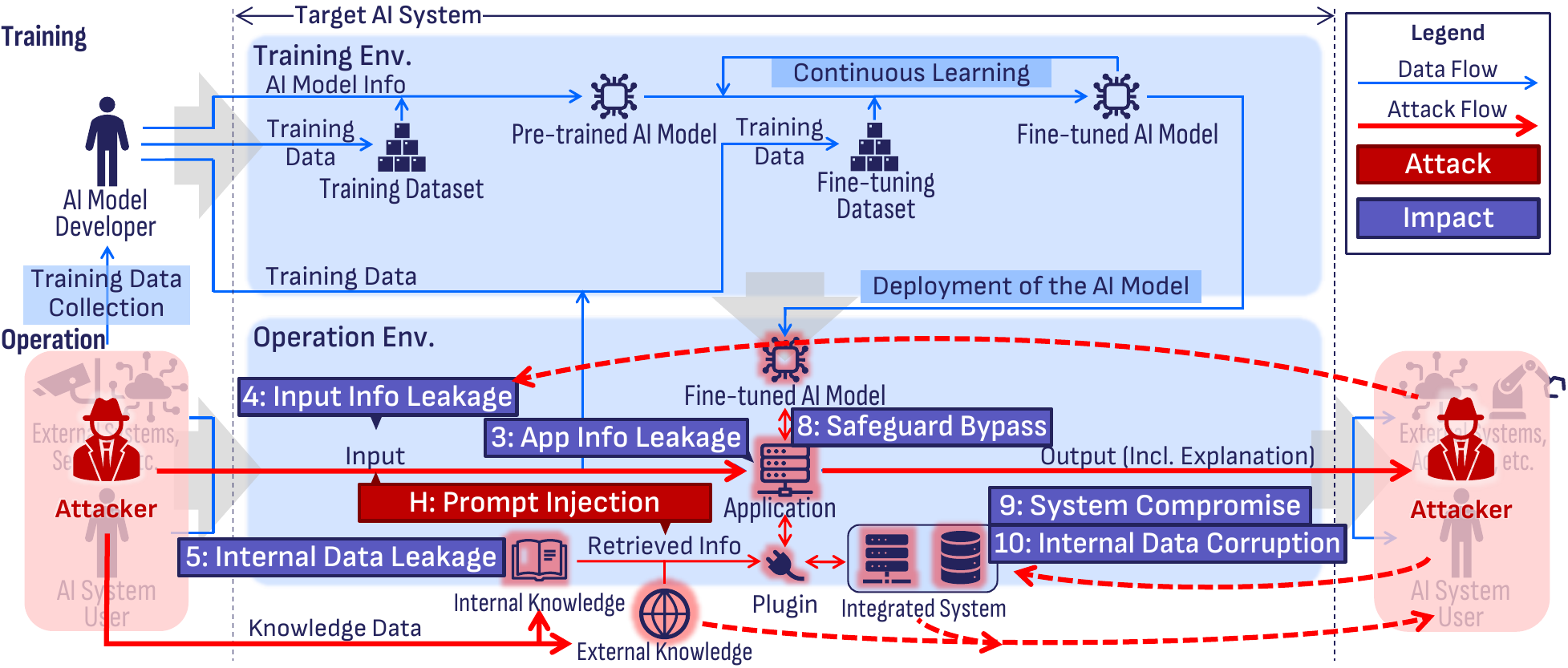}
    \caption{Attack H: Prompt Injection (Overview)}
    \Description{
    The figure shows a prompt injection attack on an AI system. An attacker injects malicious instructions directly or indirectly into input prompts or system-referenced data. This attack bypasses safeguards, causing various impacts, including application and internal data leakage, input data leakage, internal data corruption, and potential system compromise. Data flow arrows indicate the points of injection and the affected system components.
    }
    \label{fig:atk_h}
\end{figure}
Prompt injection attacks manipulate AI systems to perform unintended or harmful actions by embedding malicious instructions into input prompts (Figure~\ref{fig:atk_h}). 
These attacks can be classified into two types: 
\begin{itemize}
    \item {\bf Direct Prompt Injection:} Harmful commands are directly included in user inputs.
    \item {\bf Indirect Prompt Injection:} Attackers manipulate external sources (e.g., databases, APIs, or plugins) accessed by the AI system.
\end{itemize}

The impacts of prompt injection attacks are extensive, including bypassing security safeguards, divulging various types of information, and executing unauthorized commands (see Table~\ref{tab:big}).

A variety of mitigation strategies for prompt injection attacks have been proposed.
As external measures, which are implemented outside of AI models, include detecting and filtering malicious inputs or risky outputs~\cite{HRE+24}, controlling model behavior by system prompts~\cite{XYS+23}, and separating trusted and untrusted data in a prompt~\cite{HLH+24}.
Furthermore, the integration of defense mechanisms directly into AI models has become prevalent. 
Such methods include adversarial training~\cite{ESH+24} and other specific training~\cite{ZPW+24} to increase the models' resilience against malicious prompts.

\subsection{I: Code Injection}
\label{sec:i}
According to OWASP~\cite{ZR25}, {\bf Code Injection} is ``the general term for attack types which consist of injecting code that is then interpreted/executed by the application.'' 
In this paper, we focus on code injection into AI models, whereby malicious executable code is embedded within models and is executed upon model loading.
In contrast to prompt injection, which manipulates model inputs during inference, code injection embeds malicious code directly into the model file itself.

\begin{figure}[tb]
    \centering
    \includegraphics[width=\textwidth]{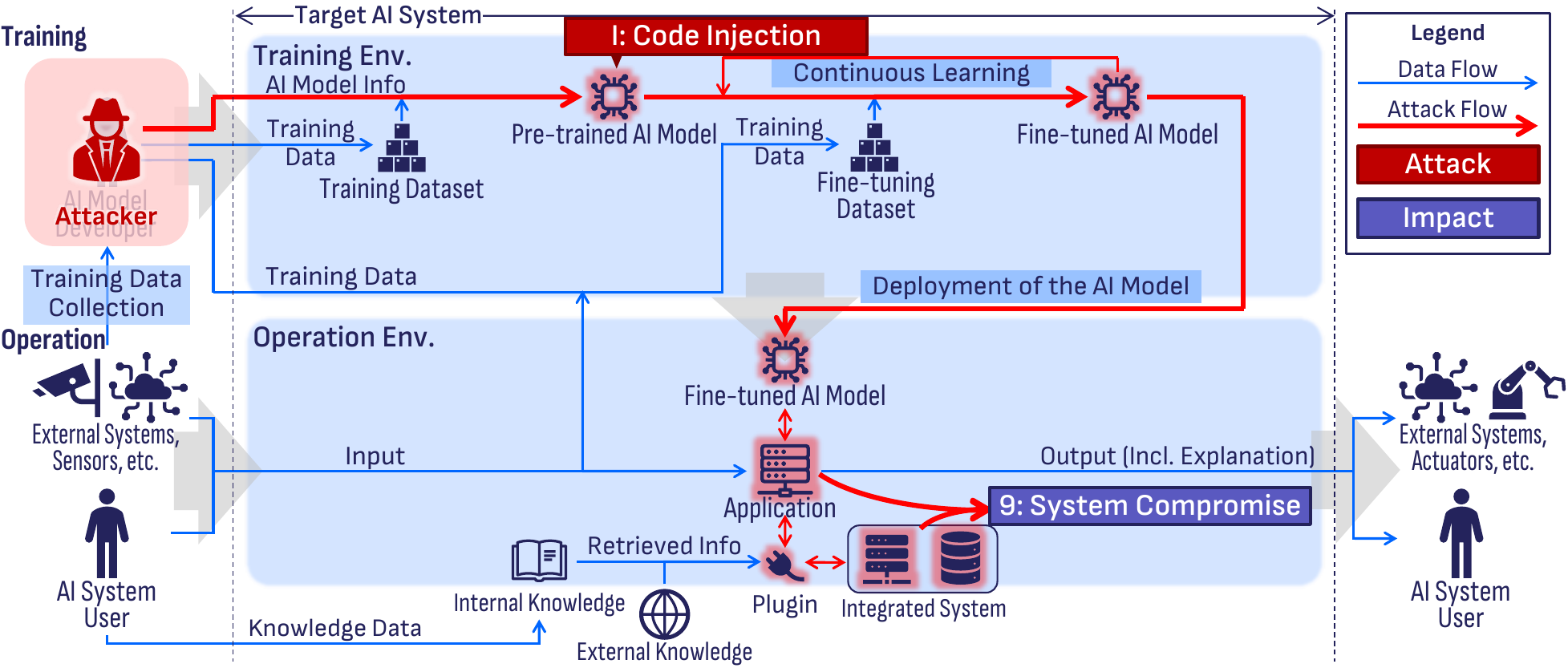}
    \caption{Attack I: Code Injection (Overview)}
    \Description{
    The figure illustrates a code injection attack on an AI system. An attacker embeds malicious executable code into AI model files. When these compromised model files are loaded into the system, unauthorized commands execute, resulting in system compromise. Data flow arrows identify the points of code injection and subsequent execution.
    }
    \label{fig:atk_i}
\end{figure}

The primary impact of code injection is system compromise, which allows attackers to execute unauthorized commands. 
In contrast to other attacks, code injection poses risks across predictive and generative AI models, regardless of data modality (Table~\ref{tab:big}).

Code injection exploits data formats such as Python's \texttt{pickle}, which are designed to support code execution as their intended design. 
Attackers embed malicious payloads into models using such formats (e.g., \texttt{pickle}, TensorFlow's \texttt{SavedModel}, and Keras's \texttt{Lambda} layers).
These malicious models are typically uploaded to public repositories by the attackers, and when downloaded and executed without proper validation, the injected code runs automatically (illustrated in Figure~\ref{fig:atk_i}). 
It is important to note that the embedding of executable code in these formats is not a vulnerability in itself; rather, it is an intentional feature that is being misused by the attackers.

The mitigation of code injection involves the use of inherently safer formats, such as \texttt{GGUF}, \texttt{ONNX}, or \texttt{Safetensors}. 
These formats exclusively contain model weights (and architecture) without embedded code. 
Other measures include validating model provenance, employing sandbox environments, and leveraging analysis tools for detecting malicious code~\cite{ZWZ+24}. 

\subsection{J: Adversarial Fine-tuning}\label{sec:j}
\begin{figure}[tb]
    \centering
    \includegraphics[width=\textwidth]{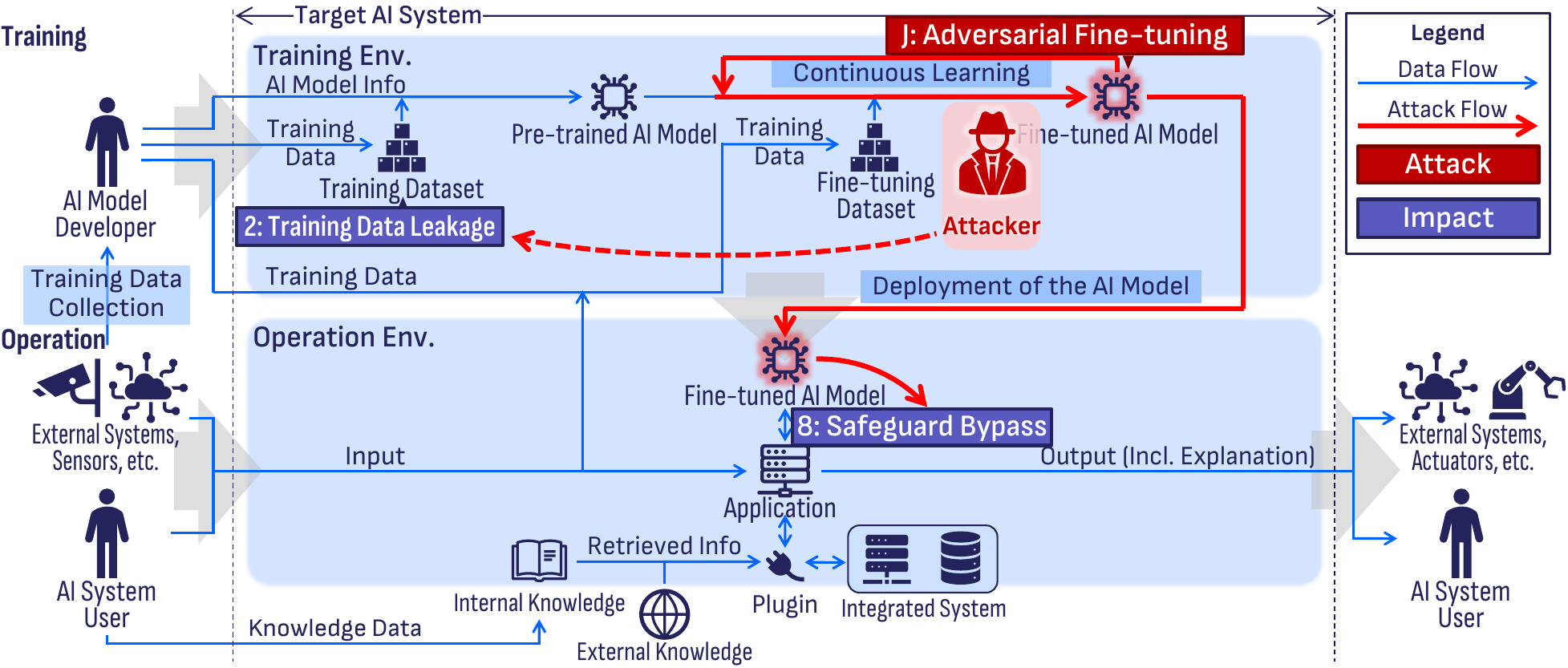}
    \caption{Attack J: Adversarial Fine-tuning (Overview)}
    \Description{
    The figure depicts an adversarial fine-tuning attack within an AI system. An attacker maliciously fine-tunes a pre-trained AI model, leading to safeguard bypass and training data leakage. Arrows indicate the data flow, highlighting the point of malicious fine-tuning and the operational components of the AI system that are affected.
    }
    \label{fig:atk_j}
\end{figure}
Adversarial fine-tuning attacks leverage the fine-tuning process of AI models to compromise their intended safety alignment. 
This category of attacks is distinguished by its focus on fine-tuning procedures, which are frequently utilized in adapting pre-trained LLMs to specialized tasks. 
These attacks are designed to bypass safeguards in order to induce harmful behaviors, undermine model safety, or recover sensitive information from previously learned data (see Figure~\ref{fig:atk_j}).
The fundamental mechanism of these attacks involves conducting specialized fine-tuning procedures that effectively reverse the safety alignment of models or reactivate previously suppressed sensitive information. 

Recent research~\cite{CTZ+24} has demonstrated that fine-tuning with a limited set of carefully selected examples can significantly amplify privacy leakage. 
Similarly, another study~\cite{YYC+24} demonstrated the ease of reversing alignment in open-access LLMs, thereby enabling models to generate harmful or unethical responses (see Table~\ref{tab:big}).

Mitigation strategies against adversarial fine-tuning include several measures.
Such measures encompass the implementation of controlled access to fine-tuning interfaces and continuous monitoring of model behavior.

\subsection{K: Rowhammer}\label{sec:k}
Rowhammer attacks exploit hardware-level vulnerabilities in dynamic random access memory (DRAM) modules, causing unauthorized bit flips through repeated memory access. 
In contrast to many other attacks, rowhammer targets the physical memory cells that store AI model parameters, resulting in the compromise of both predictive and generative models.

\begin{figure}[tb]
    \centering
    \includegraphics[width=\textwidth]{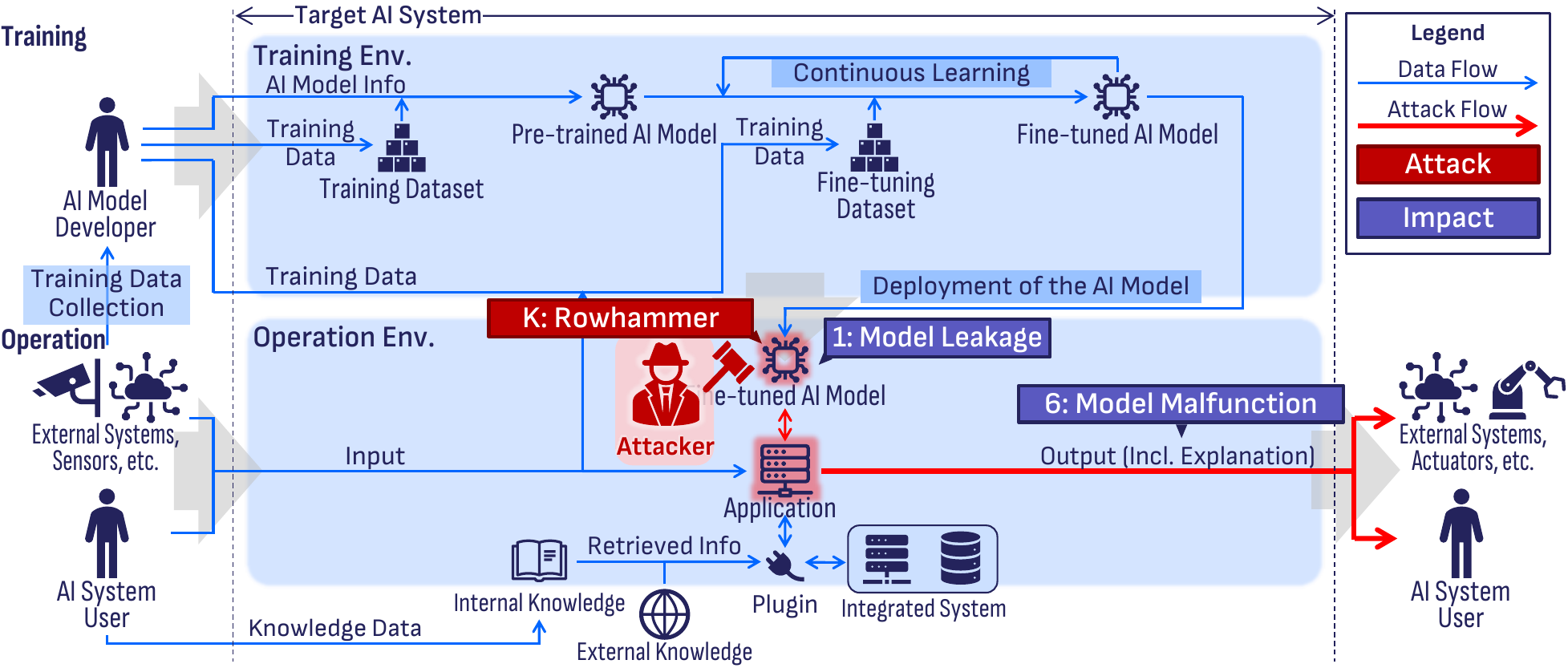}
    \caption{Attack K: Rowhammer (Overview)}
    \Description{
    The figure shows a Rowhammer attack targeting an AI system. The attacker exploits hardware vulnerabilities in memory modules storing AI model parameters, causing unintended bit flips. This manipulation results in model leakage or malfunction. Arrows indicate data flow, clearly identifying the hardware-level target and affected model components.
    }
    \label{fig:atk_k}
\end{figure}

The fundamental mechanism of rowhammer is the repeated access to specific memory rows, which induces electrical interference. 
This phenomenon results in unintended bit flips in adjacent memory rows. 
These unintentional flips have the potential to corrupt critical model parameters or execution codes during runtime (Figure~\ref{fig:atk_k}).

For instance, attacks such as DeepSteal~\cite{RCY+22} have exploited rowhammer-induced bit flips to remotely extract partial weights from neural network models. 
This has resulted in the unauthorized replication of the victim models. 
Similarly, the FrameFlip~\cite{LWX+24} attack demonstrated universal depletion of deep neural network inference performance by injecting faults into the runtime code, without requiring detailed knowledge of their internal structures.

The impacts of rowhammer include model malfunction, or model leakage (see Table~\ref{tab:big}).
Recent research demonstrated that even a single bit flip can significantly degrade or alter the performance of neural networks, highlighting the sensitivity of AI model inference to hardware-level perturbations~\cite{LWX+24}.

Mitigation strategies for rowhammer attacks range from the hardware to the software level.
On the hardware side, solutions include improved DRAM designs with stronger error correction, such as target row refresh. 
Software-based mitigations focus on robust memory management practices, such as randomizing memory allocation. 
Recently proposed techniques, such as the strategic ``rewiring'' of model neurons to hide critical weights, have also demonstrated effectiveness in enhancing the resilience of transformer-based models against bit-flip attacks~\cite{NMF+24}.

\section{Conclusions}
\label{sec:conclusions}
In this paper, we have provided an overview of known adversarial attacks specifically targeting AI systems, detailing their underlying mechanisms, impacts, and the attacker's capabilities. 
By categorizing these attacks according to their technical methodologies, we established a clear and intuitive taxonomy that facilitates both understanding and practical application, especially for researchers, developers, security practitioners, and policymakers who may lack specialized expertise in AI security.

Our analysis underscored the critical need to secure AI systems by highlighting their inherent vulnerabilities across multiple stages, from training to deployment. 
We have systematically illustrated the connections between individual attacks and their potential impacts on confidentiality, integrity, and availability---the core components of the cybersecurity CIA triad. 
The intuitive visual mappings provided throughout the paper are intended to help  practitioners in quickly identify vulnerabilities relevant to their AI deployments and formulate targeted mitigation strategies.

In addition, we extensively reviewed the existing works, including NIST AI 100-2e2025 and MITRE ATLAS\texttrademark, integrating foundational research with cutting-edge developments, and identified the limitations of these works. 
By providing additional insights, our work not only complements existing frameworks, but also provides practical guidance that could apply to actual security management scenarios, such as red-teaming exercises.

As AI technologies continue to rapidly advance, new threats will inevitably emerge, requiring continuous updates and extensions to threat modeling frameworks and defense strategies. 
Future research should focus on further exploring emerging AI vulnerabilities, refining defense methodologies, and expanding knowledge dissemination efforts to ensure that stakeholders at all levels are equipped to proactively address evolving AI security challenges.

Ultimately, improving AI security is not just a technical imperative but a broader societal responsibility. 
It requires an integrated approach that includes corporate governance, robust technological defenses, comprehensive policymaking, and ongoing international collaboration. 
We believe this paper contributes to the fundamental knowledge needed to address these challenges and promote safer, more reliable, and responsible AI systems.

\begin{acks}
The authors would like to thank all the J-AISI and IPA colleagues for their continuous support. 
This work would not have been possible without the valuable advice provided by Takehito Akima, Hiromu (Kit) Kitamura, Yuma Kurihara, Katsuhisa Nakazato, and Hikaru Matsuoka. 
Generative AI tools, specifically OpenAI ChatGPT and Google Gemini, were utilized to write part of the draft text of this paper efficiently. 
The authors have reviewed and verified all content in this paper.
\end{acks}

\bibliographystyle{ACM-Reference-Format}
\bibliography{aml}


\begin{thebibliography}{106}


\ifx \showCODEN    \undefined \def \showCODEN     #1{\unskip}     \fi
\ifx \showISBNx    \undefined \def \showISBNx     #1{\unskip}     \fi
\ifx \showISBNxiii \undefined \def \showISBNxiii  #1{\unskip}     \fi
\ifx \showISSN     \undefined \def \showISSN      #1{\unskip}     \fi
\ifx \showLCCN     \undefined \def \showLCCN      #1{\unskip}     \fi
\ifx \shownote     \undefined \def \shownote      #1{#1}          \fi
\ifx \showarticletitle \undefined \def \showarticletitle #1{#1}   \fi
\ifx \showURL      \undefined \def \showURL       {\relax}        \fi
\providecommand\bibfield[2]{#2}
\providecommand\bibinfo[2]{#2}
\providecommand\natexlab[1]{#1}
\providecommand\showeprint[2][]{arXiv:#2}

\bibitem[Adamczyk et~al\mbox{.}(2023)]%
        {APP+23}
\bibfield{author}{\bibinfo{person}{Monika Adamczyk}, \bibinfo{person}{Nineta Polemi}, \bibinfo{person}{Isabel Praça}, {and} \bibinfo{person}{Konstantinos Moulinos}.} \bibinfo{year}{2023}\natexlab{}.
\newblock \bibinfo{booktitle}{\emph{A multilayer framework for good cybersecurity practices for AI : security and resilience for smart health services and infrastructures}}.
\newblock \bibinfo{type}{{T}echnical {R}eport}. \bibinfo{institution}{European Union Agency for Cybersecurity}.
\newblock
\href{https://doi.org/10.2824/588830}{doi:\nolinkurl{10.2824/588830}}


\bibitem[Ahmed et~al\mbox{.}(2019)]%
        {AYM+19}
\bibfield{author}{\bibinfo{person}{Salem Ahmed}, \bibinfo{person}{Zhang Yang}, \bibinfo{person}{Humbert Mathias}, \bibinfo{person}{Berrang Pascal}, \bibinfo{person}{Fritz Mario}, {and} \bibinfo{person}{Backes Michael}.} \bibinfo{year}{2019}\natexlab{}.
\newblock \showarticletitle{ML-Leaks: Model and Data Independent Membership Inference Attacks and Defenses on Machine Learning Models}. In \bibinfo{booktitle}{\emph{Proceedings 2019 Network and Distributed System Security Symposium}} (San Diego, CA, USA) \emph{(\bibinfo{series}{NDSS 2019})}. \bibinfo{publisher}{Internet Society}, \bibinfo{address}{Reston, VA, USA}.
\newblock
\href{https://doi.org/10.14722/ndss.2019.23119}{doi:\nolinkurl{10.14722/ndss.2019.23119}}


\bibitem[Aldahdooh et~al\mbox{.}(2022)]%
        {AHF+22}
\bibfield{author}{\bibinfo{person}{Ahmed Aldahdooh}, \bibinfo{person}{Wassim Hamidouche}, \bibinfo{person}{Sid~Ahmed Fezza}, {and} \bibinfo{person}{Olivier D\'{e}forges}.} \bibinfo{year}{2022}\natexlab{}.
\newblock \showarticletitle{Adversarial Example Detection for DNN Models: A Review and Experimental Comparison}.
\newblock \bibinfo{journal}{\emph{Artificial Intelligence Review}} \bibinfo{volume}{55}, \bibinfo{number}{6} (\bibinfo{date}{Aug.} \bibinfo{year}{2022}), \bibinfo{pages}{4403–4462}.
\newblock
\showISSN{0269-2821}
\href{https://doi.org/10.1007/s10462-021-10125-w}{doi:\nolinkurl{10.1007/s10462-021-10125-w}}


\bibitem[Ateniese et~al\mbox{.}(2015)]%
        {AMS+15}
\bibfield{author}{\bibinfo{person}{Giuseppe Ateniese}, \bibinfo{person}{Luigi~V. Mancini}, \bibinfo{person}{Angelo Spognardi}, \bibinfo{person}{Antonio Villani}, \bibinfo{person}{Domenico Vitali}, {and} \bibinfo{person}{Giovanni Felici}.} \bibinfo{year}{2015}\natexlab{}.
\newblock \showarticletitle{Hacking Smart Machines with Smarter Ones: How to Extract Meaningful Data from Machine Learning Classifiers}.
\newblock \bibinfo{journal}{\emph{International Journal of Security and Networks}} \bibinfo{volume}{10}, \bibinfo{number}{3} (\bibinfo{date}{Sept.} \bibinfo{year}{2015}), \bibinfo{pages}{137–150}.
\newblock
\showISSN{1747-8405}
\href{https://doi.org/10.1504/IJSN.2015.071829}{doi:\nolinkurl{10.1504/IJSN.2015.071829}}


\bibitem[Balle et~al\mbox{.}(2022)]%
        {BCH22}
\bibfield{author}{\bibinfo{person}{Borja Balle}, \bibinfo{person}{Giovanni Cherubin}, {and} \bibinfo{person}{Jamie Hayes}.} \bibinfo{year}{2022}\natexlab{}.
\newblock \showarticletitle{Reconstructing Training Data with Informed Adversaries}. In \bibinfo{booktitle}{\emph{2022 IEEE Symposium on Security and Privacy (SP)}} (San Francisco, CA, USA). \bibinfo{publisher}{IEEE Computer Society}, \bibinfo{address}{Los Alamitos, CA, USA}, \bibinfo{pages}{1138--1156}.
\newblock
\href{https://doi.org/10.1109/SP46214.2022.9833677}{doi:\nolinkurl{10.1109/SP46214.2022.9833677}}


\bibitem[Boucher et~al\mbox{.}(2022)]%
        {BSA+22}
\bibfield{author}{\bibinfo{person}{Nicholas Boucher}, \bibinfo{person}{Ilia Shumailov}, \bibinfo{person}{Ross Anderson}, {and} \bibinfo{person}{Nicolas Papernot}.} \bibinfo{year}{2022}\natexlab{}.
\newblock \showarticletitle{Bad Characters: Imperceptible NLP Attacks}. In \bibinfo{booktitle}{\emph{2022 IEEE Symposium on Security and Privacy (SP)}} (San Francisco, CA, USA). \bibinfo{publisher}{IEEE Computer Society}, \bibinfo{address}{Los Alamitos, CA, USA}, \bibinfo{pages}{1987--2004}.
\newblock
\href{https://doi.org/10.1109/SP46214.2022.9833641}{doi:\nolinkurl{10.1109/SP46214.2022.9833641}}


\bibitem[Buzaglo et~al\mbox{.}(2023)]%
        {BHY+23}
\bibfield{author}{\bibinfo{person}{Gon Buzaglo}, \bibinfo{person}{Niv Haim}, \bibinfo{person}{Gilad Yehudai}, \bibinfo{person}{Gal Vardi}, {and} \bibinfo{person}{Michal Irani}.} \bibinfo{year}{2023}\natexlab{}.
\newblock \bibinfo{title}{Reconstructing Training Data from Multiclass Neural Networks}.
\newblock
\showeprint[arxiv]{2305.03350}~[cs.LG]


\bibitem[Cao et~al\mbox{.}(2021)]%
        {CJG21}
\bibfield{author}{\bibinfo{person}{Xiaoyu Cao}, \bibinfo{person}{Jinyuan Jia}, {and} \bibinfo{person}{Neil~Zhenqiang Gong}.} \bibinfo{year}{2021}\natexlab{}.
\newblock \showarticletitle{IPGuard: Protecting Intellectual Property of Deep Neural Networks via Fingerprinting the Classification Boundary}. In \bibinfo{booktitle}{\emph{Proceedings of the 2021 ACM Asia Conference on Computer and Communications Security}} (Virtual Event, Hong Kong) \emph{(\bibinfo{series}{ASIA CCS '21})}. \bibinfo{publisher}{Association for Computing Machinery}, \bibinfo{address}{New York, NY, USA}, \bibinfo{pages}{14–25}.
\newblock
\showISBNx{9781450382878}
\href{https://doi.org/10.1145/3433210.3437526}{doi:\nolinkurl{10.1145/3433210.3437526}}


\bibitem[Carlini(2021)]%
        {Car21}
\bibfield{author}{\bibinfo{person}{Nicholas Carlini}.} \bibinfo{year}{2021}\natexlab{}.
\newblock \showarticletitle{Poisoning the Unlabeled Dataset of {Semi-Supervised} Learning}. In \bibinfo{booktitle}{\emph{30th USENIX Security Symposium (USENIX Security 21)}}. \bibinfo{publisher}{USENIX Association}, \bibinfo{address}{Berkeley, CA, USA}, \bibinfo{pages}{1577--1592}.
\newblock
\showISBNx{978-1-939133-24-3}
\urldef\tempurl%
\url{https://www.usenix.org/conference/usenixsecurity21/presentation/carlini-poisoning}
\showURL{%
\tempurl}


\bibitem[Carlini et~al\mbox{.}(2023a)]%
        {CHN+23}
\bibfield{author}{\bibinfo{person}{Nicolas Carlini}, \bibinfo{person}{Jamie Hayes}, \bibinfo{person}{Milad Nasr}, \bibinfo{person}{Matthew Jagielski}, \bibinfo{person}{Vikash Sehwag}, \bibinfo{person}{Florian Tram{\`e}r}, \bibinfo{person}{Borja Balle}, \bibinfo{person}{Daphne Ippolito}, {and} \bibinfo{person}{Eric Wallace}.} \bibinfo{year}{2023}\natexlab{a}.
\newblock \showarticletitle{Extracting Training Data from Diffusion Models}. In \bibinfo{booktitle}{\emph{32nd USENIX Security Symposium (USENIX Security 23)}} (Anaheim, CA). \bibinfo{publisher}{USENIX Association}, \bibinfo{address}{Berkeley, CA, USA}, \bibinfo{pages}{5253--5270}.
\newblock
\showISBNx{978-1-939133-37-3}
\urldef\tempurl%
\url{https://www.usenix.org/conference/usenixsecurity23/presentation/carlini}
\showURL{%
\tempurl}


\bibitem[Carlini et~al\mbox{.}(2019)]%
        {CLE+19}
\bibfield{author}{\bibinfo{person}{Nicholas Carlini}, \bibinfo{person}{Chang Liu}, \bibinfo{person}{{\'U}lfar Erlingsson}, \bibinfo{person}{Jernej Kos}, {and} \bibinfo{person}{Dawn Song}.} \bibinfo{year}{2019}\natexlab{}.
\newblock \showarticletitle{The Secret Sharer: Evaluating and Testing Unintended Memorization in Neural Networks}. In \bibinfo{booktitle}{\emph{28th USENIX Security Symposium (USENIX Security 19)}} (Santa Clara, CA, USA). \bibinfo{publisher}{USENIX Association}, \bibinfo{address}{Berkeley, CA, USA}, \bibinfo{pages}{267--284}.
\newblock
\showISBNx{978-1-939133-06-9}
\urldef\tempurl%
\url{https://www.usenix.org/conference/usenixsecurity19/presentation/carlini}
\showURL{%
\tempurl}


\bibitem[Carlini et~al\mbox{.}(2023b)]%
        {CNC+23}
\bibfield{author}{\bibinfo{person}{Nicholas Carlini}, \bibinfo{person}{Milad Nasr}, \bibinfo{person}{Christopher~A. Choquette-Choo}, \bibinfo{person}{Matthew Jagielski}, \bibinfo{person}{Irena Gao}, \bibinfo{person}{Pang Wei~W Koh}, \bibinfo{person}{Daphne Ippolito}, \bibinfo{person}{Florian Tramer}, {and} \bibinfo{person}{Ludwig Schmidt}.} \bibinfo{year}{2023}\natexlab{b}.
\newblock \showarticletitle{Are aligned neural networks adversarially aligned?}. In \bibinfo{booktitle}{\emph{Advances in Neural Information Processing Systems}} (New Orleans, LA, USA) \emph{(\bibinfo{series}{NIPS '23}, Vol.~\bibinfo{volume}{36})}, \bibfield{editor}{\bibinfo{person}{A.~Oh}, \bibinfo{person}{T.~Naumann}, \bibinfo{person}{A.~Globerson}, \bibinfo{person}{K.~Saenko}, \bibinfo{person}{M.~Hardt}, {and} \bibinfo{person}{S.~Levine}} (Eds.). \bibinfo{publisher}{Curran Associates, Inc.}, \bibinfo{address}{Red Hook, NY, USA}, \bibinfo{pages}{61478--61500}.
\newblock
\urldef\tempurl%
\url{https://proceedings.neurips.cc/paper_files/paper/2023/hash/c1f0b856a35986348ab3414177266f75-Abstract-Conference.html}
\showURL{%
\tempurl}


\bibitem[Carlini et~al\mbox{.}(2021)]%
        {CTW+21}
\bibfield{author}{\bibinfo{person}{Nicholas Carlini}, \bibinfo{person}{Florian Tram{\`e}r}, \bibinfo{person}{Eric Wallace}, \bibinfo{person}{Matthew Jagielski}, \bibinfo{person}{Ariel Herbert-Voss}, \bibinfo{person}{Katherine Lee}, \bibinfo{person}{Adam Roberts}, \bibinfo{person}{Tom Brown}, \bibinfo{person}{Dawn Song}, \bibinfo{person}{{\'U}lfar Erlingsson}, \bibinfo{person}{Alina Oprea}, {and} \bibinfo{person}{Colin Raffel}.} \bibinfo{year}{2021}\natexlab{}.
\newblock \showarticletitle{Extracting Training Data from Large Language Models}. In \bibinfo{booktitle}{\emph{30th USENIX Security Symposium (USENIX Security 21)}}. \bibinfo{publisher}{USENIX Association}, \bibinfo{address}{Berkeley, CA, USA}, \bibinfo{pages}{2633--2650}.
\newblock
\showISBNx{978-1-939133-24-3}
\urldef\tempurl%
\url{https://www.usenix.org/conference/usenixsecurity21/presentation/carlini-extracting}
\showURL{%
\tempurl}


\bibitem[Centre(2024)]%
        {Nat24}
\bibfield{author}{\bibinfo{person}{National Cyber~Security Centre}.} \bibinfo{year}{2024}\natexlab{}.
\newblock \bibinfo{title}{Machine learning principles}.
\newblock
\urldef\tempurl%
\url{https://www.ncsc.gov.uk/collection/machine-learning-principles}
\showURL{%
\tempurl}


\bibitem[Centre et~al\mbox{.}(2023)]%
        {NCI+23}
\bibfield{author}{\bibinfo{person}{National Cyber~Security Centre}, \bibinfo{person}{{Cybersecurity and Infrastructure Security Agency}}, \bibinfo{person}{National~Security Agency}, \bibinfo{person}{Federal~Bureau of Investigation}, \bibinfo{person}{Australian Signals Directorate's Australian Cyber~Security Centre}, \bibinfo{person}{Canadian~Centre for Cyber~Security}, \bibinfo{person}{New Zealand National Cyber~Security Centre}, \bibinfo{person}{Chile's~Government CSIRT}, \bibinfo{person}{{National Cyber and Information Security Agency of the Czech Republic}}, \bibinfo{person}{Information System~Authority of Estonia}, \bibinfo{person}{National Cyber Security~Centre of Estonia}, \bibinfo{person}{French~Cybersecurity Agency}, \bibinfo{person}{Germany's Federal~Office for Information~Security}, \bibinfo{person}{Israeli National~Cyber Directorate}, \bibinfo{person}{Italian National~Cybersecurity Agency}, \bibinfo{person}{{Japan's National center of Incident readiness and Strategy for Cybersecurity}},
  \bibinfo{person}{{Japan's Secretariat of Science, Technology and Innovation Policy, Cabinet Office}}, \bibinfo{person}{Nigeria's National Information Technology~Development Agency}, \bibinfo{person}{Norwegian National Cyber~Security Centre}, \bibinfo{person}{Poland~Ministry of Digital~Affairs}, \bibinfo{person}{Poland's NASK National~Research Institute}, \bibinfo{person}{Republic of Korea National Intelligence~Service}, {and} \bibinfo{person}{Cyber Security~Agency of Singapore}.} \bibinfo{year}{2023}\natexlab{}.
\newblock \bibinfo{title}{Guidelines for secure AI system development}.
\newblock
\urldef\tempurl%
\url{https://www.ncsc.gov.uk/collection/guidelines-secure-ai-system-development}
\showURL{%
\tempurl}


\bibitem[Chao et~al\mbox{.}(2024)]%
        {CRD+24}
\bibfield{author}{\bibinfo{person}{Patrick Chao}, \bibinfo{person}{Alexander Robey}, \bibinfo{person}{Edgar Dobriban}, \bibinfo{person}{Hamed Hassani}, \bibinfo{person}{George~J. Pappas}, {and} \bibinfo{person}{Eric Wong}.} \bibinfo{year}{2024}\natexlab{}.
\newblock \bibinfo{title}{Jailbreaking Black Box Large Language Models in Twenty Queries}.
\newblock
\showeprint[arxiv]{2310.08419}~[cs.LG]


\bibitem[Chen and Babar(2024)]%
        {CB24}
\bibfield{author}{\bibinfo{person}{Huaming Chen} {and} \bibinfo{person}{M.~Ali Babar}.} \bibinfo{year}{2024}\natexlab{}.
\newblock \showarticletitle{Security for Machine Learning-based Software Systems: A Survey of Threats, Practices, and Challenges}.
\newblock \bibinfo{journal}{\emph{ACM Comput. Surv.}} \bibinfo{volume}{56}, \bibinfo{number}{6}, Article \bibinfo{articleno}{151} (\bibinfo{date}{Feb.} \bibinfo{year}{2024}), \bibinfo{numpages}{38}~pages.
\newblock
\showISSN{0360-0300}
\href{https://doi.org/10.1145/3638531}{doi:\nolinkurl{10.1145/3638531}}


\bibitem[Chen et~al\mbox{.}(2024)]%
        {CTZ+24}
\bibfield{author}{\bibinfo{person}{Xiaoyi Chen}, \bibinfo{person}{Siyuan Tang}, \bibinfo{person}{Rui Zhu}, \bibinfo{person}{Shijun Yan}, \bibinfo{person}{Lei Jin}, \bibinfo{person}{Zihao Wang}, \bibinfo{person}{Liya Su}, \bibinfo{person}{Zhikun Zhang}, \bibinfo{person}{XiaoFeng Wang}, {and} \bibinfo{person}{Haixu Tang}.} \bibinfo{year}{2024}\natexlab{}.
\newblock \showarticletitle{The Janus Interface: How Fine-Tuning in Large Language Models Amplifies the Privacy Risks}. In \bibinfo{booktitle}{\emph{Proceedings of the 2024 on ACM SIGSAC Conference on Computer and Communications Security}} (Salt Lake City, UT, USA) \emph{(\bibinfo{series}{CCS '24})}. \bibinfo{publisher}{Association for Computing Machinery}, \bibinfo{address}{New York, NY, USA}, \bibinfo{pages}{1285–1299}.
\newblock
\showISBNx{9798400706363}
\href{https://doi.org/10.1145/3658644.3690325}{doi:\nolinkurl{10.1145/3658644.3690325}}


\bibitem[Cinà et~al\mbox{.}(2023)]%
        {CDB+23}
\bibfield{author}{\bibinfo{person}{Antonio~Emanuele Cinà}, \bibinfo{person}{Ambra Demontis}, \bibinfo{person}{Battista Biggio}, \bibinfo{person}{Fabio Roli}, {and} \bibinfo{person}{Marcello Pelillo}.} \bibinfo{year}{2023}\natexlab{}.
\newblock \bibinfo{title}{Energy-Latency Attacks via Sponge Poisoning}.
\newblock
\showeprint[arxiv]{2203.08147}~[cs.CR]


\bibitem[Correia-Silva et~al\mbox{.}(2018)]%
        {CBB+18}
\bibfield{author}{\bibinfo{person}{Jacson~Rodrigues Correia-Silva}, \bibinfo{person}{Rodrigo~F. Berriel}, \bibinfo{person}{Claudine Badue}, \bibinfo{person}{Alberto~F. de Souza}, {and} \bibinfo{person}{Thiago Oliveira-Santos}.} \bibinfo{year}{2018}\natexlab{}.
\newblock \showarticletitle{Copycat CNN: Stealing Knowledge by Persuading Confession with Random Non-Labeled Data}. In \bibinfo{booktitle}{\emph{2018 International Joint Conference on Neural Networks (IJCNN)}} (Rio de Janeiro, Brazil). \bibinfo{publisher}{IEEE}, \bibinfo{address}{New York, NY, USA}, \bibinfo{pages}{1--8}.
\newblock
\href{https://doi.org/10.1109/IJCNN.2018.8489592}{doi:\nolinkurl{10.1109/IJCNN.2018.8489592}}


\bibitem[Croce et~al\mbox{.}(2021)]%
        {CAS+21}
\bibfield{author}{\bibinfo{person}{Francesco Croce}, \bibinfo{person}{Maksym Andriushchenko}, \bibinfo{person}{Vikash Sehwag}, \bibinfo{person}{Edoardo Debenedetti}, \bibinfo{person}{Nicolas Flammarion}, \bibinfo{person}{Mung Chiang}, \bibinfo{person}{Prateek Mittal}, {and} \bibinfo{person}{Matthias Hein}.} \bibinfo{year}{2021}\natexlab{}.
\newblock \showarticletitle{RobustBench: a standardized adversarial robustness benchmark}. In \bibinfo{booktitle}{\emph{Thirty-fifth Conference on Neural Information Processing Systems (NeurIPS 2021) Track on Datasets and Benchmarks}}, \bibfield{editor}{\bibinfo{person}{J.~Vanschoren} {and} \bibinfo{person}{S.~Yeung}} (Eds.). \bibinfo{publisher}{Neural Information Processing Systems Foundation}, \bibinfo{address}{San Diego, CA, USA}.
\newblock
\urldef\tempurl%
\url{https://datasets-benchmarks-proceedings.neurips.cc/paper_files/paper/2021/hash/a3c65c2974270fd093ee8a9bf8ae7d0b-Abstract-round2.html}
\showURL{%
\tempurl}


\bibitem[Das et~al\mbox{.}(2025)]%
        {DAW25}
\bibfield{author}{\bibinfo{person}{Badhan~Chandra Das}, \bibinfo{person}{M.~Hadi Amini}, {and} \bibinfo{person}{Yanzhao Wu}.} \bibinfo{year}{2025}\natexlab{}.
\newblock \showarticletitle{Security and Privacy Challenges of Large Language Models: A Survey}.
\newblock \bibinfo{journal}{\emph{ACM Comput. Surv.}} \bibinfo{volume}{57}, \bibinfo{number}{6}, Article \bibinfo{articleno}{152} (\bibinfo{date}{Feb.} \bibinfo{year}{2025}), \bibinfo{numpages}{39}~pages.
\newblock
\showISSN{0360-0300}
\href{https://doi.org/10.1145/3712001}{doi:\nolinkurl{10.1145/3712001}}


\bibitem[{Digital Architecture Research Center} et~al\mbox{.}(2024)]%
        {AIQM4}
\bibfield{author}{\bibinfo{person}{{Digital Architecture Research Center}}, \bibinfo{person}{{Cyber Physical Security Research Center}}, {and} \bibinfo{person}{{Artificial Intelligence Research Center}}.} \bibinfo{year}{2024}\natexlab{}.
\newblock \bibinfo{booktitle}{\emph{Machine Learning Quality Management Guidline (Revision 4.2.0)}}.
\newblock \bibinfo{type}{{T}echnical {R}eport}. \bibinfo{institution}{{National Institute of Advanced Industrial Science and Technology (AIST)}}.
\newblock
\urldef\tempurl%
\url{https://www.digiarc.aist.go.jp/publication/aiqm/guideline-rev4.html}
\showURL{%
\tempurl}
\newblock
\shownote{in Japanese}.


\bibitem[Dombrowski et~al\mbox{.}(2019)]%
        {DAA+19}
\bibfield{author}{\bibinfo{person}{Ann-Kathrin Dombrowski}, \bibinfo{person}{Maximillian Alber}, \bibinfo{person}{Christopher Anders}, \bibinfo{person}{Marcel Ackermann}, \bibinfo{person}{Klaus-Robert M\"{u}ller}, {and} \bibinfo{person}{Pan Kessel}.} \bibinfo{year}{2019}\natexlab{}.
\newblock \showarticletitle{Explanations can be manipulated and geometry is to blame}. In \bibinfo{booktitle}{\emph{Advances in Neural Information Processing Systems}} (Vancouver, Canada), \bibfield{editor}{\bibinfo{person}{H.~Wallach}, \bibinfo{person}{H.~Larochelle}, \bibinfo{person}{A.~Beygelzimer}, \bibinfo{person}{F.~d\textquotesingle Alch\'{e}-Buc}, \bibinfo{person}{E.~Fox}, {and} \bibinfo{person}{R.~Garnett}} (Eds.), Vol.~\bibinfo{volume}{32}. \bibinfo{publisher}{Curran Associates, Inc.}, \bibinfo{address}{Red Hook, NY, USA}.
\newblock
\urldef\tempurl%
\url{https://proceedings.neurips.cc/paper/2019/hash/bb836c01cdc9120a9c984c525e4b1a4a-Abstract.html}
\showURL{%
\tempurl}


\bibitem[Dong et~al\mbox{.}(2024)]%
        {DZY+24}
\bibfield{author}{\bibinfo{person}{Zhichen Dong}, \bibinfo{person}{Zhanhui Zhou}, \bibinfo{person}{Chao Yang}, \bibinfo{person}{Jing Shao}, {and} \bibinfo{person}{Yu Qiao}.} \bibinfo{year}{2024}\natexlab{}.
\newblock \showarticletitle{Attacks, Defenses and Evaluations for {LLM} Conversation Safety: A Survey}. In \bibinfo{booktitle}{\emph{Proceedings of the 2024 Conference of the North American Chapter of the Association for Computational Linguistics: Human Language Technologies (Volume 1: Long Papers)}} (Mexico City, Mexico), \bibfield{editor}{\bibinfo{person}{Kevin Duh}, \bibinfo{person}{Helena Gomez}, {and} \bibinfo{person}{Steven Bethard}} (Eds.). \bibinfo{publisher}{Association for Computational Linguistics}, \bibinfo{address}{Stroudsburg, PA, USA}, \bibinfo{pages}{6734--6747}.
\newblock
\href{https://doi.org/10.18653/v1/2024.naacl-long.375}{doi:\nolinkurl{10.18653/v1/2024.naacl-long.375}}


\bibitem[Ebrahimi et~al\mbox{.}(2018)]%
        {ERL+18}
\bibfield{author}{\bibinfo{person}{Javid Ebrahimi}, \bibinfo{person}{Anyi Rao}, \bibinfo{person}{Daniel Lowd}, {and} \bibinfo{person}{Dejing Dou}.} \bibinfo{year}{2018}\natexlab{}.
\newblock \showarticletitle{{H}ot{F}lip: White-Box Adversarial Examples for Text Classification}. In \bibinfo{booktitle}{\emph{Proceedings of the 56th Annual Meeting of the Association for Computational Linguistics (Volume 2: Short Papers)}} (Melbourne, Australia), \bibfield{editor}{\bibinfo{person}{Iryna Gurevych} {and} \bibinfo{person}{Yusuke Miyao}} (Eds.). \bibinfo{publisher}{Association for Computational Linguistics}, \bibinfo{address}{Stroudsburg, PA, USA}, \bibinfo{pages}{31--36}.
\newblock
\href{https://doi.org/10.18653/v1/P18-2006}{doi:\nolinkurl{10.18653/v1/P18-2006}}


\bibitem[Ewart et~al\mbox{.}(2024)]%
        {ESH+24}
\bibfield{author}{\bibinfo{person}{Aidan Ewart}, \bibinfo{person}{Abhay Sheshadri}, \bibinfo{person}{Phillip~Huang Guo}, \bibinfo{person}{Aengus Lynch}, \bibinfo{person}{Cindy Wu}, \bibinfo{person}{Vivek Hebbar}, \bibinfo{person}{Henry Sleight}, \bibinfo{person}{Asa~Cooper Stickland}, \bibinfo{person}{Ethan Perez}, \bibinfo{person}{Dylan Hadfield-Menell}, {and} \bibinfo{person}{Stephen Casper}.} \bibinfo{year}{2024}\natexlab{}.
\newblock \bibinfo{title}{Latent Adversarial Training Improves Robustness to Persistent Harmful Behaviors in {LLM}s}.  (\bibinfo{year}{2024}).
\newblock
\urldef\tempurl%
\url{https://neurips.cc/virtual/2024/103310}
\showURL{%
\tempurl}
\newblock
\shownote{Workshop on Socially Responsible Language Modelling Research (SoLaR) at NeurIPS}.


\bibitem[Fredrikson et~al\mbox{.}(2015)]%
        {FJR15}
\bibfield{author}{\bibinfo{person}{Matt Fredrikson}, \bibinfo{person}{Somesh Jha}, {and} \bibinfo{person}{Thomas Ristenpart}.} \bibinfo{year}{2015}\natexlab{}.
\newblock \showarticletitle{Model Inversion Attacks that Exploit Confidence Information and Basic Countermeasures}. In \bibinfo{booktitle}{\emph{Proceedings of the 22nd ACM SIGSAC Conference on Computer and Communications Security}} (Denver, Colorado, USA) \emph{(\bibinfo{series}{CCS '15})}. \bibinfo{publisher}{Association for Computing Machinery}, \bibinfo{address}{New York, NY, USA}, \bibinfo{pages}{1322–1333}.
\newblock
\showISBNx{9781450338325}
\href{https://doi.org/10.1145/2810103.2813677}{doi:\nolinkurl{10.1145/2810103.2813677}}


\bibitem[Ganju et~al\mbox{.}(2018)]%
        {GWY+18}
\bibfield{author}{\bibinfo{person}{Karan Ganju}, \bibinfo{person}{Qi Wang}, \bibinfo{person}{Wei Yang}, \bibinfo{person}{Carl~A. Gunter}, {and} \bibinfo{person}{Nikita Borisov}.} \bibinfo{year}{2018}\natexlab{}.
\newblock \showarticletitle{Property Inference Attacks on Fully Connected Neural Networks using Permutation Invariant Representations}. In \bibinfo{booktitle}{\emph{Proceedings of the 2018 ACM SIGSAC Conference on Computer and Communications Security}} (Toronto, Canada) \emph{(\bibinfo{series}{CCS '18})}. \bibinfo{publisher}{Association for Computing Machinery}, \bibinfo{address}{New York, NY, USA}, \bibinfo{pages}{619–633}.
\newblock
\showISBNx{9781450356930}
\href{https://doi.org/10.1145/3243734.3243834}{doi:\nolinkurl{10.1145/3243734.3243834}}


\bibitem[Goodfellow et~al\mbox{.}(2015)]%
        {GSS15}
\bibfield{author}{\bibinfo{person}{Ian~J. Goodfellow}, \bibinfo{person}{Jonathon Shlens}, {and} \bibinfo{person}{Christian Szegedy}.} \bibinfo{year}{2015}\natexlab{}.
\newblock \bibinfo{title}{Explaining and Harnessing Adversarial Examples}.
\newblock
\showeprint[arxiv]{1412.6572}~[stat.ML]
\newblock
\shownote{3rd International Conference on Learning Representations, {ICLR} 2015, San Diego, CA, USA, May 7-9, 2015, Conference Track Proceedings}.


\bibitem[Grana(2020)]%
        {Gra20}
\bibfield{author}{\bibinfo{person}{Justin Grana}.} \bibinfo{year}{2020}\natexlab{}.
\newblock \showarticletitle{Perturbing Inputs to Prevent Model Stealing}. In \bibinfo{booktitle}{\emph{2020 IEEE Conference on Communications and Network Security (CNS)}} (Avignon, France). \bibinfo{publisher}{IEEE}, \bibinfo{address}{New York, NY, USA}, \bibinfo{pages}{1--9}.
\newblock
\href{https://doi.org/10.1109/CNS48642.2020.9162336}{doi:\nolinkurl{10.1109/CNS48642.2020.9162336}}


\bibitem[Greshake et~al\mbox{.}(2023)]%
        {GAM+23}
\bibfield{author}{\bibinfo{person}{Kai Greshake}, \bibinfo{person}{Sahar Abdelnabi}, \bibinfo{person}{Shailesh Mishra}, \bibinfo{person}{Christoph Endres}, \bibinfo{person}{Thorsten Holz}, {and} \bibinfo{person}{Mario Fritz}.} \bibinfo{year}{2023}\natexlab{}.
\newblock \showarticletitle{Not What You've Signed Up For: Compromising Real-World LLM-Integrated Applications with Indirect Prompt Injection}. In \bibinfo{booktitle}{\emph{Proceedings of the 16th ACM Workshop on Artificial Intelligence and Security}} (Copenhagen, Denmark) \emph{(\bibinfo{series}{AISec '23})}. \bibinfo{publisher}{Association for Computing Machinery}, \bibinfo{address}{New York, NY, USA}, \bibinfo{pages}{79–90}.
\newblock
\showISBNx{9798400702600}
\href{https://doi.org/10.1145/3605764.3623985}{doi:\nolinkurl{10.1145/3605764.3623985}}


\bibitem[Guo et~al\mbox{.}(2021)]%
        {GSJ+21}
\bibfield{author}{\bibinfo{person}{Chuan Guo}, \bibinfo{person}{Alexandre Sablayrolles}, \bibinfo{person}{Herv\'e J\'egou}, {and} \bibinfo{person}{Douwe Kiela}.} \bibinfo{year}{2021}\natexlab{}.
\newblock \showarticletitle{Gradient-based Adversarial Attacks against Text Transformers}. In \bibinfo{booktitle}{\emph{Proceedings of the 2021 Conference on Empirical Methods in Natural Language Processing}} (Punta Cana, Dominican Republic), \bibfield{editor}{\bibinfo{person}{Marie-Francine Moens}, \bibinfo{person}{Xuanjing Huang}, \bibinfo{person}{Lucia Specia}, {and} \bibinfo{person}{Scott Wen-tau Yih}} (Eds.). \bibinfo{publisher}{Association for Computational Linguistics}, \bibinfo{address}{Stroudsburg, PA, USA}, \bibinfo{pages}{5747--5757}.
\newblock
\href{https://doi.org/10.18653/v1/2021.emnlp-main.464}{doi:\nolinkurl{10.18653/v1/2021.emnlp-main.464}}


\bibitem[Haim et~al\mbox{.}(2022)]%
        {HVY+22}
\bibfield{author}{\bibinfo{person}{Niv Haim}, \bibinfo{person}{Gal Vardi}, \bibinfo{person}{Gilad Yehudai}, \bibinfo{person}{Ohad Shamir}, {and} \bibinfo{person}{Michal Irani}.} \bibinfo{year}{2022}\natexlab{}.
\newblock \showarticletitle{Reconstructing Training Data from Trained Neural Networks}. In \bibinfo{booktitle}{\emph{Proceedings of the 36th International Conference on Neural Information Processing Systems}} (New Orleans, LA, USA) \emph{(\bibinfo{series}{NIPS '22})}. \bibinfo{publisher}{Curran Associates Inc.}, \bibinfo{address}{Red Hook, NY, USA}, Article \bibinfo{articleno}{1665}, \bibinfo{numpages}{14}~pages.
\newblock
\showISBNx{9781713871088}
\urldef\tempurl%
\url{https://proceedings.neurips.cc/paper_files/paper/2022/file/906927370cbeb537781100623cca6fa6-Paper-Conference.pdf}
\showURL{%
\tempurl}


\bibitem[Han et~al\mbox{.}(2024)]%
        {HRE+24}
\bibfield{author}{\bibinfo{person}{Seungju Han}, \bibinfo{person}{Kavel Rao}, \bibinfo{person}{Allyson Ettinger}, \bibinfo{person}{Liwei Jiang}, \bibinfo{person}{Bill~Yuchen Lin}, \bibinfo{person}{Nathan Lambert}, \bibinfo{person}{Yejin Choi}, {and} \bibinfo{person}{Nouha Dziri}.} \bibinfo{year}{2024}\natexlab{}.
\newblock \showarticletitle{Wild{G}uard: {O}pen One-stop Moderation Tools for Safety Risks, Jailbreaks, and Refusals of {LLM}s}. In \bibinfo{booktitle}{\emph{The Thirty-eight Conference on Neural Information Processing Systems (NeurIPS 2024) Track on Datasets and Benchmarks}} (Vancouver, BC, Canada), \bibfield{editor}{\bibinfo{person}{A.~Globerson}, \bibinfo{person}{L.~Mackey}, \bibinfo{person}{D.~Belgrave}, \bibinfo{person}{A.~Fan}, \bibinfo{person}{U.~Paquet}, \bibinfo{person}{J.~Tomczak}, {and} \bibinfo{person}{C.~Zhang}} (Eds.), Vol.~\bibinfo{volume}{37}. \bibinfo{publisher}{Curran Associates, Inc.}, \bibinfo{address}{Red Hook, NY, USA}, \bibinfo{pages}{8093--8131}.
\newblock
\urldef\tempurl%
\url{https://proceedings.neurips.cc/paper_files/paper/2024/hash/0f69b4b96a46f284b726fbd70f74fb3b-Abstract-Datasets_and_Benchmarks_Track.html}
\showURL{%
\tempurl}


\bibitem[Hines et~al\mbox{.}(2024)]%
        {HLH+24}
\bibfield{author}{\bibinfo{person}{Keegan Hines}, \bibinfo{person}{Gary Lopez}, \bibinfo{person}{Matthew Hall}, \bibinfo{person}{Federico Zarfati}, \bibinfo{person}{Yonatan Zunger}, {and} \bibinfo{person}{Emre K{\i}c{\i}man}.} \bibinfo{year}{2024}\natexlab{}.
\newblock \showarticletitle{Defending Against Indirect Prompt Injection Attacks With Spotlighting}. In \bibinfo{booktitle}{\emph{Conference on Applied Machine Learning for Information Security (CAMLIS 2024)}} (Arlington, VA, USA), \bibfield{editor}{\bibinfo{person}{Rachel Allen}, \bibinfo{person}{Sagar Samtani}, \bibinfo{person}{Edward Raff}, {and} \bibinfo{person}{Ethan Rudd}} (Eds.), Vol.~\bibinfo{volume}{3920}. \bibinfo{publisher}{CEUR-WS.org}, \bibinfo{address}{Aachen, Germany}, \bibinfo{pages}{48--62}.
\newblock
\urldef\tempurl%
\url{https://ceur-ws.org/Vol-3920/paper03.pdf}
\showURL{%
\tempurl}


\bibitem[Hu et~al\mbox{.}(2021)]%
        {HKQ+21}
\bibfield{author}{\bibinfo{person}{Yupeng Hu}, \bibinfo{person}{Wenxin Kuang}, \bibinfo{person}{Zheng Qin}, \bibinfo{person}{Kenli Li}, \bibinfo{person}{Jiliang Zhang}, \bibinfo{person}{Yansong Gao}, \bibinfo{person}{Wenjia Li}, {and} \bibinfo{person}{Keqin Li}.} \bibinfo{year}{2021}\natexlab{}.
\newblock \showarticletitle{Artificial Intelligence Security: Threats and Countermeasures}.
\newblock \bibinfo{journal}{\emph{ACM Comput. Surv.}} \bibinfo{volume}{55}, \bibinfo{number}{1}, Article \bibinfo{articleno}{20} (\bibinfo{date}{Nov.} \bibinfo{year}{2021}), \bibinfo{numpages}{36}~pages.
\newblock
\showISSN{0360-0300}
\href{https://doi.org/10.1145/3487890}{doi:\nolinkurl{10.1145/3487890}}


\bibitem[(J-AISI)(2024a)]%
        {JAISI24b}
\bibfield{author}{\bibinfo{person}{Japan AI Safety~Institute (J-AISI)}.} \bibinfo{year}{2024}\natexlab{a}.
\newblock \bibinfo{title}{Guide to Red Teaming Methodology on AI Safety}.
\newblock
\urldef\tempurl%
\url{https://aisi.go.jp/output/output_framework/guide_to_red_teaming_methodology_on_ai_safety/}
\showURL{%
\tempurl}


\bibitem[(J-AISI)(2024b)]%
        {JAISI24a}
\bibfield{author}{\bibinfo{person}{Japan AI Safety~Institute (J-AISI)}.} \bibinfo{year}{2024}\natexlab{b}.
\newblock \bibinfo{title}{Overview of the Japan AI Safety Institute (J-AISI)}.
\newblock
\urldef\tempurl%
\url{https://aisi.go.jp/about/}
\showURL{%
\tempurl}


\bibitem[(J-AISI)(2025)]%
        {JAISI25a}
\bibfield{author}{\bibinfo{person}{Japan AI Safety~Institute (J-AISI)}.} \bibinfo{year}{2025}\natexlab{}.
\newblock \bibinfo{title}{Known Attacks and Their Impacts on AI Systems}.
\newblock
\urldef\tempurl%
\url{https://aisi.go.jp/output/output_security/known_attacks_and_impacts/}
\showURL{%
\tempurl}


\bibitem[Jagielski et~al\mbox{.}(2018)]%
        {JOB+18}
\bibfield{author}{\bibinfo{person}{Matthew Jagielski}, \bibinfo{person}{Alina Oprea}, \bibinfo{person}{Battista Biggio}, \bibinfo{person}{Chang Liu}, \bibinfo{person}{Cristina Nita-Rotaru}, {and} \bibinfo{person}{Bo Li}.} \bibinfo{year}{2018}\natexlab{}.
\newblock \showarticletitle{Manipulating Machine Learning: Poisoning Attacks and Countermeasures for Regression Learning}. In \bibinfo{booktitle}{\emph{2018 IEEE Symposium on Security and Privacy (SP)}} (San Francisco, CA, USA). \bibinfo{publisher}{IEEE}, \bibinfo{address}{New York, NY, USA}, \bibinfo{pages}{19--35}.
\newblock
\href{https://doi.org/10.1109/SP.2018.00057}{doi:\nolinkurl{10.1109/SP.2018.00057}}


\bibitem[Juuti et~al\mbox{.}(2019)]%
        {JSM+19}
\bibfield{author}{\bibinfo{person}{Mika Juuti}, \bibinfo{person}{Sebastian Szyller}, \bibinfo{person}{Samuel Marchal}, {and} \bibinfo{person}{N. Asokan}.} \bibinfo{year}{2019}\natexlab{}.
\newblock \showarticletitle{PRADA: Protecting Against DNN Model Stealing Attacks}. In \bibinfo{booktitle}{\emph{2019 IEEE European Symposium on Security and Privacy (EuroS\&P)}} (Stockholm, Sweden). \bibinfo{publisher}{IEEE Computer Society}, \bibinfo{address}{Los Alamitos, CA, USA}, \bibinfo{pages}{512--527}.
\newblock
\href{https://doi.org/10.1109/EuroSP.2019.00044}{doi:\nolinkurl{10.1109/EuroSP.2019.00044}}


\bibitem[Kawamoto et~al\mbox{.}(2023)]%
        {KMK+23}
\bibfield{author}{\bibinfo{person}{Yusuke Kawamoto}, \bibinfo{person}{Kazumasa Miyake}, \bibinfo{person}{Koichi Konishi}, {and} \bibinfo{person}{Yutaka Oiwa}.} \bibinfo{year}{2023}\natexlab{}.
\newblock \bibinfo{title}{Threats, Vulnerabilities, and Controls of Machine Learning Based Systems: A Survey and Taxonomy}.
\newblock
\showeprint[arxiv]{2301.07474}~[cs.CR]
\urldef\tempurl%
\url{https://arxiv.org/abs/2301.07474}
\showURL{%
\tempurl}


\bibitem[Kong et~al\mbox{.}(2021)]%
        {KXW+21}
\bibfield{author}{\bibinfo{person}{Zixiao Kong}, \bibinfo{person}{Jingfeng Xue}, \bibinfo{person}{Yong Wang}, \bibinfo{person}{Lu Huang}, \bibinfo{person}{Zequn Niu}, {and} \bibinfo{person}{Feng Li}.} \bibinfo{year}{2021}\natexlab{}.
\newblock \showarticletitle{A Survey on Adversarial Attack in the Age of Artificial Intelligence}.
\newblock \bibinfo{journal}{\emph{Wireless Communications and Mobile Computing}} \bibinfo{volume}{2021}, \bibinfo{number}{1} (\bibinfo{year}{2021}), \bibinfo{pages}{4907754}.
\newblock
\href{https://doi.org/10.1155/2021/4907754}{doi:\nolinkurl{10.1155/2021/4907754}}


\bibitem[Kyle(2025)]%
        {DIST25}
\bibfield{author}{\bibinfo{person}{Peter Kyle}.} \bibinfo{year}{2025}\natexlab{}.
\newblock \bibinfo{title}{Tackling AI security risks to unleash growth and deliver Plan for Change}.
\newblock
\urldef\tempurl%
\url{https://www.gov.uk/government/news/tackling-ai-security-risks-to-unleash-growth-and-deliver-plan-for-change}
\showURL{%
\tempurl}
\newblock
\shownote{Secretary of State for Science, Innovation and Technology, UK Government}.


\bibitem[Li et~al\mbox{.}(2024b)]%
        {LWX+24}
\bibfield{author}{\bibinfo{person}{Shaofeng Li}, \bibinfo{person}{Xinyu Wang}, \bibinfo{person}{Minhui Xue}, \bibinfo{person}{Haojin Zhu}, \bibinfo{person}{Zhi Zhang}, \bibinfo{person}{Yansong Gao}, \bibinfo{person}{Wen Wu}, {and} \bibinfo{person}{Xuemin~(Sherman) Shen}.} \bibinfo{year}{2024}\natexlab{b}.
\newblock \showarticletitle{Yes, {One-Bit-Flip} Matters! Universal {DNN} Model Inference Depletion with Runtime Code Fault Injection}. In \bibinfo{booktitle}{\emph{33rd USENIX Security Symposium (USENIX Security 24)}} (Philadelphia, PA, USA). \bibinfo{publisher}{USENIX Association}, \bibinfo{address}{Berkeley, CA, USA}, \bibinfo{pages}{1315--1330}.
\newblock
\showISBNx{978-1-939133-44-1}
\urldef\tempurl%
\url{https://www.usenix.org/conference/usenixsecurity24/presentation/li-shaofeng}
\showURL{%
\tempurl}


\bibitem[Li et~al\mbox{.}(2024a)]%
        {LJL+24}
\bibfield{author}{\bibinfo{person}{Yiming Li}, \bibinfo{person}{Yong Jiang}, \bibinfo{person}{Zhifeng Li}, {and} \bibinfo{person}{Shu-Tao Xia}.} \bibinfo{year}{2024}\natexlab{a}.
\newblock \showarticletitle{Backdoor Learning: A Survey}.
\newblock \bibinfo{journal}{\emph{IEEE Transactions on Neural Networks and Learning Systems}} \bibinfo{volume}{35}, \bibinfo{number}{1} (\bibinfo{year}{2024}), \bibinfo{pages}{5--22}.
\newblock
\href{https://doi.org/10.1109/TNNLS.2022.3182979}{doi:\nolinkurl{10.1109/TNNLS.2022.3182979}}
\showeprint[arxiv]{2007.08745}~[cs.CR]


\bibitem[Li et~al\mbox{.}(2023)]%
        {LWW+23}
\bibfield{author}{\bibinfo{person}{Zongjie Li}, \bibinfo{person}{Chaozheng Wang}, \bibinfo{person}{Shuai Wang}, {and} \bibinfo{person}{Cuiyun Gao}.} \bibinfo{year}{2023}\natexlab{}.
\newblock \showarticletitle{Protecting Intellectual Property of Large Language Model-Based Code Generation APIs via Watermarks}. In \bibinfo{booktitle}{\emph{Proceedings of the 2023 ACM SIGSAC Conference on Computer and Communications Security}} (Copenhagen, Denmark) \emph{(\bibinfo{series}{CCS '23})}. \bibinfo{publisher}{Association for Computing Machinery}, \bibinfo{address}{New York, NY, USA}, \bibinfo{pages}{2336–2350}.
\newblock
\showISBNx{9798400700507}
\href{https://doi.org/10.1145/3576915.3623120}{doi:\nolinkurl{10.1145/3576915.3623120}}


\bibitem[Liu et~al\mbox{.}(2024)]%
        {LDL+24}
\bibfield{author}{\bibinfo{person}{Yi Liu}, \bibinfo{person}{Gelei Deng}, \bibinfo{person}{Yuekang Li}, \bibinfo{person}{Kailong Wang}, \bibinfo{person}{Zihao Wang}, \bibinfo{person}{Xiaofeng Wang}, \bibinfo{person}{Tianwei Zhang}, \bibinfo{person}{Yepang Liu}, \bibinfo{person}{Haoyu Wang}, \bibinfo{person}{Yan Zheng}, {and} \bibinfo{person}{Yang Liu}.} \bibinfo{year}{2024}\natexlab{}.
\newblock \bibinfo{title}{Prompt Injection attack against LLM-integrated Applications}.
\newblock
\showeprint[arxiv]{2306.05499}~[cs.CR]


\bibitem[Liu et~al\mbox{.}(2022)]%
        {LWH+22}
\bibfield{author}{\bibinfo{person}{Yugeng Liu}, \bibinfo{person}{Rui Wen}, \bibinfo{person}{Xinlei He}, \bibinfo{person}{Ahmed Salem}, \bibinfo{person}{Zhikun Zhang}, \bibinfo{person}{Michael Backes}, \bibinfo{person}{Emiliano~De Cristofaro}, \bibinfo{person}{Mario Fritz}, {and} \bibinfo{person}{Yang Zhang}.} \bibinfo{year}{2022}\natexlab{}.
\newblock \showarticletitle{{ML-Doctor}: Holistic Risk Assessment of Inference Attacks Against Machine Learning Models}. In \bibinfo{booktitle}{\emph{31st USENIX Security Symposium (USENIX Security 22)}} (Boston, MA, USA). \bibinfo{publisher}{USENIX Association}, \bibinfo{address}{Berkeley, CA, USA}, \bibinfo{pages}{4525--4542}.
\newblock
\showISBNx{978-1-939133-31-1}
\urldef\tempurl%
\url{https://www.usenix.org/conference/usenixsecurity22/presentation/liu-yugeng}
\showURL{%
\tempurl}


\bibitem[Long et~al\mbox{.}(2018)]%
        {LBW+18}
\bibfield{author}{\bibinfo{person}{Yunhui Long}, \bibinfo{person}{Vincent Bindschaedler}, \bibinfo{person}{Lei Wang}, \bibinfo{person}{Diyue Bu}, \bibinfo{person}{Xiaofeng Wang}, \bibinfo{person}{Haixu Tang}, \bibinfo{person}{Carl~A. Gunter}, {and} \bibinfo{person}{Kai Chen}.} \bibinfo{year}{2018}\natexlab{}.
\newblock \bibinfo{title}{Understanding Membership Inferences on Well-Generalized Learning Models}.
\newblock
\showeprint[arxiv]{1802.04889}~[cs.CR]


\bibitem[Lukas et~al\mbox{.}(2023)]%
        {LSS+23}
\bibfield{author}{\bibinfo{person}{Nils Lukas}, \bibinfo{person}{Ahmed Salem}, \bibinfo{person}{Robert Sim}, \bibinfo{person}{Shruti Tople}, \bibinfo{person}{Lukas Wutschitz}, {and} \bibinfo{person}{Santiago Zanella-Béguelin}.} \bibinfo{year}{2023}\natexlab{}.
\newblock \showarticletitle{Analyzing Leakage of Personally Identifiable Information in Language Models}. In \bibinfo{booktitle}{\emph{2023 IEEE Symposium on Security and Privacy (SP)}} (San Francisco, CA, USA). \bibinfo{publisher}{IEEE}, \bibinfo{address}{New York, NY, USA}, \bibinfo{pages}{346--363}.
\newblock
\href{https://doi.org/10.1109/SP46215.2023.10179300}{doi:\nolinkurl{10.1109/SP46215.2023.10179300}}


\bibitem[Ma et~al\mbox{.}(2019)]%
        {MLT+19}
\bibfield{author}{\bibinfo{person}{Shiqing Ma}, \bibinfo{person}{Yingqi Liu}, \bibinfo{person}{Guanhong Tao}, \bibinfo{person}{Wen-Chuan Lee}, {and} \bibinfo{person}{Xiangyu Zhang}.} \bibinfo{year}{2019}\natexlab{}.
\newblock \showarticletitle{NIC: Detecting Adversarial Samples with Neural Network Invariant Checking}. In \bibinfo{booktitle}{\emph{Proceedings 2019 Network and Distributed System Security Symposium}} (San Diego, CA, USA) \emph{(\bibinfo{series}{NDSS 2019})}. \bibinfo{publisher}{Internet Society}, \bibinfo{address}{Reston, VA, USA}.
\newblock
\href{https://doi.org/10.14722/ndss.2019.23415}{doi:\nolinkurl{10.14722/ndss.2019.23415}}


\bibitem[Mahloujifar et~al\mbox{.}(2022)]%
        {MGC22}
\bibfield{author}{\bibinfo{person}{Saeed Mahloujifar}, \bibinfo{person}{Esha Ghosh}, {and} \bibinfo{person}{Melissa Chase}.} \bibinfo{year}{2022}\natexlab{}.
\newblock \showarticletitle{Property Inference from Poisoning}. In \bibinfo{booktitle}{\emph{2022 IEEE Symposium on Security and Privacy (SP)}} (San Francisco, CA, USA). \bibinfo{publisher}{IEEE}, \bibinfo{address}{New York, NY, USA}, \bibinfo{pages}{1120--1137}.
\newblock
\href{https://doi.org/10.1109/SP46214.2022.9833623}{doi:\nolinkurl{10.1109/SP46214.2022.9833623}}


\bibitem[Mehrotra et~al\mbox{.}(2024)]%
        {MZK+24}
\bibfield{author}{\bibinfo{person}{Anay Mehrotra}, \bibinfo{person}{Manolis Zampetakis}, \bibinfo{person}{Paul Kassianik}, \bibinfo{person}{Blaine Nelson}, \bibinfo{person}{Hyrum Anderson}, \bibinfo{person}{Yaron Singer}, {and} \bibinfo{person}{Amin Karbasi}.} \bibinfo{year}{2024}\natexlab{}.
\newblock \bibinfo{title}{Tree of Attacks: Jailbreaking Black-Box LLMs Automatically}.
\newblock
\showeprint[arxiv]{2312.02119}~[cs.LG]


\bibitem[MITRE(2013)]%
        {Mit13}
\bibfield{author}{\bibinfo{person}{MITRE}.} \bibinfo{year}{2013}\natexlab{}.
\newblock \bibinfo{title}{MITRE ATT\&CK\textsuperscript{\textregistered}}.
\newblock
\urldef\tempurl%
\url{https://attack.mitre.org/}
\showURL{%
\tempurl}


\bibitem[MITRE(2021)]%
        {Mit21}
\bibfield{author}{\bibinfo{person}{MITRE}.} \bibinfo{year}{2021}\natexlab{}.
\newblock \bibinfo{title}{MITRE ATLAS\texttrademark}.
\newblock
\urldef\tempurl%
\url{https://atlas.mitre.org/}
\showURL{%
\tempurl}


\bibitem[Murakonda and Shokri(2020)]%
        {MS20}
\bibfield{author}{\bibinfo{person}{Sasi~Kumar Murakonda} {and} \bibinfo{person}{Reza Shokri}.} \bibinfo{year}{2020}\natexlab{}.
\newblock \bibinfo{title}{{ML} Privacy Meter: Aiding Regulatory Compliance by Quantifying the Privacy Risks of Machine Learning}.  (\bibinfo{year}{2020}).
\newblock
\showeprint[arxiv]{2007.09339}~[cs.CR]
\newblock
\shownote{Workshop on Hot Topics in Privacy Enhancing Technologies (HotPETs)}.


\bibitem[Nasr et~al\mbox{.}(2019)]%
        {NSH19}
\bibfield{author}{\bibinfo{person}{Milad Nasr}, \bibinfo{person}{Reza Shokri}, {and} \bibinfo{person}{Amir Houmansadr}.} \bibinfo{year}{2019}\natexlab{}.
\newblock \showarticletitle{Comprehensive Privacy Analysis of Deep Learning: Passive and Active White-box Inference Attacks against Centralized and Federated Learning}. In \bibinfo{booktitle}{\emph{2019 IEEE Symposium on Security and Privacy (SP)}} (San Francisco, CA, USA). \bibinfo{publisher}{IEEE}, \bibinfo{address}{New York, NY, USA}, \bibinfo{pages}{739--753}.
\newblock
\href{https://doi.org/10.1109/SP.2019.00065}{doi:\nolinkurl{10.1109/SP.2019.00065}}


\bibitem[Nazari et~al\mbox{.}(2024)]%
        {NMF+24}
\bibfield{author}{\bibinfo{person}{Najmeh Nazari}, \bibinfo{person}{Hosein~Mohammadi Makrani}, \bibinfo{person}{Chongzhou Fang}, \bibinfo{person}{Hossein Sayadi}, \bibinfo{person}{Setareh Rafatirad}, \bibinfo{person}{Khaled~N. Khasawneh}, {and} \bibinfo{person}{Houman Homayoun}.} \bibinfo{year}{2024}\natexlab{}.
\newblock \showarticletitle{Forget and Rewire: Enhancing the Resilience of Transformer-based Models against {Bit-Flip} Attacks}. In \bibinfo{booktitle}{\emph{33rd USENIX Security Symposium (USENIX Security 24)}} (Philadelphia, PA, USA). \bibinfo{publisher}{USENIX Association}, \bibinfo{address}{Berkeley, CA, USA}, \bibinfo{pages}{1349--1366}.
\newblock
\showISBNx{978-1-939133-44-1}
\urldef\tempurl%
\url{https://www.usenix.org/conference/usenixsecurity24/presentation/nazari}
\showURL{%
\tempurl}


\bibitem[Nicolae et~al\mbox{.}(2019)]%
        {NSN+19}
\bibfield{author}{\bibinfo{person}{Maria-Irina Nicolae}, \bibinfo{person}{Mathieu Sinn}, \bibinfo{person}{Minh~Ngoc Tran}, \bibinfo{person}{Beat Buesser}, \bibinfo{person}{Ambrish Rawat}, \bibinfo{person}{Martin Wistuba}, \bibinfo{person}{Valentina Zantedeschi}, \bibinfo{person}{Nathalie Baracaldo}, \bibinfo{person}{Bryant Chen}, \bibinfo{person}{Heiko Ludwig}, \bibinfo{person}{Ian~M. Molloy}, {and} \bibinfo{person}{Ben Edwards}.} \bibinfo{year}{2019}\natexlab{}.
\newblock \bibinfo{title}{Adversarial Robustness Toolbox v1.0.0}.
\newblock
\showeprint[arxiv]{1807.01069}~[cs.LG]


\bibitem[Ntalampiras et~al\mbox{.}(2023)]%
        {NPB+23}
\bibfield{author}{\bibinfo{person}{Stavros Ntalampiras}, \bibinfo{person}{Corina Pascu}, \bibinfo{person}{Marco~Barros Lourenco}, \bibinfo{person}{Gianluca Misuraca}, {and} \bibinfo{person}{Pierre Rossel}.} \bibinfo{year}{2023}\natexlab{}.
\newblock \bibinfo{booktitle}{\emph{Artificial intelligence and cybersecurity research}}.
\newblock \bibinfo{type}{{T}echnical {R}eport}. \bibinfo{institution}{European Union Agency for Cybersecurity}.
\newblock
\href{https://doi.org/10.2824/808362}{doi:\nolinkurl{10.2824/808362}}


\bibitem[of~Commerce(2025)]%
        {USD25}
\bibfield{author}{\bibinfo{person}{U.S.~Department of Commerce}.} \bibinfo{year}{2025}\natexlab{}.
\newblock \bibinfo{title}{Statement from U.S. Secretary of Commerce Howard Lutnick on Transforming the U.S. AI Safety Institute into the Pro-Innovation, Pro-Science U.S. Center for AI Standards and Innovation}.
\newblock
\urldef\tempurl%
\url{https://www.commerce.gov/news/press-releases/2025/06/statement-us-secretary-commerce-howard-lutnick-transforming-us-ai}
\showURL{%
\tempurl}


\bibitem[of~Standards and Technology(2023a)]%
        {Nat23a}
\bibfield{author}{\bibinfo{person}{National~Institute of Standards} {and} \bibinfo{person}{Technology}.} \bibinfo{year}{2023}\natexlab{a}.
\newblock \bibinfo{title}{Artificial Intelligence Risk Management Framework (AI RMF 1.0), NIST AI 100-1}.
\newblock
\href{https://doi.org/10.6028/NIST.AI.100-1}{doi:\nolinkurl{10.6028/NIST.AI.100-1}}


\bibitem[of~Standards and Technology(2023b)]%
        {Nat23b}
\bibfield{author}{\bibinfo{person}{National~Institute of Standards} {and} \bibinfo{person}{Technology}.} \bibinfo{year}{2023}\natexlab{b}.
\newblock \bibinfo{title}{NIST AI RMF Playbook}.
\newblock
\urldef\tempurl%
\url{https://airc.nist.gov/airmf-resources/playbook/}
\showURL{%
\tempurl}


\bibitem[of~the Group~of Seven~(G7)(2023)]%
        {G723}
\bibfield{author}{\bibinfo{person}{The~Leaders of~the Group~of Seven~(G7)}.} \bibinfo{year}{2023}\natexlab{}.
\newblock \bibinfo{title}{Hiroshima AI Process}.
\newblock
\urldef\tempurl%
\url{https://www.soumu.go.jp/hiroshimaaiprocess/en/}
\showURL{%
\tempurl}


\bibitem[Oh et~al\mbox{.}(2019)]%
        {OSF19a}
\bibfield{author}{\bibinfo{person}{Seong~Joon Oh}, \bibinfo{person}{Bernt Schiele}, {and} \bibinfo{person}{Mario Fritz}.} \bibinfo{year}{2019}\natexlab{}.
\newblock \showarticletitle{Towards Reverse-Engineering Black-Box Neural Networks}.
\newblock In \bibinfo{booktitle}{\emph{Explainable AI: Interpreting, Explaining and Visualizing Deep Learning}}, \bibfield{editor}{\bibinfo{person}{Wojciech Samek}, \bibinfo{person}{Gr{\'e}goire Montavon}, \bibinfo{person}{Andrea Vedaldi}, \bibinfo{person}{Lars~Kai Hansen}, {and} \bibinfo{person}{Klaus-Robert M{\"u}ller}} (Eds.). \bibinfo{series}{Lecture Notes in Computer Science}, Vol.~\bibinfo{volume}{11700}. \bibinfo{publisher}{Springer International Publishing}, \bibinfo{address}{Cham}, \bibinfo{pages}{121--144}.
\newblock
\showISBNx{978-3-030-28954-6}
\href{https://doi.org/10.1007/978-3-030-28954-6_7}{doi:\nolinkurl{10.1007/978-3-030-28954-6_7}}


\bibitem[Orekondy et~al\mbox{.}(2019)]%
        {OSF19b}
\bibfield{author}{\bibinfo{person}{Tribhuvanesh Orekondy}, \bibinfo{person}{Bernt Schiele}, {and} \bibinfo{person}{Mario Fritz}.} \bibinfo{year}{2019}\natexlab{}.
\newblock \showarticletitle{Knockoff Nets: Stealing Functionality of Black-Box Models}. In \bibinfo{booktitle}{\emph{2019 IEEE/CVF Conference on Computer Vision and Pattern Recognition (CVPR)}} (Long Beach, CA, USA). \bibinfo{publisher}{IEEE}, \bibinfo{address}{New York, NY, USA}, \bibinfo{pages}{4949--4958}.
\newblock
\href{https://doi.org/10.1109/CVPR.2019.00509}{doi:\nolinkurl{10.1109/CVPR.2019.00509}}


\bibitem[Oseni et~al\mbox{.}(2021)]%
        {OMJ+21}
\bibfield{author}{\bibinfo{person}{Ayodeji Oseni}, \bibinfo{person}{Nour Moustafa}, \bibinfo{person}{Helge Janicke}, \bibinfo{person}{Peng Liu}, \bibinfo{person}{Zahir Tari}, {and} \bibinfo{person}{Athanasios Vasilakos}.} \bibinfo{year}{2021}\natexlab{}.
\newblock \bibinfo{title}{Security and Privacy for Artificial Intelligence: Opportunities and Challenges}.
\newblock
\showeprint[arxiv]{2102.04661}~[cs.CR]


\bibitem[Papernot et~al\mbox{.}(2018)]%
        {PFC+18}
\bibfield{author}{\bibinfo{person}{Nicolas Papernot}, \bibinfo{person}{Fartash Faghri}, \bibinfo{person}{Nicholas Carlini}, \bibinfo{person}{Ian Goodfellow}, \bibinfo{person}{Reuben Feinman}, \bibinfo{person}{Alexey Kurakin}, \bibinfo{person}{Cihang Xie}, \bibinfo{person}{Yash Sharma}, \bibinfo{person}{Tom Brown}, \bibinfo{person}{Aurko Roy}, \bibinfo{person}{Alexander Matyasko}, \bibinfo{person}{Vahid Behzadan}, \bibinfo{person}{Karen Hambardzumyan}, \bibinfo{person}{Zhishuai Zhang}, \bibinfo{person}{Yi-Lin Juang}, \bibinfo{person}{Zhi Li}, \bibinfo{person}{Ryan Sheatsley}, \bibinfo{person}{Abhibhav Garg}, \bibinfo{person}{Jonathan Uesato}, \bibinfo{person}{Willi Gierke}, \bibinfo{person}{Yinpeng Dong}, \bibinfo{person}{David Berthelot}, \bibinfo{person}{Paul Hendricks}, \bibinfo{person}{Jonas Rauber}, \bibinfo{person}{Rujun Long}, {and} \bibinfo{person}{Patrick McDaniel}.} \bibinfo{year}{2018}\natexlab{}.
\newblock \bibinfo{title}{Technical Report on the CleverHans v2.1.0 Adversarial Examples Library}.
\newblock
\showeprint[arxiv]{1610.00768}~[cs.LG]


\bibitem[Papernot et~al\mbox{.}(2017)]%
        {PMG+17}
\bibfield{author}{\bibinfo{person}{Nicolas Papernot}, \bibinfo{person}{Patrick McDaniel}, \bibinfo{person}{Ian Goodfellow}, \bibinfo{person}{Somesh Jha}, \bibinfo{person}{Z.~Berkay Celik}, {and} \bibinfo{person}{Ananthram Swami}.} \bibinfo{year}{2017}\natexlab{}.
\newblock \showarticletitle{Practical Black-Box Attacks against Machine Learning}. In \bibinfo{booktitle}{\emph{Proceedings of the 2017 ACM on Asia Conference on Computer and Communications Security}} (Abu Dhabi, United Arab Emirates) \emph{(\bibinfo{series}{ASIA CCS '17})}. \bibinfo{publisher}{Association for Computing Machinery}, \bibinfo{address}{New York, NY, USA}, \bibinfo{pages}{506–519}.
\newblock
\showISBNx{9781450349444}
\href{https://doi.org/10.1145/3052973.3053009}{doi:\nolinkurl{10.1145/3052973.3053009}}


\bibitem[Pedro et~al\mbox{.}(2025)]%
        {PEC+25}
\bibfield{author}{\bibinfo{person}{Rodrigo Pedro}, \bibinfo{person}{Miguel E.~Coimbra}, \bibinfo{person}{Daniel Castro}, \bibinfo{person}{Paulo Carreira}, {and} \bibinfo{person}{Nuno Santos}.} \bibinfo{year}{2025}\natexlab{}.
\newblock \showarticletitle{{Prompt-to-SQL Injections in LLM-Integrated Web Applications: Risks and Defenses}}. In \bibinfo{booktitle}{\emph{2025 IEEE/ACM 47th International Conference on Software Engineering (ICSE)}} (Ottawa, ON, Canada). \bibinfo{publisher}{IEEE Computer Society}, \bibinfo{address}{Los Alamitos, CA, USA}, \bibinfo{pages}{76--88}.
\newblock
\showISSN{1558-1225}
\href{https://doi.org/10.1109/ICSE55347.2025.00007}{doi:\nolinkurl{10.1109/ICSE55347.2025.00007}}


\bibitem[Perez et~al\mbox{.}(2022)]%
        {PHS+22}
\bibfield{author}{\bibinfo{person}{Ethan Perez}, \bibinfo{person}{Saffron Huang}, \bibinfo{person}{Francis Song}, \bibinfo{person}{Trevor Cai}, \bibinfo{person}{Roman Ring}, \bibinfo{person}{John Aslanides}, \bibinfo{person}{Amelia Glaese}, \bibinfo{person}{Nat McAleese}, {and} \bibinfo{person}{Geoffrey Irving}.} \bibinfo{year}{2022}\natexlab{}.
\newblock \showarticletitle{Red Teaming Language Models with Language Models}. In \bibinfo{booktitle}{\emph{Proceedings of the 2022 Conference on Empirical Methods in Natural Language Processing}} (Abu Dhabi, United Arab Emirates), \bibfield{editor}{\bibinfo{person}{Yoav Goldberg}, \bibinfo{person}{Zornitsa Kozareva}, {and} \bibinfo{person}{Yue Zhang}} (Eds.). \bibinfo{publisher}{Association for Computational Linguistics}, \bibinfo{address}{Stroudsburg, PA, USA}, \bibinfo{pages}{3419--3448}.
\newblock
\href{https://doi.org/10.18653/v1/2022.emnlp-main.225}{doi:\nolinkurl{10.18653/v1/2022.emnlp-main.225}}


\bibitem[Rakin et~al\mbox{.}(2022)]%
        {RCY+22}
\bibfield{author}{\bibinfo{person}{Adnan~Siraj Rakin}, \bibinfo{person}{Md~Hafizul~Islam Chowdhuryy}, \bibinfo{person}{Fan Yao}, {and} \bibinfo{person}{Deliang Fan}.} \bibinfo{year}{2022}\natexlab{}.
\newblock \showarticletitle{DeepSteal: Advanced Model Extractions Leveraging Efficient Weight Stealing in Memories}. In \bibinfo{booktitle}{\emph{2022 IEEE Symposium on Security and Privacy (SP)}} (San Francisco, CA, USA). \bibinfo{publisher}{IEEE}, \bibinfo{address}{New York, NY, USA}, \bibinfo{pages}{1157--1174}.
\newblock
\href{https://doi.org/10.1109/SP46214.2022.9833743}{doi:\nolinkurl{10.1109/SP46214.2022.9833743}}


\bibitem[Samoilenko(2023)]%
        {Sam23}
\bibfield{author}{\bibinfo{person}{Roman Samoilenko}.} \bibinfo{year}{2023}\natexlab{}.
\newblock \bibinfo{booktitle}{\emph{New prompt injection attack on ChatGPT web version. Reckless copy-pasting may lead to serious privacy issues in your chat.}}
\newblock Netograph.
\newblock
\urldef\tempurl%
\url{https://kajojify.github.io/articles/1_chatgpt_attack.pdf}
\showURL{%
Retrieved April 25, 2025 from \tempurl}


\bibitem[Schwartz(2019)]%
        {Sch19}
\bibfield{author}{\bibinfo{person}{Oscar Schwartz}.} \bibinfo{year}{2019}\natexlab{}.
\newblock \showarticletitle{In 2016, Microsoft’s Racist Chatbot Revealed the Dangers of Online Conversation}.
\newblock \bibinfo{journal}{\emph{IEEE Spectrum}}  \bibinfo{volume}{Nov.} (\bibinfo{year}{2019}).
\newblock
\urldef\tempurl%
\url{https://spectrum.ieee.org/in-2016-microsofts-racist-chatbot-revealed-the-dangers-of-online-conversation}
\showURL{%
\tempurl}
\newblock
\shownote{Updated: Jan. 2024}.


\bibitem[Shayegani et~al\mbox{.}(2023)]%
        {SMF+23}
\bibfield{author}{\bibinfo{person}{Erfan Shayegani}, \bibinfo{person}{Md~Abdullah~Al Mamun}, \bibinfo{person}{Yu Fu}, \bibinfo{person}{Pedram Zaree}, \bibinfo{person}{Yue Dong}, {and} \bibinfo{person}{Nael Abu-Ghazaleh}.} \bibinfo{year}{2023}\natexlab{}.
\newblock \bibinfo{title}{Survey of Vulnerabilities in Large Language Models Revealed by Adversarial Attacks}.
\newblock
\showeprint[arxiv]{2310.10844}~[cs.CL]


\bibitem[Shen et~al\mbox{.}(2024a)]%
        {SCB+24}
\bibfield{author}{\bibinfo{person}{Xinyue Shen}, \bibinfo{person}{Zeyuan Chen}, \bibinfo{person}{Michael Backes}, \bibinfo{person}{Yun Shen}, {and} \bibinfo{person}{Yang Zhang}.} \bibinfo{year}{2024}\natexlab{a}.
\newblock \showarticletitle{"{D}o {A}nything {N}ow": Characterizing and Evaluating In-The-Wild Jailbreak Prompts on Large Language Models}. In \bibinfo{booktitle}{\emph{Proceedings of the 2024 on ACM SIGSAC Conference on Computer and Communications Security}} (Salt Lake City, UT, USA) \emph{(\bibinfo{series}{CCS '24})}. \bibinfo{publisher}{Association for Computing Machinery}, \bibinfo{address}{New York, NY, USA}, \bibinfo{pages}{1671–1685}.
\newblock
\showISBNx{9798400706363}
\href{https://doi.org/10.1145/3658644.3670388}{doi:\nolinkurl{10.1145/3658644.3670388}}


\bibitem[Shen et~al\mbox{.}(2024b)]%
        {SQB+24}
\bibfield{author}{\bibinfo{person}{Xinyue Shen}, \bibinfo{person}{Yiting Qu}, \bibinfo{person}{Michael Backes}, {and} \bibinfo{person}{Yang Zhang}.} \bibinfo{year}{2024}\natexlab{b}.
\newblock \showarticletitle{Prompt Stealing Attacks Against {Text-to-Image} Generation Models}. In \bibinfo{booktitle}{\emph{33rd USENIX Security Symposium (USENIX Security 24)}} (Philadelphia, PA, USA). \bibinfo{publisher}{USENIX Association}, \bibinfo{address}{Berkeley, CA, USA}, \bibinfo{pages}{5823--5840}.
\newblock
\showISBNx{978-1-939133-44-1}
\urldef\tempurl%
\url{https://www.usenix.org/conference/usenixsecurity24/presentation/shen-xinyue}
\showURL{%
\tempurl}


\bibitem[Shokri et~al\mbox{.}(2017)]%
        {SSS+17}
\bibfield{author}{\bibinfo{person}{Reza Shokri}, \bibinfo{person}{Marco Stronati}, \bibinfo{person}{Congzheng Song}, {and} \bibinfo{person}{Vitaly Shmatikov}.} \bibinfo{year}{2017}\natexlab{}.
\newblock \showarticletitle{Membership Inference Attacks Against Machine Learning Models}. In \bibinfo{booktitle}{\emph{2017 IEEE Symposium on Security and Privacy (SP)}} (San Jose, CA, USA). \bibinfo{publisher}{IEEE}, \bibinfo{address}{New York, NY, USA}, \bibinfo{pages}{3--18}.
\newblock
\href{https://doi.org/10.1109/SP.2017.41}{doi:\nolinkurl{10.1109/SP.2017.41}}


\bibitem[Shumailov et~al\mbox{.}(2021)]%
        {SZB+21}
\bibfield{author}{\bibinfo{person}{Ilia Shumailov}, \bibinfo{person}{Yiren Zhao}, \bibinfo{person}{Daniel Bates}, \bibinfo{person}{Nicolas Papernot}, \bibinfo{person}{Robert Mullins}, {and} \bibinfo{person}{Ross Anderson}.} \bibinfo{year}{2021}\natexlab{}.
\newblock \showarticletitle{Sponge Examples: Energy-Latency Attacks on Neural Networks}. In \bibinfo{booktitle}{\emph{2021 IEEE European Symposium on Security and Privacy (EuroS\&P)}} (Vienna, Austria). \bibinfo{publisher}{IEEE}, \bibinfo{address}{New York, NY, USA}, \bibinfo{pages}{212--231}.
\newblock
\href{https://doi.org/10.1109/EuroSP51992.2021.00024}{doi:\nolinkurl{10.1109/EuroSP51992.2021.00024}}


\bibitem[Slack et~al\mbox{.}(2020)]%
        {SHJ+20}
\bibfield{author}{\bibinfo{person}{Dylan Slack}, \bibinfo{person}{Sophie Hilgard}, \bibinfo{person}{Emily Jia}, \bibinfo{person}{Sameer Singh}, {and} \bibinfo{person}{Himabindu Lakkaraju}.} \bibinfo{year}{2020}\natexlab{}.
\newblock \showarticletitle{Fooling LIME and SHAP: Adversarial Attacks on Post hoc Explanation Methods}. In \bibinfo{booktitle}{\emph{Proceedings of the AAAI/ACM Conference on AI, Ethics, and Society}} (New York, NY, USA) \emph{(\bibinfo{series}{AIES '20})}. \bibinfo{publisher}{Association for Computing Machinery}, \bibinfo{address}{New York, NY, USA}, \bibinfo{pages}{180–186}.
\newblock
\showISBNx{9781450371100}
\href{https://doi.org/10.1145/3375627.3375830}{doi:\nolinkurl{10.1145/3375627.3375830}}


\bibitem[Song et~al\mbox{.}(2017)]%
        {SRS17}
\bibfield{author}{\bibinfo{person}{Congzheng Song}, \bibinfo{person}{Thomas Ristenpart}, {and} \bibinfo{person}{Vitaly Shmatikov}.} \bibinfo{year}{2017}\natexlab{}.
\newblock \showarticletitle{Machine Learning Models that Remember Too Much}. In \bibinfo{booktitle}{\emph{Proceedings of the 2017 ACM SIGSAC Conference on Computer and Communications Security}} (Dallas, Texas, USA) \emph{(\bibinfo{series}{CCS '17})}. \bibinfo{publisher}{Association for Computing Machinery}, \bibinfo{address}{New York, NY, USA}, \bibinfo{pages}{587–601}.
\newblock
\showISBNx{9781450349468}
\href{https://doi.org/10.1145/3133956.3134077}{doi:\nolinkurl{10.1145/3133956.3134077}}


\bibitem[Song and Shmatikov(2020)]%
        {SS20}
\bibfield{author}{\bibinfo{person}{Congzheng Song} {and} \bibinfo{person}{Vitaly Shmatikov}.} \bibinfo{year}{2020}\natexlab{}.
\newblock \bibinfo{title}{Overlearning Reveals Sensitive Attributes}.
\newblock
\showeprint[arxiv]{1905.11742}~[cs.LG]
\urldef\tempurl%
\url{https://openreview.net/forum?id=SJeNz04tDS}
\showURL{%
\tempurl}
\newblock
\shownote{8th International Conference on Learning Representations, {ICLR} 2020, Addis Ababa, Ethiopia, April 26-30, 2020}.


\bibitem[Sun et~al\mbox{.}(2023)]%
        {SZZ+23}
\bibfield{author}{\bibinfo{person}{Hui Sun}, \bibinfo{person}{Tianqing Zhu}, \bibinfo{person}{Zhiqiu Zhang}, \bibinfo{person}{Dawei Jin}, \bibinfo{person}{Ping Xiong}, {and} \bibinfo{person}{Wanlei Zhou}.} \bibinfo{year}{2023}\natexlab{}.
\newblock \showarticletitle{{Adversarial Attacks Against Deep Generative Models on Data: A Survey}}.
\newblock \bibinfo{journal}{\emph{IEEE Transactions on Knowledge \& Data Engineering}} \bibinfo{volume}{35}, \bibinfo{number}{04} (\bibinfo{date}{April} \bibinfo{year}{2023}), \bibinfo{pages}{3367--3388}.
\newblock
\showISSN{1558-2191}
\href{https://doi.org/10.1109/TKDE.2021.3130903}{doi:\nolinkurl{10.1109/TKDE.2021.3130903}}


\bibitem[Szegedy et~al\mbox{.}(2014)]%
        {SZS+14}
\bibfield{author}{\bibinfo{person}{Christian Szegedy}, \bibinfo{person}{Wojciech Zaremba}, \bibinfo{person}{Ilya Sutskever}, \bibinfo{person}{Joan Bruna}, \bibinfo{person}{Dumitru Erhan}, \bibinfo{person}{Ian~J. Goodfellow}, {and} \bibinfo{person}{Rob Fergus}.} \bibinfo{year}{2014}\natexlab{}.
\newblock \bibinfo{title}{Intriguing properties of neural networks}.
\newblock
\showeprint[arxiv]{1312.6199}~[cs.CV]
\newblock
\shownote{2nd International Conference on Learning Representations, {ICLR} 2014, Banff, AB, Canada, April 14-16, 2014, Conference Track Proceedings}.


\bibitem[Tram{\`e}r et~al\mbox{.}(2016)]%
        {TZJ+16}
\bibfield{author}{\bibinfo{person}{Florian Tram{\`e}r}, \bibinfo{person}{Fan Zhang}, \bibinfo{person}{Ari Juels}, \bibinfo{person}{Michael~K. Reiter}, {and} \bibinfo{person}{Thomas Ristenpart}.} \bibinfo{year}{2016}\natexlab{}.
\newblock \showarticletitle{Stealing Machine Learning Models via Prediction {APIs}}. In \bibinfo{booktitle}{\emph{25th USENIX Security Symposium (USENIX Security 16)}} (Austin, TX, USA). \bibinfo{publisher}{USENIX Association}, \bibinfo{address}{Berkeley, CA, USA}, \bibinfo{pages}{601--618}.
\newblock
\showISBNx{978-1-931971-32-4}
\urldef\tempurl%
\url{https://www.usenix.org/conference/usenixsecurity16/technical-sessions/presentation/tramer}
\showURL{%
\tempurl}


\bibitem[van~der Stock et~al\mbox{.}(2021)]%
        {SGS+21}
\bibfield{author}{\bibinfo{person}{Andrew van~der Stock}, \bibinfo{person}{Brian Glas}, \bibinfo{person}{Neil Smithline}, {and} \bibinfo{person}{Torsten Gigler}.} \bibinfo{year}{2021}\natexlab{}.
\newblock \bibinfo{title}{OWASP Top 10 (Version 2021)}.
\newblock
\urldef\tempurl%
\url{https://owasp.org/Top10/}
\showURL{%
\tempurl}


\bibitem[Vassilev et~al\mbox{.}(2024)]%
        {VOF+24}
\bibfield{author}{\bibinfo{person}{Apostol Vassilev}, \bibinfo{person}{Alina Oprea}, \bibinfo{person}{Alie Fordyce}, {and} \bibinfo{person}{Hyrum Anderson}.} \bibinfo{year}{2024}\natexlab{}.
\newblock \bibinfo{booktitle}{\emph{Adversarial Machine Learning: A Taxonomy and Terminology of Attacks and Mitigations}}.
\newblock \bibinfo{type}{{T}echnical {R}eport} NIST Artifcial Intelligence (AI) Report, NIST Trustworthy and Responsible AI NIST AI 100-2e2023. \bibinfo{institution}{National Institute of Standards and Technology}, \bibinfo{address}{Gaithersburg, MD}.
\newblock
\urldef\tempurl%
\url{https://doi.org/10.6028/NIST.AI.100-2e2023}
\showURL{%
\tempurl}


\bibitem[Vassilev et~al\mbox{.}(2025)]%
        {VOF+25}
\bibfield{author}{\bibinfo{person}{Apostol Vassilev}, \bibinfo{person}{Alina Oprea}, \bibinfo{person}{Alie Fordyce}, \bibinfo{person}{Hyrum Anderson}, \bibinfo{person}{Xander Davies}, {and} \bibinfo{person}{Maia Hamin}.} \bibinfo{year}{2025}\natexlab{}.
\newblock \bibinfo{booktitle}{\emph{Adversarial Machine Learning: A Taxonomy and Terminology of Attacks and Mitigations}}.
\newblock \bibinfo{type}{{T}echnical {R}eport} NIST Trustworthy and Responsible AI, NIST AI 100-2e2025. \bibinfo{institution}{National Institute of Standards and Technology}, \bibinfo{address}{Gaithersburg, MD}.
\newblock
\urldef\tempurl%
\url{https://doi.org/10.6028/NIST.AI.100-2e2025}
\showURL{%
\tempurl}


\bibitem[Wallace et~al\mbox{.}(2019)]%
        {WFK+19}
\bibfield{author}{\bibinfo{person}{Eric Wallace}, \bibinfo{person}{Shi Feng}, \bibinfo{person}{Nikhil Kandpal}, \bibinfo{person}{Matt Gardner}, {and} \bibinfo{person}{Sameer Singh}.} \bibinfo{year}{2019}\natexlab{}.
\newblock \showarticletitle{Universal Adversarial Triggers for Attacking and Analyzing {NLP}}. In \bibinfo{booktitle}{\emph{Proceedings of the 2019 Conference on Empirical Methods in Natural Language Processing and the 9th International Joint Conference on Natural Language Processing (EMNLP-IJCNLP)}} (Hong Kong, China), \bibfield{editor}{\bibinfo{person}{Kentaro Inui}, \bibinfo{person}{Jing Jiang}, \bibinfo{person}{Vincent Ng}, {and} \bibinfo{person}{Xiaojun Wan}} (Eds.). \bibinfo{publisher}{Association for Computational Linguistics}, \bibinfo{address}{Stroudsburg, PA, USA}, \bibinfo{pages}{2153--2162}.
\newblock
\href{https://doi.org/10.18653/v1/D19-1221}{doi:\nolinkurl{10.18653/v1/D19-1221}}


\bibitem[Wang and Gong(2018)]%
        {WG18}
\bibfield{author}{\bibinfo{person}{Binghui Wang} {and} \bibinfo{person}{Neil~Zhenqiang Gong}.} \bibinfo{year}{2018}\natexlab{}.
\newblock \showarticletitle{Stealing Hyperparameters in Machine Learning}. In \bibinfo{booktitle}{\emph{2018 IEEE Symposium on Security and Privacy (SP)}} (San Francisco, CA, USA). \bibinfo{publisher}{IEEE}, \bibinfo{address}{New York, NY, USA}, \bibinfo{pages}{36--52}.
\newblock
\href{https://doi.org/10.1109/SP.2018.00038}{doi:\nolinkurl{10.1109/SP.2018.00038}}


\bibitem[Wei et~al\mbox{.}(2023)]%
        {WHS23}
\bibfield{author}{\bibinfo{person}{Alexander Wei}, \bibinfo{person}{Nika Haghtalab}, {and} \bibinfo{person}{Jacob Steinhardt}.} \bibinfo{year}{2023}\natexlab{}.
\newblock \showarticletitle{Jailbroken: How Does LLM Safety Training Fail?}. In \bibinfo{booktitle}{\emph{Advances in Neural Information Processing Systems}} (New Orleans, LA, USA), \bibfield{editor}{\bibinfo{person}{A.~Oh}, \bibinfo{person}{T.~Naumann}, \bibinfo{person}{A.~Globerson}, \bibinfo{person}{K.~Saenko}, \bibinfo{person}{M.~Hardt}, {and} \bibinfo{person}{S.~Levine}} (Eds.), Vol.~\bibinfo{volume}{36}. \bibinfo{publisher}{Curran Associates, Inc.}, \bibinfo{address}{Red Hook, NY, USA}, \bibinfo{pages}{80079--80110}.
\newblock
\urldef\tempurl%
\url{https://proceedings.neurips.cc/paper_files/paper/2023/hash/fd6613131889a4b656206c50a8bd7790-Abstract-Conference.html}
\showURL{%
\tempurl}


\bibitem[Wilson and Dawson(2024)]%
        {WD24}
\bibfield{author}{\bibinfo{person}{Steve Wilson} {and} \bibinfo{person}{Ads Dawson}.} \bibinfo{year}{2024}\natexlab{}.
\newblock \bibinfo{title}{OWASP Top 10 for Large Language Model Applications (Version 2025)}.
\newblock
\urldef\tempurl%
\url{https://owasp.org/www-project-top-10-for-large-language-model-applications/}
\showURL{%
\tempurl}


\bibitem[Xie et~al\mbox{.}(2023)]%
        {XYS+23}
\bibfield{author}{\bibinfo{person}{Yueqi Xie}, \bibinfo{person}{Jingwei Yi}, \bibinfo{person}{Jiawei Shao}, \bibinfo{person}{Justin Curl}, \bibinfo{person}{Lingjuan Lyu}, \bibinfo{person}{Qifeng Chen}, \bibinfo{person}{Xing Xie}, {and} \bibinfo{person}{Fangzhao Wu}.} \bibinfo{year}{2023}\natexlab{}.
\newblock \showarticletitle{Defending chat{GPT} against Jailbreak Attack via Self-Reminders}.
\newblock \bibinfo{journal}{\emph{Nature Machine Intelligence}} \bibinfo{volume}{5}, \bibinfo{number}{12} (\bibinfo{year}{2023}), \bibinfo{pages}{1486--1496}.
\newblock
\href{https://doi.org/10.1038/s42256-023-00765-8}{doi:\nolinkurl{10.1038/s42256-023-00765-8}}


\bibitem[Xu et~al\mbox{.}(2018)]%
        {XEQ18}
\bibfield{author}{\bibinfo{person}{Weilin Xu}, \bibinfo{person}{David Evans}, {and} \bibinfo{person}{Yanjun Qi}.} \bibinfo{year}{2018}\natexlab{}.
\newblock \showarticletitle{Feature Squeezing: Detecting Adversarial Examples in Deep Neural Networks}. In \bibinfo{booktitle}{\emph{Proceedings 2018 Network and Distributed System Security Symposium}} (San Diego, CA, USA) \emph{(\bibinfo{series}{NDSS 2018})}. \bibinfo{publisher}{Internet Society}, \bibinfo{address}{Reston, VA, USA}.
\newblock
\href{https://doi.org/10.14722/ndss.2018.23198}{doi:\nolinkurl{10.14722/ndss.2018.23198}}


\bibitem[Yan et~al\mbox{.}(2025)]%
        {YLX+25}
\bibfield{author}{\bibinfo{person}{Biwei Yan}, \bibinfo{person}{Kun Li}, \bibinfo{person}{Minghui Xu}, \bibinfo{person}{Yueyan Dong}, \bibinfo{person}{Yue Zhang}, \bibinfo{person}{Zhaochun Ren}, {and} \bibinfo{person}{Xiuzhen Cheng}.} \bibinfo{year}{2025}\natexlab{}.
\newblock \showarticletitle{On protecting the data privacy of Large Language Models (LLMs) and LLM agents: A literature review}.
\newblock \bibinfo{journal}{\emph{High-Confidence Computing}} \bibinfo{volume}{5}, \bibinfo{number}{2} (\bibinfo{year}{2025}), \bibinfo{pages}{100300}.
\newblock
\showISSN{2667-2952}
\href{https://doi.org/10.1016/j.hcc.2025.100300}{doi:\nolinkurl{10.1016/j.hcc.2025.100300}}


\bibitem[Yao et~al\mbox{.}(2024)]%
        {YDX+24}
\bibfield{author}{\bibinfo{person}{Yifan Yao}, \bibinfo{person}{Jinhao Duan}, \bibinfo{person}{Kaidi Xu}, \bibinfo{person}{Yuanfang Cai}, \bibinfo{person}{Zhibo Sun}, {and} \bibinfo{person}{Yue Zhang}.} \bibinfo{year}{2024}\natexlab{}.
\newblock \showarticletitle{A survey on large language model (LLM) security and privacy: The Good, The Bad, and The Ugly}.
\newblock \bibinfo{journal}{\emph{High-Confidence Computing}} \bibinfo{volume}{4}, \bibinfo{number}{2} (\bibinfo{year}{2024}), \bibinfo{pages}{100211}.
\newblock
\showISSN{2667-2952}
\href{https://doi.org/10.1016/j.hcc.2024.100211}{doi:\nolinkurl{10.1016/j.hcc.2024.100211}}


\bibitem[Yeom et~al\mbox{.}(2018)]%
        {YGF+18}
\bibfield{author}{\bibinfo{person}{Samuel Yeom}, \bibinfo{person}{Irene Giacomelli}, \bibinfo{person}{Matt Fredrikson}, {and} \bibinfo{person}{Somesh Jha}.} \bibinfo{year}{2018}\natexlab{}.
\newblock \showarticletitle{Privacy Risk in Machine Learning: Analyzing the Connection to Overfitting}. In \bibinfo{booktitle}{\emph{2018 IEEE 31st Computer Security Foundations Symposium (CSF)}} (Oxford, UK). \bibinfo{publisher}{IEEE}, \bibinfo{address}{New York, NY, USA}, \bibinfo{pages}{268--282}.
\newblock
\href{https://doi.org/10.1109/CSF.2018.00027}{doi:\nolinkurl{10.1109/CSF.2018.00027}}


\bibitem[Yi et~al\mbox{.}(2024)]%
        {YYC+24}
\bibfield{author}{\bibinfo{person}{Jingwei Yi}, \bibinfo{person}{Rui Ye}, \bibinfo{person}{Qisi Chen}, \bibinfo{person}{Bin Zhu}, \bibinfo{person}{Siheng Chen}, \bibinfo{person}{Defu Lian}, \bibinfo{person}{Guangzhong Sun}, \bibinfo{person}{Xing Xie}, {and} \bibinfo{person}{Fangzhao Wu}.} \bibinfo{year}{2024}\natexlab{}.
\newblock \showarticletitle{On the Vulnerability of Safety Alignment in Open-Access {LLM}s}. In \bibinfo{booktitle}{\emph{Findings of the Association for Computational Linguistics: ACL 2024}} (Bangkok, Thailand), \bibfield{editor}{\bibinfo{person}{Lun-Wei Ku}, \bibinfo{person}{Andre Martins}, {and} \bibinfo{person}{Vivek Srikumar}} (Eds.). \bibinfo{publisher}{Association for Computational Linguistics}, \bibinfo{address}{Stroudsburg, PA, USA}, \bibinfo{pages}{9236--9260}.
\newblock
\href{https://doi.org/10.18653/v1/2024.findings-acl.549}{doi:\nolinkurl{10.18653/v1/2024.findings-acl.549}}


\bibitem[Zhang et~al\mbox{.}(2022)]%
        {ZHK22}
\bibfield{author}{\bibinfo{person}{Ruisi Zhang}, \bibinfo{person}{Seira Hidano}, {and} \bibinfo{person}{Farinaz Koushanfar}.} \bibinfo{year}{2022}\natexlab{}.
\newblock \bibinfo{title}{Text Revealer: Private Text Reconstruction via Model Inversion Attacks against Transformers}.
\newblock
\showeprint[arxiv]{2209.10505}~[cs.CL]


\bibitem[Zhang et~al\mbox{.}(2024)]%
        {ZCI24}
\bibfield{author}{\bibinfo{person}{Yiming Zhang}, \bibinfo{person}{Nicholas Carlini}, {and} \bibinfo{person}{Daphne Ippolito}.} \bibinfo{year}{2024}\natexlab{}.
\newblock \bibinfo{title}{Effective Prompt Extraction from Language Models}.
\newblock
\showeprint[arxiv]{2307.06865}~[cs.CL]
\urldef\tempurl%
\url{https://openreview.net/forum?id=0o95CVdNuz}
\showURL{%
\tempurl}
\newblock
\shownote{First Conference on Language Modeling (COLM 2024) (Philadelphia, PA, USA)}.


\bibitem[Zhao et~al\mbox{.}(2024)]%
        {ZWZ+24}
\bibfield{author}{\bibinfo{person}{Jian Zhao}, \bibinfo{person}{Shenao Wang}, \bibinfo{person}{Yanjie Zhao}, \bibinfo{person}{Xinyi Hou}, \bibinfo{person}{Kailong Wang}, \bibinfo{person}{Peiming Gao}, \bibinfo{person}{Yuanchao Zhang}, \bibinfo{person}{Chen Wei}, {and} \bibinfo{person}{Haoyu Wang}.} \bibinfo{year}{2024}\natexlab{}.
\newblock \showarticletitle{Models Are Codes: Towards Measuring Malicious Code Poisoning Attacks on Pre-trained Model Hubs}. In \bibinfo{booktitle}{\emph{Proceedings of the 39th IEEE/ACM International Conference on Automated Software Engineering}} (Sacramento, CA, USA) \emph{(\bibinfo{series}{ASE '24})}. \bibinfo{publisher}{Association for Computing Machinery}, \bibinfo{address}{New York, NY, USA}, \bibinfo{pages}{2087–2098}.
\newblock
\showISBNx{9798400712487}
\href{https://doi.org/10.1145/3691620.3695271}{doi:\nolinkurl{10.1145/3691620.3695271}}


\bibitem[Zhong and Rezos(2021)]%
        {ZR25}
\bibfield{author}{\bibinfo{person}{Weilin Zhong} {and} \bibinfo{person}{Rezos}.} \bibinfo{year}{2021}\natexlab{}.
\newblock \bibinfo{booktitle}{\emph{Code Injection}}.
\newblock OWASP.
\newblock
\urldef\tempurl%
\url{https://owasp.org/www-community/attacks/Code_Injection}
\showURL{%
Retrieved May 8, 2025 from \tempurl}


\bibitem[Zou et~al\mbox{.}(2024)]%
        {ZPW+24}
\bibfield{author}{\bibinfo{person}{Andy Zou}, \bibinfo{person}{Long Phan}, \bibinfo{person}{Justin Wang}, \bibinfo{person}{Derek Duenas}, \bibinfo{person}{Maxwell Lin}, \bibinfo{person}{Maksym Andriushchenko}, \bibinfo{person}{Rowan Wang}, \bibinfo{person}{Zico Kolter}, \bibinfo{person}{Matt Fredrikson}, {and} \bibinfo{person}{Dan Hendrycks}.} \bibinfo{year}{2024}\natexlab{}.
\newblock \showarticletitle{Improving Alignment and Robustness with Circuit Breakers}. In \bibinfo{booktitle}{\emph{Advances in Neural Information Processing Systems}} (Vancouver, BC, Canada), \bibfield{editor}{\bibinfo{person}{A.~Globerson}, \bibinfo{person}{L.~Mackey}, \bibinfo{person}{D.~Belgrave}, \bibinfo{person}{A.~Fan}, \bibinfo{person}{U.~Paquet}, \bibinfo{person}{J.~Tomczak}, {and} \bibinfo{person}{C.~Zhang}} (Eds.), Vol.~\bibinfo{volume}{37}. \bibinfo{publisher}{Curran Associates, Inc.}, \bibinfo{address}{Red Hook, NY, USA}, \bibinfo{pages}{83345--83373}.
\newblock
\urldef\tempurl%
\url{https://proceedings.neurips.cc/paper_files/paper/2024/file/97ca7168c2c333df5ea61ece3b3276e1-Paper-Conference.pdf}
\showURL{%
\tempurl}


\bibitem[Zou et~al\mbox{.}(2023)]%
        {ZWC+23}
\bibfield{author}{\bibinfo{person}{Andy Zou}, \bibinfo{person}{Zifan Wang}, \bibinfo{person}{Nicholas Carlini}, \bibinfo{person}{Milad Nasr}, \bibinfo{person}{J.~Zico Kolter}, {and} \bibinfo{person}{Matt Fredrikson}.} \bibinfo{year}{2023}\natexlab{}.
\newblock \bibinfo{title}{Universal and Transferable Adversarial Attacks on Aligned Language Models}.
\newblock
\showeprint[arxiv]{2307.15043}~[cs.CL]


\end{thebibliography}

\appendix
\section{Terminology and Description}
\subsection{Components of AI System Architecture and Related Terms}
\label{appendix:components}
The subsequent list describes key components and terms depicted in Figure~\ref{fig:system}, clarifying their roles within the assumed AI system architecture.
\begin{itemize}
    \item {\bf AI Model:} A model based on artificial intelligence techniques that is incorporated as a component within an AI system. This term encompasses both pre-trained models and fine-tuned models.
    \item {\bf AI Model Developer:} An entity involved in the Training Environment, responsible for aspects such as providing AI Model Info and Training Data for the creation and fine-tuning of AI models.
    \item {\bf AI Model Info:} Information detailing the internal characteristics and configurations of AI models, including model architectures, hyperparameters, training algorithms, and other technical specifications necessary for developing AI models. 
    \item {\bf AI System:} An information system that incorporates AI models as components.
    \item {\bf AI System User:} An entity (e.g., human user) that interacts with the deployed AI model in the Operation Environment. This interaction can involve providing Input (including Knowledge Data) to the system or receiving Output.
    \item {\bf Application:} The framework within the Operation Environment where the trained (Fine-tuned) AI model is deployed and integrated to process data and produce outputs. It may also interact with Internal Knowledge, External Knowledge, and Plugins.
    \item {\bf External Knowledge:} Information sources outside the AI system that are accessed during operation, such as the Internet or external systems. These sources are not maintained within the system but are queried when needed to obtain up-to-date or supplemental information.
    \item {\bf External Systems, Sensors, etc.:} External sources that provide data inputs to the AI system in the Operation Environment, or external entities that receive or are affected by the AI system's outputs.
    \item {\bf Fine-tuned AI Model:} An AI model that has undergone a fine-tuning process using a Fine-tuning Dataset to adapt or specialize its capabilities. This model is then typically deployed in the Operation Environment.
    \item {\bf Fine-tuning Dataset:} A specific collection of Training Data used in the Training Environment specifically for the fine-tuning process of a Pre-trained AI Model.
    \item {\bf Input (to Operation Environment):} Data provided from external systems, sensors, or human users that are processed by the AI model within the Operation Environment.
    \item {\bf Internal Knowledge:} A collection of data and documents provided by the AI System User and stored within the AI system. The Application can access these data during its operation, potentially through plugins or retrieval-augmented generation (RAG).
    \item {\bf Knowledge Data:} Data provided by the AI System User, such as internal documents or records, that is accumulated and stored as part of the Internal Knowledge. This data is not used as a one-time input but retained to support future responses or decisions.
    \item {\bf Operation Environment:} The environment where the trained AI model is deployed and integrated into an application framework to process inputs and produce outputs.
    \item {\bf Output (from Operation Environment):} The results, including potential explanations, produced by the AI model in the Operation Environment after processing inputs.
    \item {\bf Pre-trained AI Model:} An AI model that has undergone initial training and serves as a basis for further, more specialized training (fine-tuning) using a Fine-tuning Dataset.
    \item {\bf Training Data:} Data used to train or fine-tune AI models within the Training Environment. This data might be collected from various sources, including real-world operations.
    \item {\bf Training Dataset:} A specific collection of Training Data used for the initial training or subsequent fine-tuning of an AI model within the Training Environment.
    \item {\bf Training Environment:} The environment where AI models undergo training and fine-tuning using datasets. This term includes data collected during real-world operations and scenarios with continuous or periodic fine-tuning.
\end{itemize}
\subsection{Impacts by Attacks}
\label{appendix:impacts}
The subsequent list provides detailed descriptions of each impact (1--11) resulting from the attacks on AI systems depicted in Figure~\ref{fig:impacts}, offering further clarity and supporting comprehensive understanding of their security implications.
\begin{enumerate}
    \item {\bf Model Leakage:} Unauthorized disclosure or extraction of sensitive internal details of an AI model, such as parameters, architectures, or hyperparameters, enabling attackers to replicate or exploit the model.
    \item {\bf Training Data Leakage:} Exposure or inference of sensitive training data or information derived from it, potentially violating data privacy and confidentiality.
    \item {\bf Application Information Leakage:} Disclosure of sensitive information related to the application's internal functionalities, configuration, or associated metadata, potentially enabling further attacks.
    \item {\bf Input Information Leakage:} Unauthorized extraction or inference of sensitive data provided as input by users or external systems, compromising user privacy or confidentiality.
    \item {\bf Internal Data Leakage:} Exposure or unauthorized extraction of confidential internal organizational data stored within or accessible by the AI system.
    \item {\bf Model Malfunction:} Alteration or degradation of AI model behavior resulting in incorrect, misleading, or degraded outputs, undermining reliability and trustworthiness.
    \item {\bf Interpretability Malfunction:} Deliberate distortion or corruption of the AI model’s explanation mechanisms, leading to incorrect or misleading interpretations of the model's decision-making processes.
    \item {\bf Safeguard Bypass:} Circumvention of protective measures designed to prevent misuse or harmful behavior of AI models, leading to potential unauthorized actions.
    \item {\bf System Compromise:} Complete or partial takeover of an AI system, potentially enabling attackers to execute unauthorized actions, escalate privileges, or facilitate subsequent cyberattacks.
    \item {\bf Internal Data Corruption:} Unauthorized alteration, manipulation, or deletion of internal data used by or stored within the AI system, compromising data integrity and availability, thus potentially affecting operational reliability.
    \item {\bf Computational Waste:} Intentional induction of excessive computational resource usage or processing time, causing increased operational costs, performance degradation, or denial of service conditions.
\end{enumerate}

\end{document}